\documentclass[11pt,english,twoside]{article}

\usepackage[T1]{fontenc}
\usepackage[latin1]{inputenc}
\usepackage[english]{babel}
\usepackage{lmodern}
\usepackage{a4wide}
\usepackage{amssymb, amsmath, amsthm}
\usepackage{slashed}
\usepackage{float}
\usepackage{graphicx}
\usepackage[dvips]{epsfig}
\usepackage{psfrag}
\usepackage{lscape}
\usepackage[all]{xy}
\usepackage{hyperref}
\usepackage{enumerate}
\usepackage{dsfont}
\usepackage{color}
\usepackage{mathabx}
\usepackage{mathtools}

\newcommand{\beq}{\begin{equation}}
\newcommand{\eeq}{\end{equation}}
\def\bea#1\eea{\begin{align}#1\end{align}}
\newcommand{\nn}{\nonumber}

\newcommand{\id}{\mathds{1}}
\newcommand{\ov}{\overline}

\def\del {\partial}
\def\d {{\rm d}}
\def\R {\mathcal{R}}
\def\L {\mathcal{L}}
\def\tL {\tilde{\mathcal{L}}}
\def\hhh {\mathcal{H}}

\def\tg {\tilde{g}}
\def\b {\beta}
\def\tp {\tilde{\phi}}
\def\te {\tilde{e}}

\def\p {\phi}

\def\cR {\widecheck{\cal R}}
\def\cG {\widecheck{\Gamma}}
\def\cN {\widecheck{\nabla}}
\def\g {\gamma}
\def\G {\Gamma}
\def\o {\omega}
\def\O {\Omega}

\def\hO {\hat{\O}}
\def\eps {\epsilon}
\def\la {\lambda}
\def\La {\Lambda}
\def\na {\nabla}

\def\T {\mathcal {T}}

\def\eee {\mathcal{E}}
\def\teee {\tilde{\mathcal{E}}}
\def\mmm {\mathcal{M}}
\def\reee {\mathring{\mathcal{E}}}
\def\Gg {\mathfrak{g}}

\makeatletter
\@addtoreset{equation}{section}
\makeatother

\begin{document}

\begin{titlepage}

\begin{center}

\rightline{\small MPP-2013-159}

\vskip 3cm

{\fontsize{16.1}{21}\selectfont \noindent\textbf{$\b$-supergravity: a
ten-dimensional theory\\ \vskip 0.25cm with non-geometric fluxes, and its
geometric framework} }

\vskip 2.1cm

\textbf{David Andriot, Andr\'e Betz}

\vskip 0.6cm

\textit{Max-Planck-Institut f\"ur Physik,\\F\"ohringer Ring 6, 80805
M\"unchen, Germany}

\vskip 0.2cm

{\small \texttt{andriot@mpp.mpg.de}, \texttt{abetz@mpp.mpg.de}}

\end{center}

\vskip 2.0cm

\begin{center}
{\bf Abstract}
\end{center}

\noindent We present a ten-dimensional theory, named $\b$-supergravity, that contains non-geometric fluxes and could uplift some four-dimensional gauged supergravities. Building on earlier work, we study here its NSNS sector, where $Q$- and $R$-fluxes are precisely identified. Interestingly, the $Q$-flux is captured in an analogue of the Levi-Civita spin connection, giving rise to a second curvature scalar. We reproduce the ten-dimensional Lagrangian using the Generalized Geometry formalism; this provides us with enlightening interpretations of the new structures. Then, we derive the equations of motion of our theory, and finally discuss further aspects: the dimensional reduction to four dimensions and comparison to gauged supergravities, the obtention of ten-dimensional purely NSNS solutions, the extensions to other sectors and new objects, the supergravity limit, and eventually the symmetries, in particular the $\b$ gauge transformation. We also introduce the related notion of a generalized cotangent bundle.

\vfill

\end{titlepage}

\tableofcontents

\newpage

\section{Introduction and main results}

Among the obstacles preventing to relate string theory to realistic low-energy physics, moduli stabilisation and de Sitter solutions from ten-dimensional supergravities appear as important problems. Full moduli stabilisation can be achieved, but seem to require, for generic situations, additional ingredients than those of a pure classical supergravity, such as non-perturbative corrections together with a precise control on a serie of approximations. Similarly, the standard set of ingredients makes de Sitter solutions very hard to find. When considering on top the question of stability in this context, the result is that no fully stable solution of pure classical ten-dimensional supergravity is known, while a few have been found when using the additional ingredients mentioned above. Finding a few particular backgrounds where these issues are resolved remains of great interest, but one would hope to have generically more tools at hand to find interesting solutions. Having more freedom should also be important
for the next steps on the path to a realistic model (finding the correct particle spectrum, etc.), which add more constraints.

Having non-trivial background fluxes of supergravity has already proven to help in the previous matters. They have been used extensively during the last decade to find new ten-dimensional solutions, that allow a compactification towards interesting four-dimensional supergravities. Some four-dimensional gauged supergravities nevertheless exhibit more ingredients than the ones obtained from standard compactifications. Of particular interest here are some specific terms in the super- and scalar potentials of these supergravities, generated by non-standard ``fluxes'' called the non-geometric fluxes. In the NSNS sector, these quantized objects are denoted as $Q_a{}^{bc}$ and $R^{abc}$. The position of their indices can be understood when viewing them as structure constants of the gauge algebra involved in the gauging \cite{Shelton:2005cf, Dabholkar:2002sy, Dabholkar:2005ve}, or equivalently as some components of the embedding tensor. These non-geometric fluxes and the corresponding terms in the potential cannot
be obtained from a compactification of a standard ten-dimensional supergravity. Rather, they were believed to be related to the less standard non-geometric backgrounds (ten-dimensional non-geometry \cite{Hellerman:2002ax, Dabholkar:2002sy, Flournoy:2004vn}), although no precise relation between the two concepts was established, until recent developments. For reviews and discussion on these topics, see \cite{Wecht:2007wu, Andriot:2011uh, Andriot:2013txa}.

These peculiar four-dimensional fluxes seem equally helpful for string phenomenology purposes as their counterparts that have a standard ten-dimensional origin. Indeed, some examples indicate that full moduli stabilisation can be achieved in their presence \cite{Shelton:2006fd, Micu:2007rd, Palti:2007pm}, and (metastable) de Sitter solutions can be found \cite{deCarlos:2009qm, Danielsson:2012by, Blaback:2013ht, Damian:2013dq, Damian:2013dwa, Catino:2013syn}. These non-geometric fluxes seem therefore to provide additional tools of interest for phenomenology. The caveat would be the lack of a ten-dimensional theory, allowing to get directly interesting backgrounds with such fluxes turned on, and reproducing by compactification the corresponding four-dimensional gauged supergravity. But such a theory is actually the topic of this paper.

We propose here a ten-dimensional theory that contains ten-dimensional non-geometric $Q$- and $R$-fluxes, and should provide an origin to some four-dimensional gauged supergravities. For reasons to be spelled-out, this theory can be thought of as a reformulation of the standard ten-dimensional supergravities, so we name it {\it $\b$-supergravity}. Nevertheless, we only study here its NSNS sector, which is common to all standard ten-dimensional supergravities. The completion to other sectors, discussed in section \ref{sec:beyond}, could therefore lead to distinguish several $\b$-supergravities. We denote by $\tL_{\b}$ the NSNS sector Lagrangian of this theory, and we obtain it at first using earlier work \cite{Andriot:2011uh, Andriot:2012wx, Andriot:2012an}, that we now detail.\\

Inspired by \cite{Grange:2006es, Grange:2007bp, Grana:2008yw} where Generalized Complex Geometry (GCG) tools were used to study non-geometry, we considered in \cite{Andriot:2011uh} a specific field redefinition to be performed on the standard NSNS fields. The metric $g_{mn}$, the Kalb-Ramond field $b_{mn}$, and the dilaton $\p$, get replaced by a new set of fields, given by a new metric $\tg_{mn}$, an antisymmetric bivector $\b^{mn}$, and a new dilaton $\tp$ ($m \dots z$ are the $d$-dimensional space-time curved indices). In GCG terms, this field redefinition is an $O(2d-2,2)$ transformation (more precisely here an $O(d-1,1) \times O(1,d-1)$, as detailed in appendix \ref{ap:K}) taking us from one generalized vielbein $\eee$ to another one $\teee$, while preserving the generalized metric $\hhh$, i.e. a change of generalized frame
\bea
& \eee= \begin{pmatrix} e & 0 \\ e^{-T} b & e^{-T} \end{pmatrix} \ , \ \teee= \begin{pmatrix} \te & \te \b \\ 0 & \te^{-T} \end{pmatrix} \ , \ \mathbb{I}= \begin{pmatrix} \eta_d & 0 \\ 0 & \eta_d^{-1} \end{pmatrix} \ , \label{genvielb}\\
& \hhh= \begin{pmatrix} g-b g^{-1} b & -b g^{-1} \\ g^{-1} b & g^{-1} \end{pmatrix} = \eee^T \ \mathbb{I} \ \eee = \teee^T \ \mathbb{I} \ \teee = \begin{pmatrix} \tg & \tg \b \\ - \b \tg &  \tg^{-1}-\b \tg \b \end{pmatrix} \ . \label{fieldredefH}
\eea
Here $\eta_d$ denotes the flat metric, and the vielbeins $e$ and $\te$ are associated to the respective metrics $g=e^T \eta_d e$ and $\tg=\te^T \eta_d \te$; we refer to appendix \ref{ap:conv} for more conventions.\footnote{With respect to the conventions of \cite{Andriot:2012an}, both $b$ and $\b$ get a minus sign. This does not affect at all the field redefinition: \eqref{fieldredef1} and \eqref{fieldredef2} remain unchanged. This modification allows us to match the conventions of \cite{Coimbra:2011nw}.} The field redefinition can be read from \eqref{fieldredefH} and rewritten in various manners, in particular
\beq
\begin{drcases} \tg^{-1}=(g+b)^{-1} g (g-b)^{-1}\\ \b= - (g+b)^{-1} b (g-b)^{-1} \end{drcases} \Leftrightarrow (g+b)^{-1}=(\tg^{-1}+\b) \ . \label{fieldredef1}
\eeq
We introduced additionally the new dilaton $\tp$, such that the following measure is preserved
\beq
e^{-2 \tp} \sqrt{|\tg|} = e^{-2 \p} \sqrt{|g|} = e^{-2d} \ , \label{fieldredef2}
\eeq
where $|\tg|$ denotes the absolute value of the determinant of the metric $\tg$. We will also make use of the shorthand $e^{-2d}$ just defined (the distinction with the dimension $d$ should be clear from the context).\footnote{An alternative field redefinition was proposed in \cite{Blumenhagen:2012nk, Blumenhagen:2012nt}. Both field redefinitions were then interpreted in terms of local $O(d,d)$ transformations and Lie algebroids in \cite{Blumenhagen:2013aia}.}

This field redefinition was then applied to the standard ten-dimensional NSNS Lagrangian given by
\beq
\L_{{\rm NSNS}} \equiv e^{-2\p} \sqrt{|g|} \left(\R(g) + 4(\del \p)^2 - \frac{1}{2} H^2 \right) \ , \label{LNSNS}
\eeq
where $\R$ denotes the Ricci scalar with Levi-Civita connection \eqref{def}, the $H$-flux is $H_{mnp}\equiv 3 \del_{[m}b_{np]}$, and the squares are defined in \eqref{squares}. The computation to obtain the Lagrangian $\tL$ after the field redefinition is rather involved, and has been performed in two steps. We considered in \cite{Andriot:2011uh} a simplifying assumption where the contraction $\b^{pq} \del_q$ acting on any object would be set to zero (as well as $\del_q \b^{pq}$). The new Lagrangian $\tL$ was then made of three terms: one involving the Ricci scalar of the new metric $\R(\tg)$, a standard kinetic term for the new dilaton $\tp$, and finally a square of the quantity $\del_m \b^{pq}$, which was identified with a ten-dimensional flux $Q_m{}^{pq}$. This identification was not only motivated by the correct index structure; we also verified that the dimensional reduction of this last term would generate the four-dimensional $Q$-flux term in the potential. Additionally, the use of this formula on a concrete background, called here the toroidal example, would give the expected value for the $Q$-flux (see section \ref{sec:torex}). In \cite{Andriot:2012wx, Andriot:2012an}, the full computation of $\tL$ without simplifying assumption was performed. A first method was a direct computation, starting from $\L_{{\rm NSNS}}$ and applying the field redefinition. It resulted in a new Lagrangian denoted here $\tL_0$, equal to $\L_{{\rm NSNS}}$ up to a total derivative
\bea
\tL_0 & = e^{-2\tp} \sqrt{|\tg|}\ \bigg( \R(\tg) +4(\del \tp)^2 -\frac{1}{2} R^2 \label{L0} \\
& \phantom{= e^{-2\tp} \sqrt{|\tg|}\ \ } + 4 \tg_{mn}\beta^{mp} \beta^{nq} \del_p d \ \del_q d -2 \del_p d \ \del_q \left( \tg_{mn} \b^{mp}\b^{nq} \right)  \nn \\
& \phantom{= e^{-2\tp} \sqrt{|\tg|}\ \ } -\frac{1}{4} \tg_{mp} \tg_{nq} \tg^{rs} \ \del_r \b^{pq} \ \del_s \b^{mn}+\frac{1}{2} \tg_{mn} \del_{p} \b^{qm}\ \del_q \b^{pn} \nn\\
& \phantom{= e^{-2\tp} \sqrt{|\tg|}\ \ } + \tg_{nq} \tg_{rs} \b^{nm} \big(\del_p \b^{qr}\ \del_m \tg^{ps} + \del_p \tg^{qr}\ \del_m \b^{ps} \big) \nn \\
& \phantom{= e^{-2\tp} \sqrt{|\tg|}\ \ } -\frac{1}{4} \tg_{mp} \tg_{nq} \tg_{rs} \big(\b^{ru}\beta^{sv} \del_u \tg^{pq}\ \del_v \tg^{mn} -2 \b^{mu}\b^{nv} \del_u \tg^{qr}\ \del_v \tg^{ps} \big) \bigg) \nn\\
& = \L_{{\rm NSNS}} - \del_m\left(e^{-2d}\big(\tg^{mn}\tg^{pq}\del_n\tg_{pq} - g^{mn} g^{pq} \del_n g_{pq} + \del_n(\tg^{mn}-g^{mn})\big)\right) \ . \nn
\eea
We recover the terms mentioned for \cite{Andriot:2011uh}: these are the first two terms, and the first term of the third line. There are also new terms, in particular the square defined in \eqref{squares} of a ten-dimensional $R$-flux given by
\beq
R^{mnp}\equiv 3 \b^{q[m}\del_q \b^{np]} = 3 \b^{q[m}\na_q \b^{np]} \ , \label{fluxes}
\eeq
as in \cite{Aldazabal:2011nj}. The $\na_m$ is the standard covariant derivative with Levi-Civita connection defined in \eqref{def}; this $R$-flux is therefore a tensor. A second, faster, method \cite{Andriot:2012wx, Andriot:2012an} to obtain this result \eqref{L0} made use of Double Field Theory (DFT) \cite{Hull:2009mi, Hull:2009zb, Hohm:2010jy, Hohm:2010pp}. In DFT, fields a priori depend on two sets of coordinates: the standard $x^m$ and the ``dual'' $\tilde{x}_m$. But the so-called strong constraint has then to be imposed, which effectively reduces locally the dependence to only half of the doubled coordinates. We impose here this constraint by setting to zero the derivative with respect to $\tilde{x}$, i.e. $\tilde{\del}=0$. A known result of DFT is that applying the strong constraint to its (NSNS) Lagrangian $\L_{{\rm DFT}}$ allows to recover the standard NSNS Lagrangian $\L_{{\rm NSNS}}$, up to a total derivative $\del(\dots)$. Given this relation, we performed the field redefinition directly within
DFT. This could be done easily
thanks to the
specific
form \eqref{fieldredef1} of this redefinition and the properties of DFT. On the resulting Lagrangian, we applied the same
constraint $\tilde{\del}=0$ to go back to the supergravity level, and obtained again $\tL_0$, up to a total derivative. These two methods are depicted by the two left columns and lines of the diagram \eqref{diagram}, while we now turn to its last column. Note that the plain equalities of this diagram were established in \cite{Andriot:2011uh, Andriot:2012wx, Andriot:2012an}, and the dashed ones are obtained here.
\beq
 \xymatrix{ \L_{{\rm DFT}}(g,b,\p)\ \ar@{=}[r] & \ \L_{{\rm DFT}}(\tg, \b, \tp) \ \ar@{=}[r] & \ \L_{{\rm DFT}}(\R,\cR) + \del(\dots) + \tilde{\del}(\dots) \\
 \L_{{\rm NSNS}} + \del(\dots)\ \ar@{=}[u]^{\tilde{\del}=0\ } \ar@{=}[r] &\ \tL_0 + \del(\dots) \ \ar@{=}[u]^{\tilde{\del}=0\ } \ar@{==}[r] & \ \tL_{\b} + \del(\dots) \ \ar@{==}[u]^{\tilde{\del}=0\ }  }  \label{diagram}
\eeq

The ten-dimensional theory given by the Lagrangian $\tL_0$ was proposed as a ten-dimensional uplift to some four-dimensional gauged supergravities. In particular, a (partial) dimensional reduction performed in \cite{Andriot:2012an} showed that the $Q$- and $R$-flux non-geometric terms of the four-dimensional scalar potential could be reproduced (while this cannot be achieved from the standard $\L_{{\rm NSNS}}$). The precise identification of the fluxes themselves remained nevertheless unclear. The scalar potential is essentially given by a sum of terms quadratic in the fluxes, so the identification of the latter can be difficult. It would rather require an exact reproduction of the superpotential, where the fluxes appear linearly. Nevertheless, the $R$-flux term of the scalar potential was obtained only from the $R^2$ term in $\tL_0$, i.e. from the square of the ten-dimensional $R$-flux; this makes the identification of the $R$-flux \eqref{fluxes} with its four-dimensional counterpart rather likely. On the
contrary, the $Q$-flux term of the scalar potential was more cumbersome: it was obtained mostly by the last three lines of $\tL_0$ in \eqref{L0}, which do not exhibit a clear structure, preventing the identification of the good formula for a ten-dimensional $Q$-flux. As discussed above, we know it should still involve $\del\b$, as this was the correct result for the subcase considered in \cite{Andriot:2011uh}. But we did not find a completion to a more general expression, which would make the last lines of $\tL_0$ in \eqref{L0} more structured; for instance, the square of $\na \b$ did not work. It could have allowed, through the $Q$-flux term of the scalar potential, the identification with the four-dimensional $Q$-flux.

Some progress in the structure of the Lagrangian were nevertheless obtained in \cite{Andriot:2012wx, Andriot:2012an}, at the level of DFT. Indeed, as depicted in the first line of the diagram \eqref{diagram}, the DFT Lagrangian, expressed in terms of the new fields $\tg, \b, \tp$, was reformulated in a covariant manner with respect to half of the doubled diffeomorphisms (those corresponding to transformations of the standard $x^m$). This brought a clearer structure to the various terms. The key ingredient of this reformulation was a new covariant derivative\footnote{In \cite{Andriot:2012wx, Andriot:2012an}, this covariant derivative was denoted $\tilde{\na}^m$.} that we denote here $\cN^m$, involving the derivative $\tilde{\del}^m - \b^{mn} \del_n$ and a connection $\cG_p^{mn}$. It allowed to build, at the level of DFT, various objects, in particular a new ``Ricci tensor'' $\cR^{mn}$ and associated scalar $\cR$. The latter entered the reformulated DFT Lagrangian, together with the
standard Ricci scalar $\R (\tg)$. Here, we apply the constraint $\tilde{\del}=0$ on this last DFT Lagrangian $\L_{{\rm DFT}}(\R,\cR)$ to go back to the supergravity level. Doing so, we obtain a first expression of $\tL_{\b}$, which formally looks very similar to $\L_{{\rm DFT}}(\R,\cR)$ as we inherit its structure
\beq
\tL_{\b} = e^{-2\tp} \sqrt{|\tg|}\ \bigg( \R(\tg) + \cR(\tg) +4(\del \tp)^2 -\frac{1}{2} R^2 + 4 (\b^{mp}\del_p \tp - \T^m)^2 \bigg) \ .\label{L2}
\eeq
The squares are defined in \eqref{squares}, and the objects involved here, together with the new covariant derivative on a (co)-vector $V$, are defined as
\bea
& \cR=\tg_{mn} \cR^{mn} \ , \ \cR^{mn} = -\b^{pq}\del_q \cG_p^{mn} + \b^{mq}\del_q \cG_p^{pn} + \cG_p^{mn} \cG_q^{qp} - \cG_p^{qm} \cG_q^{pn} \ ,\label{cR}\\
& \cG_p^{mn} = \frac{1}{2}\tg_{pq}\left(-\b^{mr}\del_r \tg^{nq} - \b^{nr}\del_r \tg^{mq} +  \b^{qr}\del_r \tg^{mn} \right) + \tg_{pq} \tg^{r(m} \del_r \b^{n)q}-\frac{1}{2} \del_p \b^{mn}  \ ,\label{cG}\\
& \T^n \equiv \cG_p^{pn} = \del_p \b^{np} - \frac{1}{2} \b^{nm} \tg_{pq} \del_m \tg^{pq} = \na_p \b^{np} \ , \label{tracecG} \\
& \cN^m V^p = -\b^{mn} \del_n V^p - \cG^{mp}_n V^n \ , \ \cN^m V_p = -\b^{mn} \del_n V_p + \cG^{mn}_p V_n \ . \label{defcN}
\eea
The trace $\T^n$ of the connection is a tensor, as noticed in \cite{Andriot:2012wx, Andriot:2012an}; its expression as $\na_p \b^{np}$ makes it obvious. The definition of $\cN$ is naturally extended for tensors with more indices. We come back in more details to these objects and their properties in section \ref{sec:calc}.

By construction, $\tL_{\b}$ should be equal to $\tL_0$, and to $\L_{{\rm NSNS}}$, up to total derivatives. This is depicted in the second line of the diagram \eqref{diagram}. We verify this explicitly in appendix \ref{ap:tLflat} (see \eqref{relLag}). Compared to $\tL_0$, this reformulation $\tL_{\b}$ brings a nicer structure to the ten-dimensional Lagrangian: it is manifestly diffeomorphism covariant. In addition, as noticed already at the level of DFT in \cite{Andriot:2012an}, $\cR$ captures most of the terms of the last three lines of $\tL_0$, i.e. most of the four-dimensional $Q$-flux term. Still, this interesting repackaging does not allow to identify directly the $Q$-flux. It was nevertheless pointed out in \cite{Andriot:2012wx, Andriot:2012an} that the $\del \b$ essentially appear within the new connection $\cG$. This lead us to think that the $Q$-flux should be understood as a connection coefficient. It means that it is not a tensor, on the contrary to the $R$-flux \eqref{fluxes}. This remark has
important consequences as we will see.\\

Let us recall that the four-dimensional $Q$- and $R$-flux are structure constants in an algebra of gauged supergravity. For a standard structure constant $f^a{}_{bc}$ of a Lie algebra, there is a geometric interpretation when considering the corresponding Lie group as a manifold, and using the Maurer-Cartan equations. The structure constant is then expressed in terms of vielbeins $\te^a{}_m$ and their inverse $\te^n{}_b$ as in \eqref{fabc}, and is related to the antisymmetric part of the spin connection $\o^a_{bc}\equiv \te^m{}_b\ \o_m{}^a{}_c$ (see appendix \ref{ap:conv} for more conventions). This $f$ can also appear in the algebra of four-dimensional gauged supergravity; given its geometric definition, its indices are then interpreted in ten dimensions as being the flat ones $a \dots l$. We deduce that the non-geometric fluxes should be denoted as $Q_a{}^{bc}$ and $R^{abc}$, and if they have any ten-dimensional counterparts, their indices should also be understood as being flat. For a tensor, such as the
$R$-flux \eqref{fluxes}, going from flat indices to curved ones is simply a multiplication by vielbeins. But if the ten-dimensional $Q$-flux is not a tensor as argued above, there is not such a direct relation between its expression in flat, and one in curved indices like $\del\b$ or more. Identifying the ten-dimensional $Q$-flux looks thus even more difficult. Two ideas nevertheless come to mind in order to find the expression of this ten-dimensional $Q$-flux with flat indices. First, various proposals for such an expression have been made in the literature, in particular in \cite{Grana:2008yw, Aldazabal:2011nj} where one finds\footnote{A factor $2$ is missing in this formula in \cite{Aldazabal:2011nj} (typo; private communication). It has been corrected in \cite{Geissbuhler:2013uka}.}
\beq
Q_{c}{}^{ab} \equiv \del_c \b^{ab} - 2 \b^{d[a} f^{b]}{}_{cd} \ .\label{Q}
\eeq
One can then simply rewrite the above $\tL_0$ or $\tL_{\b}$ into flat indices, using $\te^a{}_m$ and making this quantity $Q_{c}{}^{ab}$ appear, and see if the introduction of this $Q$ gives to the Lagrangian a nice structure. It turns out to be the case, as shown in appendix \ref{ap:tLflat}, and the reason for this is rather unexpected: it is related to the second idea to make a flat $Q$-flux appear, that we now detail. There is one non-tensorial quantity for which there is a principle to go from a curved expression to a flat one: the connection associated to a covariant derivative. Indeed, the spin connection is defined with respect to the standard $\na$ and its connection $\G$ as
\beq
\te^a{}_m \te^n{}_b \na_n V^m = \na_b V^a \equiv \del_b V^a + \o^a_{bc} V^c \Leftrightarrow \o^a_{bc} \equiv \te^n{}_b \te^a{}_m \left(\del_n \te^m{}_c + \te^p{}_c \G^m_{np} \right) \label{defo}
\eeq
For the Levi-Civita connection, one can then show using the definition \eqref{fabc} of $f$
\beq
\o^a_{bc} = \frac{1}{2} \left(f^{a}{}_{bc} + \eta^{ad} \eta_{ce} f^{e}{}_{db} + \eta^{ad} \eta_{be} f^{e}{}_{dc} \right) \ . \label{defof}
\eeq
Proceeding similarly for the new $\cN$ leads us to introduce $\o_Q$, (the opposite of) the flat connection associated to $\cG$
\bea
& \!\!\!\!\!\! \te^m{}_a \te^b{}_n \cN^n V_m = \cN^b V_a \equiv -\b^{bd}\del_d V_a - {\o_Q}^{bc}_a V_c \Leftrightarrow -{\o_Q}^{bc}_a \equiv \te^b{}_n \te^m{}_a \left(-\b^{nq}\del_q \te^c{}_m + \te^c{}_p \cG^{np}_m \right) \label{defoQcG}\\
& \qquad \qquad \qquad \qquad \qquad \! {\o_Q}^{bc}_a = \frac{1}{2} \left(Q_{a}{}^{bc} + \eta_{ad} \eta^{ce} Q_{e}{}^{db} + \eta_{ad} \eta^{be} Q_{e}{}^{dc} \right)  \ , \label{defoQQ}
\eea
where \eqref{defoQQ} is obtained by using the definition \eqref{cG} of $\cG$ and the $Q$-flux given by the proposal \eqref{Q}! This result explains why the introduction of this precise $Q_{c}{}^{ab}$ in $\tL_{\b}$ still gives a nice structure. In addition, \eqref{defoQcG} and \eqref{defoQQ} realise completely the idea that the $Q$-flux should be understood as a connection: by comparing with \eqref{defof}, we see that $Q_{c}{}^{ab}$ is the analogue to $f^a{}_{bc}$.

We then mimic the definition of the Ricci scalar in terms of $\o$ or $f$ and introduce the analogous quantity $\R_Q$
\bea
& \R(\tg)=2 \eta^{bc} \del_a \o^a_{bc} + \eta^{bc} \o^a_{ad} \o^d_{bc} - \eta^{bc} \o^a_{db} \o^d_{ac} \label{Ricflat}\\
& \qquad \qquad \qquad \qquad \qquad = 2 \eta^{ab} \del_a f^c{}_{bc} - \eta^{cd} f^a{}_{ac} f^b{}_{bd} - \frac{1}{4} \left( 2 \eta^{cd} f^a{}_{bc} f^b{}_{ad} + \eta_{ad} \eta^{be} \eta^{cg} f^a{}_{bc} f^d{}_{eg} \right) \ ,\nn \\
& \R_Q \equiv 2 \eta_{bc} \b^{ad} \del_d {\o_Q}_a^{bc} + \eta_{bc} {\o_Q}_a^{ad} {\o_Q}_d^{bc} - \eta_{bc} {\o_Q}_a^{db} {\o_Q}_d^{ac} \label{defRQ} \\
& \qquad \qquad \qquad \qquad = 2 \eta_{ab} \b^{ad}\del_d Q_c{}^{bc} - \eta_{cd} Q_a{}^{ac} Q_b{}^{bd} - \frac{1}{4} \left( 2 \eta_{cd} Q_a{}^{bc} Q_b{}^{ad} + \eta^{ad} \eta_{be} \eta_{cg} Q_a{}^{bc} Q_d{}^{eg} \right) \ .\nn
\eea
We show in appendix \ref{ap:tLflat} that it is related to $\cR$ as follows\footnote{\label{foot:cRRQ}It may look surprising that $\cR$ and $\R_Q$ are not the same. We comment on this point in footnote \ref{foot:cRRQap}. The difference, given by the term in $R^{acd} f^b{}_{cd}$, actually plays a crucial role in sections \ref{sec:4d} and \ref{sec:NSNSsol}.}
\beq
\cR= \R_Q - \frac{1}{2} R^{acd} f^b{}_{cd} \eta_{ab}\ . \label{cRRqRf}
\eeq
It is then straightforward to rewrite the ten-dimensional Lagrangian $\tL_{\b}$ \eqref{L2} into flat indices
\beq
\fbox{
$\tL_{\b}  = e^{-2d} \ \bigg( \R(\tg) +4(\del \tp)^2 + 4 (\b^{ab}\del_b \tp - \T^a)^2 + \R_Q - \frac{1}{2} R^{acd} f^b{}_{cd} \eta_{ab} -\frac{1}{2} R^2  \bigg) $
} \label{Lflat}
\eeq
We finally have at hand a ten-dimensional theory with a $Q$-flux precisely identified. After dimensional reduction, the comparison to four-dimensional gauged supergravities is then eased, especially with the ten-dimensional $Q$-flux \eqref{Q} and its four-dimensional counterpart matching. We come back to this comparison in section \ref{sec:4d}, and argue that the potential obtained from $\tL_{\b}$ \eqref{Lflat} should be the correct one.\\

In section \ref{sec:GGderiv} and appendix \ref{ap:GG}, we rederive this Lagrangian $\tL_{\b}$ \eqref{Lflat} from the Generalized Geometry (GG) formalism. We essentially follow the procedure of \cite{Coimbra:2011nw}, but use the generalized vielbein $\teee$ with $\b$ instead of $\eee$ with $b$, as given in \eqref{genvielb}. We study at first the $O(d,d) \times \mathbb{R}^+$ structure and associated covariant derivatives. In that context, the combination $\b^{ab} \del_b$ and the connection $\o_Q$ appear naturally, enlightening from a different perspective the origin of the new derivative $\cN^a$. Additionally, $\b^{ab}\del_b \tp - \T^a$ entering the non-standard dilaton term in $\tL_{\b}$, and the tensor $\T^a$ in particular, get the interesting interpretation of a conformal weight; this point of view is useful later on. We then consider the preservation of an $O(d-1,1) \times O(1,d-1)$ (sub)-structure. We obtain the corresponding derivatives on vectors
\beq
D_A W^B = \begin{cases} D_a w^b = \nabla_a w^b - \eta_{ad} \cN^d w^b + \frac{1}{6} \eta_{ad} \eta_{cf} R^{dbf} w^c - \frac{1}{9} (\delta^b_a \Lambda_c - \eta_{ac} \eta^{be} \Lambda_e) w^c \\
           D_a w^{\ov{b}} = \nabla_a w^{\ov{b}} - \eta_{ad} \cN^d w^{\ov{b}} - \frac{1}{2} \eta_{ad} \ov{\eta_{cf}} R^{d\ov{bf}} w^{\ov{c}} \\
           D_{\ov{a}} w^b = \nabla_{\ov{a}} w^b + \ov{\eta_{ad}} \cN^{\ov{d}} w^b - \frac{1}{2} \ov{\eta_{ad}} \eta_{cf} R^{\ov{d}bf} w^c \\
           D_{\ov{a}} w^{\ov{b}} = \nabla_{\ov{a}} w^{\ov{b}} + \ov{\eta_{ad}} \cN^{\ov{d}} w^{\ov{b}} + \frac{1}{6} \ov{\eta_{ad}} \ov{\eta_{cf}} R^{\ov{dbf}} w^{\ov{c}} - \frac{1}{9} (\delta^{\ov{b}}_{\ov{a}} \Lambda_{\ov{c}} - \ov{\eta_{ac}} \ov{\eta^{be}} \Lambda_{\ov{e}}) w^{\ov{c}}
          \end{cases}\ , \label{DWbubfinalintro}
\eeq
where the various fluxes and derivatives enter; the quantity $\Lambda_c$ is related to the dilaton and $\T^a$, and we refer to section \ref{sec:OOstruct} for more conventions. Considering a related $Spin(d-1,1)\times Spin(1,d-1)$ structure with spinors $\epsilon^{\pm}$, we deduce the corresponding spinorial derivatives; those play two important roles. First, they could provide the Killing spinor equations appearing in a future supersymmetric completion of $\b$-supergravity \cite{Future}. Second, they allow us to compute the scalar $S$ given by
\beq
S\eps^+= - 4 \left( \g^a D_a\g^b D_b- \ov{\eta^{ab}} D_{\ov{a}}D_{\ov{b}} \right)\eps^+ \  .\label{defSintro}
\eeq
This scalar has been related \cite{Coimbra:2011nw, Hohm:2011nu}, up to a total derivative, to the standard $\L_{{\rm NSNS}}$ \eqref{LNSNS} when using the generalized vielbein $\eee$ with $b$-field \eqref{genvielb}; here we eventually obtain
\beq
S= e^{2d} \ \tL_{\b} + e^{2d} \ \del(\dots) \ ,\label{SLbeta}
\eeq
where the total derivative is given in \eqref{Stotder}.

We then derive in section \ref{sec:eom} and appendix \ref{ap:eombeta} the equations of motion from $\tL_{\b}$ \eqref{L2}. This requires to establish a few more interesting differential properties of the objects introduced above. One of them, crucial for the equation of motion for $\b$, is the rewriting of the connection coefficient $\cG^{mn}_p$ \eqref{cG} as
\beq
\cG^{mn}_p= \cG_{\!\!(t)}{}_p^{mn} + \b^{ms} \G^n_{ps} \ , \qquad \cG_{\!\!(t)}{}_p^{mn}= \frac{1}{2} \tg_{pq} \left( \tg^{rm} \na_r \b^{nq} + \tg^{rn} \na_r \b^{mq} - \tg^{qr} \na_r \b^{mn} \right) \ . \label{cG_t}
\eeq
It is given by the sum of a tensor $\cG_{\!\!(t)}$ and a non-tensorial second term depending on the standard connection $\G$ defined in \eqref{def}. This has several important consequences, among which a relation \eqref{relnacN} between the covariant derivatives $\na$ and $\cN$, that allows to rewrite the $R$-flux \eqref{fluxes} as
\beq
R^{mnp} = 3\ \b^{q[m}\na_q \b^{np]} = \frac{3}{2}\ \cN^{[m} \b^{np]} \ . \label{RfluxcN}
\eeq
The dilaton, Einstein, and $\b$ equations of motion are then respectively given by
\bea
 & \! \frac{1}{4} \left(\R(\tg) + \cR(\tg) -\frac{1}{2} R^2 \right) = (\del \tp)^2 - \nabla^2 \tp + (\b^{mr}\del_r \tp - \T^m)^2 + \tg_{mn}\cN^m(\b^{nr}\del_r \tp - \T^n) \label{dileom}\\
 & \!\!\!\!\!\!\!\!\!\!\!\!\! \R_{pq} - \tg_{m(p} \tg_{q)n} \cR^{mn} + \frac{1}{4} \tg_{pm} \tg_{qn} \tg_{rs} \tg_{uv} R^{mru} R^{nsv} = - 2 \nabla_p \del_q \tp  - 2 \tg_{m(p} \tg_{q)n} \cN^m(\b^{nr} \del_r \tp - \T^n) \label{Einstein}\\
&\! \frac{1}{2} \tg_{ms} \tg_{ru} \tg_{np}  \left( e^{2\tp} \cN^m (e^{-2\tp}  R^{sun}) - 2 \T^m R^{sun}\right) \label{beom} \\
& \!\!\!\!\!\!\!\!\! = \frac{1}{2} \tg_{np} \tg_{rq} \tg^{sm} e^{2\tp} \na_m (e^{-2\tp} \na_s \b^{nq} ) + 2 \tg_{n[p} \R_{r]s} \b^{ns} - e^{-2\tp} \na_q ( e^{2\tp} \tg_{n[p} \na_{r]} \b^{nq}) + 4 \tg_{n[p} \na_{r]} (\b^{nq} \del_q \tp )  \nn \ .
\eea
The equation of motion for $\b$ is on the one hand analogous to that of the $b$-field, but has on the other hand the new feature of depending on the standard Ricci tensor. This structure is reminiscent of its two main origins, the $R$-flux and $\R_Q$.\footnote{The equations of motion \eqref{dileom} - \eqref{beom} should reproduce those obtained in \cite{Andriot:2011uh} as a subcase. We come back to this point at the end of appendix \ref{ap:eombeta}. Also, it should be possible to derive \eqref{dileom} - \eqref{beom} from the equations of motion of \cite{Blumenhagen:2013aia}. Finally, \cite{Hassler:2013wsa} gives the $Q$-brane as a solution to these equations of motion. It does not fit however the study made in section \ref{sec:NSNSsol}, as it contains a warp factor and a non-constant dilaton.}

Finally, we present in section \ref{sec:BeyondLbeta} further aspects of our theory. We first perform a dimensional reduction of $\tL_{\b}$ to four dimensions and discuss its relation to gauged supergravities. We study the possibility of getting purely NSNS ten-dimensional solutions to be used for compactification. We present ideas to extend $\tL_{\b}$ to a complete $\b$-supergravity and beyond. We then discuss the supergravity limit and the relation to non-geometry, illustrating these points with the toroidal example. We also introduce the notion of a {\it generalized cotangent bundle} $E_{T^*}$, that appears as the relevant bundle to consider here; comparing its transition functions and the symmetries of $\tL_{\b}$, among which the $\b$ gauge transformation that we present, helps to distinguish geometry from non-geometry. While we provide first building blocks to study these various points, they would deserve a more detailed investigation; this section \ref{sec:BeyondLbeta} thus offers an outlook to the present work.

\section{Generalized Geometry derivation of $\tL_{\b}$}\label{sec:GGderiv}

\subsection{Preliminary discussion on GG and DFT: formalism and geometry}\label{sec:preldisc}

The recent years have seen the development of $O(d,d)$ covariant formalisms, which describe to some extent the target space effective dynamics of strings. Generalized Geometry (GG), introduced in \cite{Coimbra:2011nw}, is one of them. It finds its origins in Generalized Complex Geometry (GCG) introduced by Hitchin \cite{Hitchin:2004ut} and Gualtieri \cite{Gualtieri:2003dx}, a mathematical set-up that has been used to study string theory and supergravity with background fluxes \cite{Koerber:2010bx}. Another formalism is Double Field Theory (DFT) \cite{Hull:2009mi, Hull:2009zb, Hohm:2010jy, Hohm:2010pp} mentioned above; a recent review can be found in \cite{Aldazabal:2013sca} (see also \cite{Berman:2013eva}).\footnote{Stringy differential geometry introduced in \cite{Jeon:2010rw, Jeon:2011cn} is a third formalism based on a projection-compatible semi-covariant derivative. Its geometric objects coincide after projection with the corresponding quantities in DFT. In addition, we refer the reader to \cite{
Coimbra:2011nw} for work related to the GG formalism in the mathematics literature.} These two formalisms consider ``generalized'' geometric objects such as the generalized metric and vielbeins, generalized covariant derivatives with associated connection, a generalized Ricci tensor and a generalized curvature scalar $S$, etc. These various objects transform under $O(d,d)$ (or a subgroup of it). The two formalisms also have a Lagrangian which can be formulated in terms of these objects. If one considers DFT in a ``conservative'' manner, meaning that one strictly enforces there the strong constraint, then the two formalisms are totally equivalent; this is the point of view we adopt here.\footnote{The strong constraint is a priori local; therefore the two formalisms are at least equivalent on each patch. For this to remain true globally, one should have the ``same'' strong constraint on every patch, for instance $\tilde{\del}=0$. This is for example possible for geometric backgrounds. In what follows, we will
mainly focus on local quantities such as the Lagrangian; we will also comment on the global aspects.} A way to see their equivalence is that their (NSNS) Lagrangian are then equal, up to a total derivative, to the standard $\L_{{\rm NSNS}}$ \eqref{LNSNS}, when choosing the particular generalized vielbein $\eee(e,b)$ given in \eqref{genvielb}. For DFT, this equality can be seen in the left column of the diagram \eqref{diagram}; for GG, the scalar $S$ is then related to $\L_{{\rm NSNS}}$.

As explained from \eqref{genvielb} to \eqref{fieldredef1}, the field redefinition considered here corresponds to a change of generalized vielbein, from $\eee (e,b)$ to $\teee (\te,\b)$. Such a change in the two above formalisms should thus lead to $\tL_{\b}$ \eqref{L2}. Although this result is already known for DFT as depicted in the diagram \eqref{diagram}, it has not been established in a formulation where one defines the ``generalized'' geometric objects just mentioned. Equivalently, this relation has not been derived within the GG formalism, that only relies on these objects. We thus show here how choosing the generalized vielbein $\teee (\te,\b)$ in the GG formalism leads to $S= e^{2d}\ (\tL_{\b} + \del(\dots))$. Doing so, the geometric objects corresponding to the choice $\teee$ are defined explicitly, and this enlightens the structures appearing. For instance, the derivative $\cN$ \eqref{defcN} and connection $\o_{Q}$ \eqref{defoQQ} appear naturally. The role of the trace of the connection $\T^m$ \eqref{tracecG} also gets clarified. Finally we obtain specific derivatives on spinors which will be useful in the study of supersymmetry \cite{Future}.

To establish our result and construct all necessary objects, we essentially follow the GG paper \cite{Coimbra:2011nw} that treats $\eee (e,b)$. An analogous DFT formalism with similar objects was developed before in \cite{Hohm:2010xe}, and its relation to \cite{Coimbra:2011nw} has been established in \cite{Hohm:2011nu}, that is also helpful here. These three papers are, to some extent, based on the early work \cite{Siegel:1993xq, Siegel:1993th}. Previous constructions of geometric objects in terms of generalized vielbeins can also be found in \cite{Grana:2008yw, Jeon:2011cn}. More recent related work for the $O(d,d)$ covariant formalisms appeared in \cite{Hohm:2011si, Hohm:2012mf, Berman:2013uda, Geissbuhler:2013uka, Garcia-Fernandez:2013gja}, where a specific form of the generalized connection is sometimes chosen. Most of this recent work remains however at the generalized or doubled level, without an explicit choice of generalized vielbein (it would break $O(d,d)$). One exception is \cite{Geissbuhler:2013uka}; we recover here some of their generalized connection components.\\

GG and DFT nevertheless differ, not at the level of the formalism, but when it comes to the underlying space or geometry. On the one hand, GG, at least with the vielbein $\eee (e,b)$, is based on the generalized tangent bundle $E_T$ defined in GCG. Given a manifold $\mmm$ of dimension $d$, this bundle is locally the direct sum $T\mmm \oplus T^* \mmm$. Globally, it should rather be thought of as a fibration of the cotangent bundle over the tangent bundle of $\mmm$, where the non-trivial fibration is encoded in the $b$-field; the latter can also be viewed as a connection on a gerbe \cite{Hitchin:1999fh}. On the other hand, DFT is based on a doubled space along the doubled coordinates $(x^m, \tilde{x}_m)$ discussed in the Introduction. This doubled geometry, a generalization of a T-fold \cite{Hull:2004in}, was at first \cite{Dabholkar:2005ve} thought of as a space of dimension $2d$ that could, in some cases, boil down to an ordinary manifold of doubled dimension with (doubled) diffeomorphisms; more generically
the latter would extend to $O(d,d)$. The
recent work \cite{Hohm:2012gk, Park:2013mpa} however indicates that the coordinate transformations are more general, and the nature of this doubled space is thus still under investigation. Interesting proposals on the geometry of DFT have also been made in \cite{Vaisman:2012ke, Vaisman:2012px, Vaisman:2013gga}.

DFT is based on this new type of space of dimension $2d$, while GG with vielbein $\eee$ has the generalized tangent bundle constructed over a manifold of dimension $d$. The underlying geometry of the two formalisms is then different, as well as the interpretation of the $O(d,d)$ action and of the ``generalized'' geometric objects introduced. We construct here these objects for the generalized vielbein $\teee$; understanding them not only formally, but also geometrically, can then be subtle given those differences. In addition, the underlying geometry for GG can be a particularly delicate question when choosing the generalized vielbein $\teee$. Indeed, as we explain in more details in section \ref{sec:GGOdd}, fixing the generalized vielbein to be $\eee$ breaks the structure group of the bundle from $O(d,d)$ to a subgroup denoted $G_{{\rm split}}$. The latter gets a priori fixed differently by different vielbeins, so the resulting bundle, constructed (globally) with this subgroup, would change accordingly.
While $\eee$ gives the generalized tangent bundle $E_T$, we propose in section \ref{sec:ET*} that $\teee$ defines a ``generalized cotangent bundle'' $E_{T^*}$: it should be viewed as the tangent bundle fibered over the cotangent bundle of $\mmm$, as already suggested in \cite{Grana:2008yw}. We only sketch this possibility though, and do not pretend to fully answer this point in the paper. These questions certainly deserve to be studied further, while so far, the underlying geometries of both formalisms are not completely understood. We come back in more details to this discussion in sections \ref{sec:GGOdd} and \ref{sec:ET*}, and give more hints about the geometry. Our first goal remains however to work with GG and DFT at a formal level, and rederive, from the objects defined, the Lagrangian $\tL_{\b}$.

\subsection{Generalized Geometry formalism: $O(d,d) \times \mathbb{R}^+$ structure and examples}\label{sec:GGOdd}

Let us first consider a manifold $\mmm$ of dimension $d$. Over each patch of the manifold, one usually defines a tangent space, equipped with a frame that we denote here $\del_a$ (a common notation is also $e_a$). When going from one patch to the next, the frames transform into each other with elements of $GL(d,\mathbb{R})$ (acting on the ``flat'' index $a$ running on the $d$ range). The consequent (global) union of the tangent spaces is the tangent bundle $T\mmm$, whose structure group is then $GL(d,\mathbb{R})$. If there is additionally a metric $\eta_d$ on these tangent spaces, and this metric is required to be the same on all of them, i.e. it is preserved globally, then the structure group is reduced to $O(d-1,1)$.\footnote{We use in this paper this standard Minkowski signature for clarity, but there is actually no restriction on it, as indicated in \cite{Coimbra:2011nw}.} We now consider a $2d$-dimensional generalization of the tangent bundle, that we call the {\it generalized bundle} $E$. It is
built
similarly over a set of patches, with (generalized) frames that transform thanks to a structure group $O(d,d)$. The latter acts on a (generalized flat) index $A$ that runs on the $2d$ range, and the $2d \times 2d$ metric
\beq
\eta_{(u/d)}=\frac{1}{2} \begin{pmatrix} 0 & \id \\ \id & 0 \end{pmatrix} \ ,\label{etaud}
\eeq
of components denoted $\eta_{AB}$, is preserved. A simple example of a local realisation of $E$ is given by the direct sum $T \mmm \oplus T^*\mmm$. Indeed, an $O(d,d)$ action is natural for this direct sum, since vectors and one-forms $(\del_a, e^b)$ couple with the metric \eqref{etaud}.

In order to take care of the dilaton, we extend the bundle $E$ by a conformal weight, following \cite{Coimbra:2011nw}. The structure group of this extended bundle is now $O(d,d) \times \mathbb{R}^+$, and the different objects involved get weighted by a conformal factor related to the dilaton. In particular, we now talk of a (generalized) conformal frame, that we denote $e^{-2d}\ \reee_A$ (the top ring notation is used here for generic quantities).

In the following, we are interested in a particular type of generalized frames: those that allow a splitting of the generalized bundle. We mean here by splitting that one can find an isomorphism $E \simeq T \mmm \oplus T^*\mmm$. As mentioned above, this is locally a natural relation, but it can be globally non-trivial; we will come back to that point. Such an isomorphism implies locally a map from a generalized frame to $(\del_a,e^a)$. In addition, one can locally choose a set of coordinates so that a vielbein $\mathring{e}$ relates $\del_a = \mathring{e}^m{}_a \del_m$, and similarly for the one-forms with $\d x^m$. In matrix notation, this relation gives $\del_a$ as $\mathring{e}^{-T} \del$ (see appendix \ref{ap:conv}). So locally, a generalized conformal frame $e^{-2d}\ \reee_A$ that allows a splitting can be denoted
\beq
e^{-2d}\ \reee^{-T} \begin{pmatrix} \del \\ \d x \end{pmatrix} \ , \label{genframes}
\eeq
where the matrix $\reee$ of components $\reee^A{}_M$ is a generalized vielbein. As in this equation, the splitting leads to $d$-dimensional blocks in matrices; equivalently, the $2d$-indices get split in two sets of $d$-indices. Because of the index placement for vectors and one-forms, we fix the following up/down (u/d) notation for these $d$-indices of objects $U$ and $V$
\beq
U_A=\begin{pmatrix} u_a \\ u^a \end{pmatrix} \ , \ V^A=\begin{pmatrix} v^a \\ v_a \end{pmatrix} \ , \qquad \eta_{AB}=\frac{1}{2} \begin{pmatrix} 0 & \delta^b_a \\ \delta^a_b & 0 \end{pmatrix} \ , \label{updownnot}
\eeq
and indicate the indices for the $O(d,d)$ metric $\eta_{(u/d)}$ \eqref{etaud}.\footnote{The $O(d,d)$ structure group considered here is a priori different from the T-duality group. Indeed, our $O(d,d)$ acts on the flat index $A$, i.e. ``on the left'' of a generalized vielbein $\reee$, while a standard T-duality acts on the ``generalized curved space'' index $M$, i.e. ``on the right'' of $\reee$; see also the generalized $\hhh$ in \eqref{fieldredefH}. The two groups may however be related. From our $O(d,d)$ metric $\eta_{AB}$, one can define a ``curved space'' metric $\eta_{MN}=\reee^A{}_M\ \eta_{AB}\ \reee^B{}_N$. The vielbeins considered in \eqref{genvielb} are elements of $O(d,d)$; for those, $\eta_{MN}$ is then equal to the $O(d,d)$ metric. One can thus consider $O(d,d)$ transformations on the curved space index. We come back to this idea in section \ref{sec:nongeo}.}

The two generalized vielbeins $\eee$ and $\teee$ of \eqref{genvielb}, that we repeat here in a more convenient manner
\bea
e^{-2d}\ \eee^{-T} &= e^{-2 \p} \sqrt{|g|} \begin{pmatrix} e^{-T} & e^{-T} b \\ 0 & e \end{pmatrix} \ , \ (\eee^{-T})_A{}^M= \eee^M{}_A =  \begin{pmatrix} e^m{}_a & e^n{}_a b_{nm} \\ 0 & e^a{}_m \end{pmatrix}\ , \label{geninvvielB} \\
e^{-2d}\ \teee^{-T} &= e^{-2 \tp} \sqrt{|\tg|} \begin{pmatrix} \te^{-T} & 0 \\ \te \b & \te \end{pmatrix} \ , \ (\teee^{-T})_A{}^M= \teee^M{}_A = \begin{pmatrix} \te^m{}_a & 0 \\ \te^a{}_n \b^{nm} & \te^a{}_m\end{pmatrix}\ , \label{geninvvielbeta}
\eea
provide interesting examples for local expressions of generalized conformal frames. Given the generic \eqref{genframes}, one has
\bea
& e^{-2d}\ \eee_A = \begin{cases} e^{-2d}\ \eee_a = e^{-2\p}\sqrt{|g|}\ (\del_a + b_{ab} e^b) \\ e^{-2d}\ \eee^a= e^{-2\p}\sqrt{|g|}\ e^a \end{cases} \ ,\label{Bsplit} \\
& e^{-2d}\ \teee_A = \begin{cases} e^{-2d}\ \teee_a= e^{-2 \tp} \sqrt{|\tg|}\ \del_a \\ e^{-2d}\ \teee^a= e^{-2\tp} \sqrt{|\tg|}\ (\te^a + \b^{ab} \del_b) \end{cases}\ .\label{betasplit}
\eea
The first example \eqref{Bsplit} was studied in \cite{Coimbra:2011nw}, while the second \eqref{betasplit} was only mentioned there, and is the one we focus on here.

These two examples are only local expressions. On each patch, they clearly provide an isomorphism to $T \mmm \oplus T^*\mmm$. Whether they actually allow for a (global) splitting requires more attention. The question is that of the transformation of a frame from one patch to the other. A global meaning can be given to the frame \eqref{Bsplit}, thanks to the gauge transformations of the $b$-field, that can be defined properly in this context \cite{Coimbra:2011nw}; this frame then allows for a splitting. Whether a similar global completion can be found for \eqref{betasplit} is less straightforward: it could amount to having a well-defined ``$\b$ gauge transformation''. We discuss this point in section \ref{sec:ET*}. In what follows, one can then consider all objects only locally (the end result, the scalar $S$, is in any case only a local quantity). Another point of view, mentioned already in section \ref{sec:preldisc}, is that we do not determine the underlying geometry here, i.e. the associated (global)
bundle. Rather, we only work at a formal level, by simply using formally the expression \eqref{betasplit} for the generalized conformal frame.

Note also that a splitting reduces the structure group to a subgroup $G_{{\rm split}}$, that preserves the form of the splitting. For \eqref{Bsplit}, it contains for example the $b$-field gauge transformations. The reduction of the structure group of the generalized bundle $E$ is equivalent to a refinement of the bundle itself: for \eqref{Bsplit}, $E$ becomes the generalized tangent bundle $E_T$ of GCG. If \eqref{betasplit} allows for a splitting, then its $G_{{\rm split}}$ and associated bundle will certainly be different: we discussed in section \ref{sec:preldisc} our proposal of obtaining the generalized cotangent bundle $E_{T^*}$, and we come back to it together with the $\b$ gauge transformation in section \ref{sec:ET*}.\\

Let us now define various ``generalized geometric objects'', that are compatible with the $O(d,d) \times \mathbb{R}^+$ structure of the extended generalized bundle. We mostly follow \cite{Coimbra:2011nw}. To start with, we introduce the bilinear product of two generalized vectors $V$ and $W$
\beq
\langle \reee_A , \reee_B \rangle \equiv \eta_{AB} \ , \quad \langle V , W \rangle = V^A \eta_{AB} W^B \ {\rm for}\ V=V^A \reee_A\ , W=W^B \reee_B\ , \label{bilin}
\eeq
where one can also multiply by a conformal factor. Then, we define a generalized (flat) covariant derivative acting on a generalized vector component $V^B$
\beq
D_A V^B = \del_A V^B + \hO_A{}^{B}{}_{C} V^C\ , \label{genconn}
\eeq
where $\del_A$ is a generalized (flat) derivative (to be specified below) and $\hO_A{}^{B}{}_{C}$ is a generalized (flat) spin connection. The compatibility of the latter with $O(d,d)\times\mathbb R^+$ requires to separate it in two pieces
\beq
\hO_A{}^{B}{}_{C}=\O_A{}^{B}{}_{C}-\La_A \delta^B_C\ , \label{tOOL}
\eeq
where $\O$ is the spin connection for $E$, and $\La$ is the part corresponding to the conformal weight (dilaton piece). Furthermore, the $O(d,d)$ structure, or equivalently compatibility with the metric \eqref{etaud}, demands the antisymmetry property (analogous to \eqref{prop})
\beq
\eta^{DC}\O_A{}^{B}{}_{C}=-\eta^{BC}\O_A{}^{D}{}_{C} \ .\label{etaO}
\eeq
On the tangent bundle, the connection is uniquely fixed (to be Levi-Civita) when requiring its compatibility with the $O(d-1,1)$ structure, and that it is torsion-free. Our goal here is to impose the same requirements on the generalized objects, but this will however not allow to fully fix the generalized connection. We will then impose further constraints, such as the recovery of the standard covariant derivative. Let us start with the definition of the generalized torsion $T$. The standard torsion can be defined as an action on two vectors giving back a vector; it is obtained by the difference of two Lie derivatives, where for one of them, the partial derivative is replaced by the covariant derivative. The generalized torsion is then defined similarly \cite{Coimbra:2011nw} in terms of the generalized Lie derivative $L$ \cite{Grana:2008yw} with the covariant derivative $D$ \eqref{genconn}
\bea
T(V,W) \equiv L_V^{D} W - L_V W\ . \label{deftorsion}
\eea
It is bilinear in $V$ and $W$. Its components on a generic frame are then defined with \eqref{bilin} as
\beq
T^A{}_{BC} \equiv \eta^{AD} \langle \reee_D , T(\reee_B, \reee_C) \rangle \ .
\eeq
They can be written using \eqref{deftorsion} as
\bea
T_{ABC}= -3 \hO_{[ABC]}+ \hO_{D}{}^{D}{}_{B}\eta_{AC}-e^{4d}\langle e^{-2d}\ \reee_A,L_{\reee_B}(e^{-2d}\ \reee_C)\rangle\ , \label{Torsion}
\eea
where indices are ``lowered with $\eta_{AB}$'' (this is the only place where we move indices in this way; this lowering is consistent with the up/down notation). Setting to zero this torsion then fixes (some of) the components of the generalized connection in terms of a given frame. Let us now exemplify all these objects with the frames \eqref{Bsplit} and \eqref{betasplit}.

In case of a splitting, the ``generalized curved index'' counterparts of the above objects can equivalently be defined, thanks to the generalized vielbein. For instance, we define in the up/down notation the generalized derivative\footnote{We set the abstract $\del^m$ to zero following \cite{Coimbra:2011nw}. It would however be tempting to keep it and study if it could serve as the $\tilde{\del}^m$ of DFT without the strong constraint.}
\beq
\del_M = \begin{cases} \del_m   \\ \del^m \equiv 0 \end{cases}\ , \qquad  \del_A=\reee^M{}_A\  \del_M \ . \label{delembed}
\eeq
While $\del_A$ is merely $\del_a$ for the frame with $b$-field \eqref{Bsplit}, we get interestingly an additional $\del^a=\beta^{ab} \del_b$ for the frame \eqref{betasplit} with $\b$, as can be seen from the generalized vielbeins in the form \eqref{geninvvielB} and \eqref{geninvvielbeta}. This gives a natural origin to the $\beta \del$ derivative discussed more in section \ref{sec:calc}, and it will lead to the new covariant derivative $\cN$ defined in \eqref{defcN}.

Using the up/down notation \eqref{updownnot}, another interesting property is the antisymmetry \eqref{etaO}, that becomes
\beq
\O_A{}_c{}^b = - \O_A{}^b{}_c \ , \qquad \O_A{}^{bc} = -\O_A{}^{cb} \ , \ \O_{Abc} = -\O_{Acb} \ .
\eeq
We can then work-out more concretely the generalized derivative \eqref{genconn} of a generalized vector expanded on a conformal frame
\bea
& (D_A V^B) e^{-2d}\ \reee_B \label{DVdvp} \\
& \qquad = e^{-2d} \left( (\del_A v^b + \O_A{}^b{}_c v^c + \O_A{}^{bc} v_c ) \reee_b + (\del_A v_b - \O_A{}^c{}_b v_c  + \O_A{}_{bc} v^c ) \reee^b - \La_A V^B \reee_B \right)\ .\nn
\eea
Some components of the connection will be fixed by the torsion-free condition, but not all. For the frame \eqref{Bsplit}, another natural requirement \cite{Coimbra:2011nw} is that $D_A$ reproduces the standard covariant derivative $\na_a$ \eqref{defo}. More precisely, given the generalized derivative \eqref{delembed} and the comment made below, it is natural for this frame that $D_a$ reproduces $\na_a$ while $D^a$ gets no contribution. Let us detail this point. To start with, $D_a$ involves three types of connection coefficients, among which: $\O_a{}^b{}_c\ ,\ \O_a{}^{bc}$. The first one enters (twice) in a similar fashion as the standard spin connection would do; in addition its antisymmetric part will be fixed by the torsion-free condition to be exactly that of the Levi-Civita spin connection $\o^b_{ac}$. To reproduce the standard $\na_a$ on both contravariant and covariant objects, namely $\na_a v^b\ \text{and}\ \na_a v_b$, it will therefore be natural to fix the symmetric part of $\O_a{}^b{}_c$ so that
it reproduces the full $\o^b_{ac}$. On the contrary, the second component $\O_a{}^{bc}$ brings a non-standard term, so we rather set it to zero (part of it may also be set to zero by the torsion-free condition). The third type of connection coefficient $\O_a{}_{bc}$ will be related later to the $H$-flux for the frame \eqref{Bsplit}. Let us now turn to $D^a$. The component $\O_a{}^{bc}$ in $D_a$ is the analogous one to $\O^a{}_{bc}$ appearing in $D^a$: we then set that one as well to zero
\beq
\O^a{}_{bc} = 0 \ , \ \O_a{}^{bc}= 0 \ .
\eeq
Most of the other contributions to $D^a$ for the frame \eqref{Bsplit} vanish thanks to the torsion-free condition. The fixing indicated then realises the requirement of reproducing the standard $\na_a$ with $D_A$. For the frame \eqref{betasplit}, we stick here to the same fixing (but the torsion-free condition will be different for $D^a$), so we are left with
\beq
\!\!\!\!\!\!\!\!\!\!\!\!\!\!\!\!\!\!\!\! \begin{cases} (D_a V^B) e^{-2d}\ \reee_B = e^{-2d} \left( (\del_a v^b + \O_a{}^b{}_c v^c  ) \reee_b + (\del_a v_b - \O_a{}^c{}_b v_c  ) \reee^b + \O_{abc} v^c \reee^b  - \lambda_a V^B \reee_B \right) \\
(D^a V^B) e^{-2d}\ \reee_B = e^{-2d} \left( (\del^a v^b - \O^{a}{}_c{}^b v^c  ) \reee_b + (\del^a v_b + \O^{a}{}_b{}^c v_c  )\reee^b + \O^{abc} v_c \reee_b  - \xi^a V^B \reee_B \right)  \end{cases}  \label{DAud}
\eeq
where we denote
\beq
\La_a \equiv \lambda_a \ , \ \La^a \equiv \xi^a \ . \label{Laud}
\eeq
On the contrary to the frame \eqref{Bsplit}, the $\del^a$ is non-trivial for the frame \eqref{betasplit} as mentioned below \eqref{delembed}, which makes us expect some contribution to $D^a$. The form of the derivatives  in \eqref{DAud} is also rather suggestive. We mentioned that $D_a$ reproduces $\na_a$ at the cost of fixing the symmetric part of the connection coefficient. The torsion-free condition will give equivalently here that the antisymmetric part of $\O^{a}{}_b{}^c$ is given by that of $\o_Q$ \eqref{defoQQ}; we will then fix similarly its symmetric part, so that the full $\o_Q$ is reproduced. This will lead to the new covariant derivative $\cN^a$ \eqref{defoQcG} being reproduced by $D^a$.\footnote{For the frame \eqref{Bsplit}, the symmetric part of $\O^{a}{}_b{}^c$ should rather be set to zero to get eventually $D^a=0$.}

Let us now work-out the torsion-free condition for the frame \eqref{betasplit}. We first compute
\bea
e^{4d}\langle e^{-2d}\ \teee_A,L_{ \teee_B}e^{-2d}\ \teee_C\rangle= & \tfrac{1}{2}(f^{a}{}_{bc}+f^{c}{}_{ab}+f^{b}{}_{ca}+Q_{a}{}^{bc}+Q_{b}{}^{ca}+Q_{c}{}^{ab}-R^{abc}) \label{comptorsion}\\
&\! +(f^{d}{}_{db}-2 \del_b\tp+ Q_d{}^{bd}+\b^{dg}f^{b}{}_{dg}-2 \b^{bd} \del_d \tp)\eta_{AC} \ ,\nn
\eea
where the terms in the right-hand side should be thought to contribute only when their indices match the position up or down of the indices on the left-hand side. The fluxes $f$, $Q$ and $R$ appearing here are those (in flat indices) defined in \eqref{fabc}, \eqref{Q} and \eqref{fluxes} respectively. It is a nice results to get precisely these fluxes here. Then, using the connection coefficients of \eqref{DAud}, and setting the torsion to zero in \eqref{Torsion}, one first obtains\footnote{These results as well as \eqref{xi} are in agreement with those of \cite{Geissbuhler:2013uka}.}
\bea
& f^{a}{}_{bc}=2\O_{[b}{}^{a}{}_{c]}\ ,\quad f^{c}{}_{ab}=2\O_{[a}{}^{c}{}_{b]}\ ,\quad f^{b}{}_{ca}=2\O_{[c}{}^{b}{}_{a]}\ ,\\
& Q_{a}{}^{bc}=2\O^{[b}{}_{a}{}^{c]}\ ,\quad Q_{b}{}^{ca}=2\O^{[c}{}_{b}{}^{a]}\ , \quad Q_{c}{}^{ab}=2\O^{[a}{}_{c}{}^{b]}\ ,\\
& R^{abc} = 3\O^{[abc]} \ , \quad  \O_{[abc]}=0 \ .
\eea
As discussed below \eqref{DVdvp} and \eqref{DAud}, and given the properties \eqref{prop} and \eqref{troQ} compared to the relations just derived, we identify for the frame \eqref{betasplit}
\beq
\O_{b}{}^{a}{}_{c}= \o^a_{bc} \ , \quad \O^{b}{}_{a}{}^{c}= {\o_Q}^{bc}_a \ .\label{identifO}
\eeq
From those, we deduce $f^{d}{}_{db}=\O_{d}{}^{d}{}_{b}$ and $Q_{d}{}^{db}=\O^{d}{}_{d}{}^{b}$. The torsion-free condition then finally gives
\beq
\lambda_{b}=2 \del_b \tp  \ , \quad \xi^{b} = -2Q_d{}^{bd}-\b^{cd}f^{b}{}_{cd} + 2 \b^{bd}\del_d\tp \ . \label{xi}
\eeq
The sign of $Q_{d}{}^{bd}$ in \eqref{comptorsion} may look surprising, as it differs from that of $f^{d}{}_{db}$ when viewed as the trace of a connection, and leads to the $-2 Q_{d}{}^{bd} $ in $\xi^b$. It is however the correct result, and one gets a better understanding by noticing that the tensor $\T^m$ \eqref{tracecG} can be expressed in flat indices (using \eqref{tricks}) as
\beq
\T^a = -Q_b{}^{ba}+\frac{1}{2}\beta^{cd}f^{a}{}_{cd} \ .\label{Ta}
\eeq
One can then rewrite
\beq
\lambda_a=2 \del_a \tp \ , \quad \xi^{a} =  2 ( \b^{ad}\del_d\tp - \T^a) \ , \label{xiT}
\eeq
which give precisely the two dilaton terms in the Lagrangian $\tL_{\b}$ in \eqref{L2} or \eqref{Lflat}! In view of this relation, the role of $\T^a$ gets clarified: it plays the role of the conformal weight together with the dilaton, and appears in the corresponding combination given by $\xi^a$. This is the ``non-standard'' conformal weight, obtained with the frame \eqref{betasplit}, that matches with the ``non-standard'' dilaton term in the Lagrangian. The standard one (obtained similarly for the frame \eqref{Bsplit}) is $\lambda_a$, corresponding to the standard dilaton kinetic term.

The components of the connection unfixed by the torsion-free condition are those that do not appear when computing $-3 \O_{[ABC]}+\O_{D}{}^{D}{}_{B}\eta_{AC}$. Here we are left with the parts of $\O_{abc}$ or $\O^{abc}$ that do not contribute to this quantity, for instance the pieces that are not fully antisymmetric. In \cite{Coimbra:2011nw} with the frame \eqref{Bsplit}, these remaining unfixed components are kept, and are found to eventually not contribute to the scalar $S$. Inspired by this situation, we choose here for simplicity to set them to zero: this restriction is still enough for our purposes, and will allow us to reproduce $\tL_{\b}$ from $S$.

We finally obtain, for the frame \eqref{Bsplit} (the torsion-free condition can be worked-out similarly \cite{Coimbra:2011nw}, and $\lambda_a= 2 \del_a \p$)
\beq
\begin{cases} (D_a V^B) e^{-2d}\ \eee_B =  e^{-2d} \left( (\na_a v^b) \eee_b + (\na_a v_b) \eee^b - \frac{1}{3} H_{abc} v^c \eee^b  - \lambda_a  V^B \eee_B \right)  \\
(D^a V^B) e^{-2d}\ \eee_B = 0  \end{cases} \ ,
\eeq
while the frame \eqref{betasplit} leads to
\beq
\begin{cases} (D_a V^B) e^{-2d}\ \teee_B =  e^{-2d} \left(\  (\na_a v^b) \teee_b + (\na_a v_b) \teee^b  - \lambda_a  V^B \teee_B \ \right)  \\
(D^a V^B) e^{-2d}\ \teee_B = e^{-2d} \left( -(\cN^a v^b) \teee_b - (\cN^a v_b) \teee^b + \frac{1}{3} R^{abc} v_c \teee_b  - \xi^a  V^B \teee_B \right)   \end{cases} \ ,\label{betacovder}
\eeq
where we read from \eqref{defcN} and \eqref{defoQcG}
\beq
\cN^a v^b = -\b^{ac}\del_c v^b + {\o_Q}_{c}^{ab} v^c \ , \quad \cN^a v_b = -\b^{ac}\del_c v_b - {\o_Q}_{b}^{ac} v_c\ . \label{cNwup}
\eeq

\subsection{Preserving an $O(d-1,1) \times O(1,d-1)$ structure and recovery of $\tL_{\b}$}\label{sec:OOstruct}

After having presented the $O(d,d) \times \mathbb{R}^+$ structure group of the extension of $E$, the associated generalized geometric objects, and given some concrete examples for those, we are now interested in a specific $O(d-1,1) \times O(1,d-1)$ subgroup. This last structure was considered for ${\cal N}=1$ supersymmetry \cite{Hohm:2011nu, Jeon:2011sq}, but also allowed to reproduce (uniquely) the type II supergravities \cite{Hohm:2010xe, Coimbra:2011nw, Jeon:2012kd, Jeon:2012hp}. Preserving such a structure brings in more constraints. For instance, the metric and dilaton are fixed by this structure, meaning that the conformal weight is globally defined, and one then only focuses on the bundle $E$. Another example is that the two orthogonal groups lead to two $Spin$ groups, and associated spinors turned out to be related to the two supersymmetries of type II theories; we will come back to that point. Finally, the generalized curvature scalar $S$ defined in terms of these spinors was shown to be related to the standard NSNS Lagrangian \eqref{LNSNS}. Here, we are interested in the frame \eqref{betasplit} with
$\b$ and the consequent derivative obtained in \eqref{betacovder}; preserving this $O(d-1,1) \times O(1,d-1)$ structure will then allow us to reproduce analogously the Lagrangian $\tL_{\b}$ \eqref{Lflat} from $S$. We will also make use of these results to study supersymmetry for $\b$-supergravity \cite{Future}.

We define the subgroup $O(d-1,1) \times O(1,d-1)$ as follows. The $O(d,d)$ metric $\eta_{{\rm (u/d)}}$ \eqref{etaud}, preserved on the generalized bundle $E$, has two sets of eigenvalues, positive and negative. One can separate them in two sets of signature $(d-1,1)$ and $(1,d-1)$, as given by the diagonalised $O(d,d)$ metric $\eta$
\beq
\eta=\begin{pmatrix} \eta_d & 0 \\ 0 & -\ov{\eta}_d \end{pmatrix} \ , \ \eta_{AB}=\begin{pmatrix} \eta_{ab} & 0 \\ 0 & -\ov{\eta_{ab}} \end{pmatrix} \ , \label{etabub}
\eeq
where we consider $\eta_{ab}$ and $\ov{\eta_{ab}}$ to be the same in value, with $(d-1,1)$ signature; the ``unbarred - barred'' notation allows us to distinguish the two sets. A conformal generalized frame can then locally be separated into these two sets, and denoted accordingly $e^{-2d}\ \reee_a\ ,\ e^{-2d}\ \reee_{\ov{a}}$. Whether these two sets remain separated spaces globally is however not guaranteed by the $O(d,d)$ structure group: preserving this is equivalent to reducing $O(d,d)$ to $O(d-1,1) \times O(1,d-1)$, since the precise form of the metric \eqref{etabub} is left invariant by this subgroup. If this $O(d-1,1) \times O(1,d-1)$ structure is preserved, the generalized bundle $E$ is then (globally) isomorphic to the direct sum of these two spaces, that are two sub-bundles: this is denoted as
\beq
E \simeq C_+ \oplus C_- \ .
\eeq
$C_+$, resp. $C_-$, is the sub-bundle on which $O(d-1,1)$, resp. $O(1,d-1)$, acts, and the corresponding indices are unbarred, resp. barred.

Of the various quantities defined in section \ref{sec:GGOdd}, we would now like to consider those that preserve this $O(d-1,1) \times O(1,d-1)$ structure. This can be done in two steps: first, one has to rotate the quantities defined above in the up/down basis (where the metric is $\eta_{{\rm (u/d)}}$) to the unbarred - barred basis (where the metric is $\eta$). In other words, the previous $O(d,d)$ representation should be rotated to a new one where the embedding of $O(d-1,1) \times O(1,d-1)$ is diagonal. Secondly, one has to project-out the quantities that do not preserve the $O(d-1,1) \times O(1,d-1)$ structure; it will be very simple to determine those in the unbarred - barred basis.

We will essentially go through this procedure for the frame \eqref{betasplit} with $\b$. One could be slightly more general and consider, for this frame or the one \eqref{Bsplit} with $b$, that each sub-bundle $C_{\pm}$ has a different set of vielbeins, $\te^a{}_m$ or $\te^{\ov{a}}{}_m$, giving however the same metric $\tg_{mn}$ \cite{Coimbra:2011nw}. At the end of the day however, it is useful to consider only one, same, vielbein. So here, although we denote them differently via the index, we will quickly consider this alignment of vielbeins, i.e. that $\te^a{}_m=\te^{\ov{a}}{}_m$ for $a=\ov{a}$ in value.

Let us now present the procedure more concretely; we will leave some details to the appendix \ref{ap:GG}. To rotate from one representation of $O(d,d)$ to the other, we introduce the matrix $P$ that transforms the previous up/down $\eta_{{\rm u/d}}$ \eqref{etaud} into the unbarred - barred diagonal $\eta$ \eqref{etabub}
\beq
\eta= P \eta_{{\rm u/d}} P^T \ , \qquad P=\begin{pmatrix} \id & \eta_d \\ \id & -\ov{\eta}_d \end{pmatrix} \ , \ P^{-1}=\frac{1}{2} \begin{pmatrix} \id & \id \\ \eta_d^{-1} & -\ov{\eta}_d^{-1} \end{pmatrix} \ .
\eeq
Any object in the fundamental representation of $O(d,d)$ (i.e. carrying an index $A$) should then be rotated as follows, from the up/down (u/d) to the unbarred - barred (not denoted)
\bea
& V_B=P_B{}^A V_{(u/d)A} \ , \quad V^B= V^A_{(u/d)} (P^{-1})_A{}^B = (P^{-T})^B{}_A V^A_{(u/d)} \label{rot} \\
{\rm with}\ & P_B{}^A=\begin{pmatrix} \delta^a_b & \eta_{bc} \delta^c_a \\ \delta^a_{\ov{b}} & - \ov{\eta_{bc}} \delta^{\ov{c}}_a \end{pmatrix} \ , \ (P^{-T})^B{}_A=\frac{1}{2}\begin{pmatrix} \delta^b_a & \eta^{bc} \delta_c^a \\ \delta^{\ov{b}}_a & - \ov{\eta^{bc}} \delta_{\ov{c}}^a \end{pmatrix}
\eea
so that bilinears are preserved. The $\delta$'s in $P$ and $P^{-T}$ allow to pass from the up/down to the unbarred - barred indices, by identifying the actual value of the index. In what follows and in appendix \ref{ap:GG}, we will mostly stop writing these $\delta$'s to simplify formulas; the notations should be clear enough. An important example of this rotation is obtained by acting on the frames \eqref{Bsplit} and \eqref{betasplit}, towards
\bea
& e^{-2d}\ \eee_A = \begin{cases} e^{-2d}\ \eee_a= e^{-2\p}\sqrt{|g|} (\del_a + b_{ab} e^b + \eta_{ab} e^b) \\
e^{-2d}\ \eee_{\ov{a}}= e^{-2\p}\sqrt{|g|} (\del_{\ov{a}} + b_{\ov{ab}} e^{\ov{b}} - \ov{\eta_{ab}} e^{\ov{b}})  \end{cases} \ , \\
& e^{-2d}\ \teee_A = \begin{cases} e^{-2d}\ \teee_a= e^{-2\tp}\sqrt{|\tg|} (\del_a + \eta_{ab} \b^{bc}\del_c + \eta_{ab} \te^b )  \\ e^{-2d}\ \teee_{\ov{a}}= e^{-2\tp}\sqrt{|\tg|} (\del_{\ov{a}} - \ov{\eta_{ab}} \b^{\ov{bc}}\del_{\ov{c}} - \ov{\eta_{ab}} \te^{\ov{b}} ) \end{cases}\ ,
\eea
where we did not write out the $\delta$'s, and used the alignment of unbarred and barred vielbeins.

We then redefine a covariant derivative $D_A (W^B) e^{-2d}\ \reee_B$, where $A,B$ are now unbarred - barred indices. A priori, one would have for the unbarred $A=a$
\beq
D_a (W^B) e^{-2d}\ \reee_B = e^{-2d} \left( \del_a (w^b) \reee_b + \del_a (w^{\ov{b}}) \reee_{\ov{b}} + \hO_a{}^B{}_C W^C \reee_B \right) \ , \label{Da1}
\eeq
where $\del_a$, resp. $\hO_a{}^B{}_C$, are defined as the unbarred component obtained from the rotation of the up/down $\del$, resp. $\hO$. As mentioned earlier, one could also define $D$, $\del$ and $\hO$ with a first generalized curved index $M$, and then rotate the (inverse) generalized vielbein acting on it. In any case, the last term in \eqref{Da1} gets developed into four terms, according to the unbarred or barred choices for $B,C$. This is however where the projection preserving the $O(d-1,1)\times O(1,d-1)$ structure enters: the objects $\hO_M{}^{\ov{b}}{}_c$ and $\hO_M{}^{b}{}_{\ov{c}}$ are in a bi-fundamental representation of $O(d,d)$ but are off-diagonal with respect to the $O(d-1,1)\times O(1,d-1)$ diagonal structure. Having them in the covariant derivative \eqref{Da1} would then allow mixed contributions from $C_{\pm}$: for instance, when considering the components on $\reee_b$, the covariant derivative of $w^b$ would get a contribution from $w^{\ov{c}}$ thanks to $\hO_M{}^{b}{}_{\ov{c}}$. The
projection allowing to preserve the $O(d-1,1)\times O(1,d-1)$ structure therefore sets these components to zero. One is then left with only two terms. Separating the components on $\reee_b$ and $\reee_{\ov{b}}$, one gets the following $O(d-1,1)\times O(1,d-1)$ derivative
\beq
D_A W^B = \begin{cases} D_a w^b = \del_a w^b + \hO_a{}^b{}_c w^c \\
           D_a w^{\ov{b}} = \del_a w^{\ov{b}} + \hO_a{}^{\ov{b}}{}_{\ov{c}} w^{\ov{c}} \\
           D_{\ov{a}} w^b = \del_{\ov{a}} w^b + \hO_{\ov{a}}{}^b{}_c w^c \\
           D_{\ov{a}} w^{\ov{b}} = \del_{\ov{a}} w^{\ov{b}} + \hO_{\ov{a}}{}^{\ov{b}}{}_{\ov{c}} w^{\ov{c}}
          \end{cases}\ , \label{DWbubgen}
\eeq
where again all indices are unbarred or barred.

We now leave to the appendix \ref{ap:OOder} the determination of the various pieces in this derivative, for the frame \eqref{betasplit} (the same procedure applied to the frame \eqref{Bsplit} allows to reproduce the result of \cite{Coimbra:2011nw}). These pieces are the derivative $\del$, the connection $\O$, and the piece due to the conformal weight. The definition of the latter is slightly changed \cite{Coimbra:2011nw} with respect to \eqref{tOOL} and we discuss this in the appendix. Determining these various pieces essentially amounts to rotate the contributions to the $O(d,d) \times \mathbb{R}^+$ derivative obtained in \eqref{betacovder}. We get eventually for the frame \eqref{betasplit}
\beq
D_A W^B = \begin{cases} D_a w^b = \nabla_a w^b - \eta_{ad} \cN^d w^b + \frac{1}{6} \eta_{ad} \eta_{cf} R^{dbf} w^c - \frac{1}{9} (\delta^b_a \Lambda_c - \eta_{ac} \eta^{be} \Lambda_e) w^c \\
           D_a w^{\ov{b}} = \nabla_a w^{\ov{b}} - \eta_{ad} \cN^d w^{\ov{b}} - \frac{1}{2} \eta_{ad} \ov{\eta_{cf}} R^{d\ov{bf}} w^{\ov{c}} \\
           D_{\ov{a}} w^b = \nabla_{\ov{a}} w^b + \ov{\eta_{ad}} \cN^{\ov{d}} w^b - \frac{1}{2} \ov{\eta_{ad}} \eta_{cf} R^{\ov{d}bf} w^c \\
           D_{\ov{a}} w^{\ov{b}} = \nabla_{\ov{a}} w^{\ov{b}} + \ov{\eta_{ad}} \cN^{\ov{d}} w^{\ov{b}} + \frac{1}{6} \ov{\eta_{ad}} \ov{\eta_{cf}} R^{\ov{dbf}} w^{\ov{c}} - \frac{1}{9} (\delta^{\ov{b}}_{\ov{a}} \Lambda_{\ov{c}} - \ov{\eta_{ac}} \ov{\eta^{be}} \Lambda_{\ov{e}}) w^{\ov{c}}
          \end{cases}\ , \label{DWbubfinal}
\eeq
as given in \eqref{DWbubfinalintro}, where
\beq
\Lambda_C = \begin{cases} \Lambda_c = \lambda_c + \eta_{cd} \xi^d \\
             \Lambda_{\ov{c}} = \lambda_{\ov{c}} - \ov{\eta_{cd}} \xi^{\ov{d}}
            \end{cases} \ , \label{Lambda}
\eeq
with $\la$ and $\xi$ given in \eqref{xi}.\\

From the above $O(d-1,1)\times O(1,d-1)$ structure, we now consider, following \cite{Coimbra:2011nw, Hohm:2011nu}, an associated $Spin(d-1,1)\times Spin(1,d-1)$ structure, whose spinors we denote respectively $\epsilon^+$ and $\epsilon^-$. Spinorial derivatives corresponding to the previous \eqref{DWbubgen} are defined naturally as
\bea
D_M \epsilon^+ &= \del_M \epsilon^+ + \frac{1}{4} \hO_M{}^b{}_c \eta_{bd} \gamma^{dc} \epsilon^+ \ , \label{Deps+}\\
D_M \epsilon^- &= \del_M \epsilon^- + \frac{1}{4} \hO_M{}^{\ov{b}}{}_{\ov{c}} \ov{\eta_{bd}} \gamma^{\ov{dc}} \epsilon^- \ ,\label{Deps-}
\eea
where the $\gamma$ matrices and their properties are discussed in appendix \ref{ap:conv}. Interestingly, one can build from these derivatives the generalized curvature scalar $S$ mentioned above, that is, in the standard NSNS case, related to the Lagrangian $\L_{{\rm NSNS}}$ up to a total derivative. As in \eqref{defSintro}, the scalar $S$ can be defined by
\beq
-\frac{1}{4}S\eps^+= \left( \g^a D_a\g^b D_b- \ov{\eta^{ab}} D_{\ov{a}}D_{\ov{b}} \right)\eps^+ \ ,\label{defS}
\eeq
or equivalently with the spinor $\eps^-$ by then exchanging unbarred and barred indices. Here we want to compute this quantity \eqref{defS} for the frame \eqref{betasplit} with $\b$, i.e. with the derivatives obtained in \eqref{DWbubfinal}. From the latter, one reads the spin connection components of \eqref{Deps+} and \eqref{Deps-}, in flat indices. In particular, the derivatives needed to compute $S$ are given by
\bea
\gamma^a D_a \epsilon^+ & = \left( \gamma^a \nabla_a - \gamma^a \eta_{ad} \cN^d  + \frac{1}{24} \eta_{ad} \eta_{be} \eta_{cf} R^{def} \gamma^{abc} - \frac{1}{2} \gamma^c \Lambda_c \right) \epsilon^+\ , \label{gDaeps} \\
D_{\ov{a}} \epsilon^+ & = \left( \nabla_{\ov{a}} + \ov{\eta_{ad}} \cN^{\ov{d}} - \frac{1}{8} \ov{\eta_{ad}} \eta_{be} \eta_{cf} R^{\ov{d}ef} \gamma^{bc} \right) \epsilon^+ \ ,\label{Dabeps} \\
D_{\ov{a}} w^{\ov{a}} & = \nabla_{\ov{a}} w^{\ov{a}} + \ov{\eta_{ad}} \cN^{\ov{d}} w^{\ov{a}} - \Lambda_{\ov{a}} w^{\ov{a}} \ , \label{traceDwb}
\eea
where we used \eqref{g2}. $\nabla$ and $\cN$ on the spinors are the naturally defined spinorial covariant derivatives, as obtained from \eqref{DWbubfinal} and \eqref{Deps+}. Equation \eqref{traceDwb} allows to determine the correct connection to use in the covariant derivative acting on $\ov{\eta^{ab}} D_{\ov{b}} \eps^+$. We thus rewrite the above expressions as
\bea
\gamma^a D_a \epsilon^+ & = \left( \gamma^a \del_a + \gamma^a \eta_{ad} \b^{dc} \del_c + X_{abc} \gamma^{abc} + X_a \gamma^a \right) \epsilon^+\ , \\
D_{\ov{a}} \epsilon^+ & = \left( \del_{\ov{a}} - \ov{\eta_{ad}} \b^{\ov{dc}} \del_{\ov{c}} + Y_{\ov{a}bc} \gamma^{bc} \right) \epsilon^+\ , \\
D_{\ov{a}} \left(\ov{\eta^{ab}} D_{\ov{b}} \epsilon^+ \right) & = \left( \del_{\ov{a}} - \ov{\eta_{ad}} \b^{\ov{dc}} \del_{\ov{c}} + Y_{\ov{a}bc} \gamma^{bc} + Z_{\ov{a}} \right) \ov{\eta^{ab}} \left( \del_{\ov{b}} - \ov{\eta_{be}} \b^{\ov{ef}} \del_{\ov{f}} + Y_{\ov{b}ef} \gamma^{ef} \right)  \epsilon^+
\eea
with
\bea
X_{abc}&= \frac{1}{4} \eta_{be} \left(\omega^e_{ac} - \eta_{ad} {\omega_{Q}}_c^{de} + \frac{1}{6} \eta_{ad} \eta_{cf} R^{def} \right)\ , \\
X_a &= \frac{1}{2} \left( \o^d_{da} + \eta_{ac} {\o_Q}_d^{dc} - \Lambda_a  \right)\ , \\
Y_{\ov{a}bc} & = \frac{1}{4} \eta_{be} \left(\omega^e_{\ov{a}c} + \ov{\eta_{ad}} {\omega_{Q}}_c^{\ov{d}e} - \frac{1}{2} \ov{\eta_{ad}} \eta_{cf} R^{\ov{d}ef} \right)\ , \\
Z_{\ov{a}} & = \o^{\ov{d}}_{\ov{da}} - \ov{\eta_{ac}} {\o_Q}_{\ov{d}}^{\ov{dc}} - \Lambda_{\ov{a}} \ ,
\eea
where \eqref{g2} and the antisymmetry properties of $\o$ and $\o_Q$ were used. With these convenient notations (analogous to those of \cite{Hohm:2011nu}), we compute $S$ from \eqref{defS}, and obtain at first
\bea
-\frac{1}{4} S \epsilon^+ = & \Bigg[ \gamma^a \gamma^b (\del_a + \eta_{ad} \b^{de} \del_e) (\del_b + \eta_{bc} \b^{cf} \del_f) - \ov{\eta^{ab}}(\del_{\ov{a}} - \ov{\eta_{ad}} \b^{\ov{de}} \del_{\ov{e}}) (\del_{\ov{b}} - \ov{\eta_{bc}} \b^{\ov{cf}} \del_{\ov{f}}) \label{S1} \\
& + 6 \eta^{ab} \gamma^{cf} X_{[bcf]} (\del_a + \eta_{ad} \b^{de} \del_e) + 2 \eta^{ac} X_c (\del_a + \eta_{ad} \b^{de} \del_e) \nn \\
& - 2 \ov{\eta^{ab}} Y_{\ov{a}de} \gamma^{de} (\del_{\ov{b}} - \ov{\eta_{bc}} \b^{\ov{cf}} \del_{\ov{f}}) - Z_{\ov{a}} \ov{\eta^{ab}} (\del_{\ov{b}} - \ov{\eta_{bc}} \b^{\ov{cf}} \del_{\ov{f}}) \nn \\
& + \gamma^a \gamma^{bcf} (\del_a + \eta_{ad} \b^{de} \del_e) (X_{bcf}) + \gamma^a \gamma^c (\del_a + \eta_{ad} \b^{de} \del_e) (X_c)  \nn \\
&+\frac{1}{2}X_{ade}X_{bcf}\{\gamma^{ade},\gamma^{bcf}\} + X_{ade}X_c \{\gamma^{ade},\gamma^c\} + X_aX_c \gamma^a\gamma^c\nn \\
& - \ov{\eta^{ab}}(\del_{\ov{a}} - \ov{\eta_{ad}} \b^{\ov{de}} \del_{\ov{e}}) (Y_{\ov{b}cf}) \gamma^{cf}  - \frac{1}{2}\ov{\eta^{ab}} Y_{\ov{a}de} Y_{\ov{b}cf} \{\gamma^{de}, \gamma^{cf}\} - Z_{\ov{a}} \ov{\eta^{ab}} Y_{\ov{b}cf} \gamma^{cf} \Bigg] \epsilon^+  \ , \nn
\eea
where we used \eqref{g2}.

The first three lines of \eqref{S1} are acting on the spinor; therefore to get for $S$ only a scalar, they have to vanish. In addition, the last three lines of \eqref{S1} contain various orders of fully antisymmetrised products of $\gamma$ matrices; a scalar would only come from the ``zeroth order'', i.e. from the terms without any $\gamma$. The other terms should then also vanish. We show in appendix \ref{ap:S} that remarkably, all these quantities indeed vanish. So we are left with only a scalar multiplying $\epsilon^+$ as in \eqref{Sfinalap}, namely
\bea
-\frac{1}{4} S \epsilon^+ =&-\frac{1}{4} \Bigg(\R(\tg) + \R_Q - \frac{1}{2} R^{acd} f^b{}_{cd} \eta_{ab} -\frac{1}{2} R^2 \label{Sfinalsec}\\
& \phantom{-\frac{1}{4} \ \ } -4(\del \tp)^2 + 4\na^2 \tp - 4 (\b^{ab}\del_b \tp - \T^a)^2 -4\eta_{ab}\cN^a (\b^{bc}\del_c \tp - \T^b) \Bigg) \epsilon^+ \ . \nn
\eea
Another remarkable property is that this scalar contains only an even number of $\b$ in each term, i.e. all mixed terms giving one or three $\b$ get canceled. Using \eqref{tracecN} and the Leibniz rule, we rewrite the second line of \eqref{Sfinalsec} and get eventually
\bea
S=& \R(\tg) +4(\del \tp)^2 + 4 (\b^{ab}\del_b \tp - \T^a)^2  + \R_Q - \frac{1}{2} R^{acd} f^b{}_{cd} \eta_{ab} -\frac{1}{2} R^2 \label{Stotder} \\
& + e^{2d}\ \del_p\left( 4 e^{-2d} \tg^{pq} \del_q \tp -4 e^{-2d} \b^{pm} \tg_{mq} (\b^{qr}\del_r\tp -\T^q) \right) \ ,\nn
\eea
or equivalently, with \eqref{Lflat}, $S= e^{2d}\ (\tL_{\b} + \del(\dots))$ as given in \eqref{SLbeta}. This concludes our derivation of $\tL_{\b}$ from the Generalized Geometry formalism.\\

The spinorial derivatives given in \eqref{Deps+} and \eqref{Deps-}, and their explicit expressions, such as \eqref{gDaeps} and \eqref{Dabeps}, may serve further purposes. As mentioned previously, it has been noticed that these quantities, for the frame \eqref{Bsplit} with $b$-field, are those entering the supersymmetry variations of the fermions in type II supergravities. In other words, these quantities lead to the Killing spinor equations. The expressions worked-out here should therefore play the analogous, important, role in (a possible supersymmetry of) $\b$-supergravity. We leave this idea to future investigations \cite{Future}.

\section{Equations of motion}\label{sec:eom}

We derive in this section the equations of motion for $\tg_{mn}, \tp$, and $\b^{mn}$, out of $\tL_{\b}$ in curved indices, given in \eqref{L2}. To do so, we first study the new objects $\cN$, $\cG$ and $\cR$ and establish some useful differential properties. Some details of the derivation of the equation of motion for $\b$ are presented in the appendix \ref{ap:eombeta}.

\subsection{Differential calculus and useful properties}\label{sec:calc}

Let us start by giving a more detailed account on the DFT origin \cite{Andriot:2012wx, Andriot:2012an} of the objects appearing in $\tL_{\b}$ in \eqref{L2}. As discussed in the Introduction, DFT fields depend a priori on two sets of coordinates $x^m$ and $\tilde{x}_m$, and one can consider two sets of associated diffeomorphisms, parametrised by $\xi^m(x,\tilde{x})$ and $\tilde{\xi}_m(x,\tilde{x})$ respectively. Given $x^m$ and the derivative $\del_m$, one can construct as in \eqref{def} the standard covariant derivative $\nabla_m$ and Ricci scalar $\R(\del,\tilde{g})$: these behave as tensors under the standard diffeomorphisms associated to $\xi^m$. The analogous quantities can be constructed with respect to $\tilde{\del}^m$ and $\tilde{\xi}_m$, where basically all up and down indices have to be exchanged; in particular the Ricci scalar is defined there with respect to the inverse metric: $\R(\tilde{\del}, \tg^{-1}) $.
After the field redefinition \eqref{fieldredef1} and \eqref{fieldredef2}, the DFT Lagrangian could be written as
\beq
\L_{\rm {DFT}}= e^{-2d} \left(\R(\del,\tg) + \R(\tilde{\del}, \tg^{-1}) + 4 (\del \tp)^2 + 4 (\tilde{\del} \tp)^2 + \dots \right) \ ,
\eeq
i.e. the two above Ricci scalars appeared, as well as two kinetic terms for the dilaton. As just explained, the two terms involving only $\del$ are covariant with respect to $\xi$ while those two involving only $\tilde{\del}$ are covariant with respect to $\tilde{\xi}$. However, these parameters depend a priori on both sets of coordinates, so considering the transformation under the ``unusual'' diffeomorphism can be non-trivial. For instance, the two terms $\R(\tilde{\del}, \tg^{-1}) + 4 (\tilde{\del} \tp)^2$ do not transform covariantly with respect to the standard $\xi$-diffeomorphism. The structures introduced in \cite{Andriot:2012wx, Andriot:2012an} therefore had the aim to restore a manifest covariance with respect to $\xi$. The DFT action is known to be invariant under both diffeomorphisms so a covariant rewriting should exist, and was found. The basic building block was the derivative $\tilde{\del}^m - \b^{mn}\del_n$: it is the new covariant derivative for a scalar such as $\tp$.\footnote{The reason
for this combination comes from the particular form of the variation of $\b^{mn}$ under $\xi^m$: on top of the standard Lie derivative, it has terms given by $\tilde{\del} \xi$. When going back to supergravity with $\tilde{\del} = 0$, these additional pieces are set to zero and one recovers the standard transformation of $\b^{mn}$.} With this starting point, one can construct further all objects, from a completion of $\tilde{\del}$ to a covariant derivative $\cN$ with connection $\cG$, till a Ricci scalar $\cR$. The whole $\L_{\rm {DFT}}$ could finally be rewritten in a manifestly covariant form with respect to half the double diffeomorphisms, namely those associated to the standard $\xi^m$. This form is very close to that of $\tL_{\b}$ \eqref{L2}.

Here, when coming back to supergravity by removing all dependence on the $\tilde{x}_m$ coordinates (imposing $\tilde{\del}^m=0$), we loose most of the motivation just given for the new objects. Indeed, the diffeomorphism parameters $\xi^m(x)$ are now just the standard ones with only a dependence on $x^m$, and the ``troublesome'' $\tilde{\del}$ which was completed to covariant objects, is not present anymore. We nevertheless inherit the remnant of the objects introduced, and all their tensorial properties with respect to the standard diffeomorphisms. The structure gained this way, when comparing for instance to $\tL_0$ \eqref{L0}, is certainly useful. In addition, we discover here new interesting properties of these objects. Let us now present them.

The former DFT derivative $\tilde{\del}^m - \b^{mn}\del_n$ becomes simply here $- \b^{mn}\del_n$, and this is the first building block of the objects we now consider. Note that such a derivative can be viewed as an Anchor map; structures and objects associated to it can be interpreted in terms of Lie algebroids \cite{Blumenhagen:2012pc, Blumenhagen:2012nk, Blumenhagen:2012nt}. The properties to be detailed here could certainly be translated into this Lie algebroid perspective (and have already been, to some extent, in \cite{Blumenhagen:2013aia}). We do not study such a link here, but doing so would probably bring interesting insight.

This derivative $- \b^{mn}\del_n$ first appears in the connection coefficient $\cG_p^{mn}$, given here by \eqref{cG}. $\cG$ is essentially made of two pieces. The first piece is the analogue to the standard Christoffel symbol \eqref{def}, where the up and down indices are exchanged, and the derivative $- \b^{mn}\del_n$ is used. The second piece in $\del \b$ looks more unusual, one particularity being that it has an antisymmetric part, on the contrary to a Christoffel symbol. But this second piece plays an important role, for instance by sometimes compensating the presence of $\b$ in the first piece. It is also crucial in order to rewrite $\cG$ as in \eqref{cG_t} with a tensorial piece $\cG_{\!\!(t)}$. From the derivative $- \b^{mn}\del_n$ and the connection $\cG$, one constructs the covariant derivative $\cN^m$ as in \eqref{defcN}. This derivative, as well as the associated Ricci tensor $\cR^{mn}$ given in \eqref{cR}, can be understood, from their definitions, as the analogous objects to the standard
covariant derivative $\nabla_m$ and Ricci tensor $\R_{mn}$ \eqref{def} (one difference though being that $\cR^{[mn]} \neq 0$). Most of their properties, that we now turn to, follow the same analogy.

Let us emphasise once more that our $\cN^m$ and $\cR^{mn}$ are tensors with respect to the standard diffeomorphisms. Another property obtained from the DFT construction \cite{Andriot:2012wx, Andriot:2012an} is the metric compatibility
\beq
\cN^m \tg_{pq} = \cN^m \tg^{pq} = 0 \ .\label{compatibility}
\eeq
Using the above, let us show the useful generalization of the Palatini identity. Under an infinitesimal variation of the metric $\delta \tg$, the variation $\delta \G \equiv \G(\tg+\delta \tg) -\G (\tg)$ of the Christoffel symbol \eqref{def} is known to be a tensor, which will allow to take its covariant derivative; more precisely, it can be written as
\beq
\delta \G^m_{np} = \frac{1}{2} \tg^{mq} \left(\nabla_n (\delta\tg_{qp})+\nabla_p (\delta \tg_{qn})-\nabla_q (\delta \tg_{np})\right) \ .\label{deltaG}
\eeq
It is then easy to show, from the definition \eqref{def}, that the variation of the Ricci tensor is given by
\beq
\delta \R_{mn}= \nabla_p (\delta \G^p_{mn}) -  \nabla_n (\delta \G^p_{pm}) \ ,\label{PalatiniR}
\eeq
which is the Palatini identity. Combining \eqref{deltaG}, \eqref{PalatiniR} and the metric compatibility leads to
\beq
\tg^{mn} \delta \R_{mn}= \tg^{mn} \tg^{pq}\ \nabla_p \left(\nabla_m (\delta\tg_{qn}) - \nabla_q (\delta\tg_{mn}) \right) \ .\label{NderEinsteinR}
\eeq
One can show that the following analogous identities hold
\bea
& \delta \cG_p^{mn} = \frac{1}{2} \tg_{pq} \left(\cN^m (\delta\tg^{qn})+\cN^n (\delta \tg^{qm})-\cN^q (\delta \tg^{mn})\right) \ ,\label{deltacG}\\
& \delta \cR^{mn} = \cN^p (\delta \cG_p^{mn}) -  \cN^m (\delta \cG_p^{pn}) \ ,\label{PalatinicR}\\
& \tg_{mn} \delta \cR^{mn}= \tg_{mn} \tg_{pq}\ \cN^p \left(\cN^m (\delta\tg^{qn}) - \cN^q (\delta\tg^{mn}) \right) \ .\label{NderEinsteincR}
\eea
To prove the first identity, it is helpful to consider separately the contributions of the two pieces of $\cG$. One deduces from this identity that $\delta \cG_p^{mn}$ is a tensor, which allows to take its covariant derivative $\cN$ in the second identity. It is also worth noting that $\delta \cG_p^{mn}$ is symmetric in $m,n$, even if $\cG_p^{mn}$ is not. These two properties will not hold for the variation with respect to $\b$, as discussed in appendix \ref{ap:eombeta}. Using those and the definition of $\cR^{mn}$ \eqref{cR}, one can prove \eqref{PalatinicR}. One then gets the final \eqref{NderEinsteincR} with the metric compatibility \eqref{compatibility}.

From \eqref{NderEinsteinR} and \eqref{NderEinsteincR}, one can get simple derivatives, which are crucial in the Einstein equation. To do so, one needs the following properties, for a vector $V^p$ or a co-vector $V_p$
\beq
\nabla_p V^p = \frac{1}{\sqrt{|\tg|}}\ \del_p \left(\sqrt{|\tg|}\ V^p \right) \ , \ \cN^p V_p = \frac{1}{\sqrt{|\tg|}}\ \del_p \left(\sqrt{|\tg|}\ \b^{pm}\ V_m \right) + 2 \T^p V_p \label{tracecN} \ ,
\eeq
where the second one is obtained using \eqref{Tapp}. Comparing these two relations highlights the specific role of $\T^p$: the way it enters here is reminiscent of its interpretation as a conformal weight, discussed below \eqref{xiT}. We deduce from these relations\footnote{For an alternative derivation of \eqref{totderEinsteinR} and of the standard dilaton and Einstein equations, see \cite{Andriot:2011iw}.}
\bea
& \sqrt{|\tg|}\ \tg^{mn} \delta \R_{mn} = 2\ \del_p \left(\sqrt{|\tg|}\ \tg^{p[q} \tg^{m]n} \nabla_m (\delta\tg_{qn}) \right) \label{totderEinsteinR} \ ,\\
& \sqrt{|\tg|}\ \tg_{mn} \delta \cR^{mn}= 2\ \del_r \left(\sqrt{|\tg|}\ \b^{rp} \tg_{p[q} \tg_{m]n} \cN^m (\delta\tg^{qn}) \right) + \sqrt{|\tg|}\ 4 \T^p \tg_{p[q} \tg_{m]n} \cN^m \delta \tg^{qn} \ .\label{totderEinsteincR}
\eea

Finally, deriving the equation of motion for $\b$ will also require some properties. A crucial one is the rewriting \eqref{cG_t} of $\cG$ as the sum of a tensor $\cG_{\!\!(t)}$ and a non-tensorial second term. The fact that the trace $\T^n$ given in \eqref{tracecG} is a tensor then appears obvious, as the second term in $\cG$ vanishes when being traced
\beq
\T^n=\cG^{pn}_p= \cG_{\!\!(t)}{}^{pn}_p = \na_p \b^{np} \ .
\eeq
Another interesting consequence of \eqref{cG_t} is the following relation between $\na$ and $\cN$
\beq
\cN^m V^p = -\b^{mn} \na_n V^p - \cG_{\!\!(t)}{}^{mp}_n V^n \ , \ \cN^m V_p = -\b^{mn} \na_n V_p + \cG_{\!\!(t)}{}^{mn}_p V_n \ ,\label{relnacN}
\eeq
which is naturally generalized for tensors of any rank. We will only use the above at the end of the derivation of the equation of motion for $\b$, but it may have more implications. One of them is the rewriting \eqref{RfluxcN} of the $R$-flux, easily obtained using \eqref{relnacN} and the definition \eqref{cG_t}.

\subsection{Einstein and dilaton equations of motion}

Let us now turn to the derivation of the equations of motion from $\tL_{\b}$ given in \eqref{L2}. We first consider the Einstein equation. The variation of the Lagrangian with respect to the metric gives at first
\bea
e^{2d} \delta \tL_{\b}  = & -\frac{1}{2} \tg_{pq} \delta \tg^{pq} \left(\R(\tg) + \cR(\tg) +4(\del \tp)^2 -\frac{1}{2} R^2 + 4 (\b^{mr}\del_r \tp - \T^m)^2  \right) \label{dL21}\\
& + (\delta \tg^{pq}) \R_{pq} + (\delta \tg_{pq}) \cR^{pq} - \frac{1}{4} (\delta \tg_{mn}) \tg_{rs} \tg_{uv} R^{mru} R^{nsv} \nn\\
& + \tg^{pq} \delta \R_{pq} + \tg_{pq} \delta \cR^{pq} + 4 \delta (\del \tp)^2 + 4 \delta (\b^{mp}\del_p \tp - \T^m)^2 \ .\nn
\eea
The last line of this expression requires some attention. Thanks to \eqref{totderEinsteinR} and \eqref{totderEinsteincR}, the terms in $\delta \R_{pq}$ and $\delta \cR^{pq}$ give derivatives multiplied by a dilaton factor. This prevents them from being total derivatives, but we can reach this by considering additional terms in derivative of the dilaton. Those would then get mixed with the variation of the two dilaton terms present in the last line of \eqref{dL21}. Note that this computation occurs because we work in the string frame and not the Einstein frame. Introducing a further total $\nabla$, one finally gets for the standard Ricci tensor and dilaton kinetic term
\beq
\tg^{pq} \delta \R_{pq} + 4 \delta (\del \tp)^2 = e^{2d}\ \del(\dots) + 2\  \delta \tg^{pq} \left(\nabla_p \del_q \tp + \tg_{pq} (2 (\del \tp)^2 -\nabla^2 \tp ) \right) \ .\label{Riccidil}
\eeq
Using a similar procedure, one gets at first
\bea
\tg_{pq} \delta \cR^{pq} &= e^{2d}\ \del(\dots) + 4 \tg_{p[q} \tg_{m]n} (\b^{rp} \del_r \tp + \T^p) \cN^m (\delta\tg^{qn}) \\
 & = e^{2d}\ \del(\dots) - 4 (\b^{pr} \del_r \tp - \T^p) (\b^{mq} \del_q \tp - \T^m) (\delta \tg_{pm} + \tg_{pm} \tg_{sn} \delta \tg^{sn}) \nn\\
 & \phantom{= e^{2d}\ \del(\dots)}\ + 4 \tg_{p[q} \tg_{m]n} (\delta \tg^{qn}) \cN^m (\b^{pr} \del_r \tp - \T^p) \ ,\nn
\eea
where to go from the first equality to the second, we used the Leibniz rule or ``integration by parts'' on $\cN$ (valid given its definition \eqref{defcN}), as well as \eqref{tracecN}, and some rearranging of terms. We then compute $\delta (\b^{mq}\del_q \tp - \T^m)^2$; we do so using \eqref{tracecG}, an ``integration by parts'' with $\del$, as well as \eqref{tracecN} and some rearranging of terms. We finally obtain
\bea
\tg_{pq} \delta \cR^{pq} + 4 & \delta (\b^{mp}\del_p \tp -\T^m)^2 = e^{2d}\ \del(\dots)\\
&+ 2\ \delta \tg_{pq} \left(\cN^p(\b^{rq} \del_r \tp + \T^q) - \tg^{pq} \left(2 (\b^{mr}\del_r \tp-\T^m)^2 - \tg_{mn}\cN^m(\b^{rn}\del_r \tp+ \T^n) \right) \right) \ ,\nn
\eea
which is analogous to \eqref{Riccidil}. Compiling these results, including the two dilaton terms of the first line of \eqref{dL21}, one gets eventually
\bea
e^{2d} \delta \tL_{\b} + e^{2d} \del(\dots) = \delta \tg^{pq}& \Bigg( -\frac{1}{2} \tg_{pq} \left(\R(\tg) + \cR(\tg) -\frac{1}{2} R^2 \right) \label{dL22}\\
& + \R_{pq} - \tg_{pm} \tg_{qn} \cR^{mn} + \frac{1}{4} \tg_{pm} \tg_{qn} \tg_{rs} \tg_{uv} R^{mru} R^{nsv} \nn\\
& + 2 \left(\tg_{pq} (\del \tp)^2 + \nabla_p \del_q \tp - \tg_{pq} \nabla^2 \tp \right) \nn\\
& + 2 \left(\tg_{pq} (\b^{mr}\del_r \tp - \T^m)^2 + ( \tg_{pm} \tg_{qn} + \tg_{pq} \tg_{mn} ) \cN^m(\b^{nr} \del_r \tp - \T^n) \right) \Bigg)\ . \nn
\eea
Let us now turn to the variation with respect to the dilaton. The variation of the two dilaton terms requires to use the Leibniz rule. We then apply on those \eqref{tracecN}, to rearrange the result into
\bea
e^{2d} \delta \tL_{\b} + e^{2d} \del(\dots) = 4 \delta \tp \Bigg(& -\frac{1}{2} \left(\R(\tg) + \cR(\tg) -\frac{1}{2} R^2 \right) + 2 (\del \tp)^2 - 2 \nabla^2 \tp \label{dL23}\\
& + 2 (\b^{mr}\del_r \tp - \T^m)^2 + 2 \tg_{mn}\cN^m(\b^{nr}\del_r \tp - \T^n) \Bigg)\ . \nn
\eea
One can read directly from the last expression the dilaton equation of motion \eqref{dileom}. It is then customary to use it to simplify the Einstein equation that can be read from \eqref{dL22}, towards \eqref{Einstein}.\footnote{The symmetry of $\delta \tg^{pq}$ in the variation \eqref{dL22} leads to considering only the symmetric part (in $p,q$) of the whole bracket there. For the standard terms, this gives no constraint (in particular $\nabla_p \del_q \tp = \nabla_p \na_q \tp$ is symmetric). But the same does not hold for the other terms, hence the symmetric parts in \eqref{Einstein}. One could study the symmetry of $\cN^m(\b^{nr}\del_r \tp)= -\cN^m \cN^n \tp$, but $\T^n$ is not expressible with a $\cN$. The identity \eqref{identkindilT} does not seem to help either. A similar phenomenon will occur for the equation of motion of $\b$.} We are left with the equation of motion for $\b$.

\subsection{$\b$ equation of motion}\label{sec:beom}

To get this last equation, we vary $\tL_{\b}$ \eqref{L2} with respect to $\b$, which gets the following three contributions
\beq
\delta \tL_{\b} =  e^{-2d}\ \bigg( \delta \cR(\tg) -\frac{1}{2} \delta R^2 + \delta \left( 4 (\b^{mp}\del_p \tp - \T^m)^2 \right) \bigg)  \ . \label{dL2db}
\eeq

Let us start with the dilaton term. It will be useful when studying the more involved $\delta \cR(\tg)$. With $\T^m=\na_p \b^{mp}$ as in \eqref{tracecG}, one first has the identity
\beq
\b^{nq}\del_q \tp - \T^n = e^{\tp} \na_q (e^{-\tp} \b^{qn}) \ ,\label{identkindilT}
\eeq
which can be used to rewrite the dilaton term. Its variation is then simple to compute. One uses in addition the Leibniz rule (``integration by parts'') for $\na$ to get a total derivative thanks to \eqref{tracecN}, and the other following term
\beq
e^{-2d} \delta \left( 4 (\b^{mp}\del_p \tp - \T^m)^2 \right) = 8 e^{-2d} \tg_{mn} \delta \b^{mp} e^{\tp} \na_p \na_q (e^{-\tp} \b^{qn}) + \del(\dots) \ . \label{dilvarb}
\eeq

We now turn to the variation $\delta \cR(\tg)$. This one is more involved, so we detail its derivation in appendix \ref{ap:eombeta}, and only give the main steps here. We first make an important use of the rewriting of $\cG$ as in \eqref{cG_t}. $\delta \cG_{\!\!(t)}$ is a tensor, similarly to \eqref{deltaG} and \eqref{deltacG}, so one can consider its covariant derivative, from which we write the variation of \eqref{cR} $\delta \cR^{mn}$ analogously to the Palatini identity \eqref{PalatinicR}
\bea
\delta \cR^{mn} & = \cN^p \left(\delta \cG_{\!\!(t)}{}_p^{mn}\right) - \cN^m \left(\delta \cG_{\!\!(t)}{}_p^{pn} \right) \label{dcRmndb} \\
& -(\delta \b^{pr})\del_r \cG_p^{mn}+(\delta \b^{mr})\del_r \cG_p^{pn} + 2\cG_r^{pn} \delta \cG_p^{[mr]} \nn \\
& - \cG^{pm}_r \G^n_{ps} \delta \b^{rs} - \left(\cG^{pn}_r \G^r_{ps} - \cG^{rp}_r \G^n_{ps} \right) \delta \b^{ms} - \b^{pr} \del_r \left(\G^n_{ps} \delta \b^{ms} \right) \ . \nn
\eea
We perform various manipulations on this expression, described in the appendix. We use in particular the explicit expression of $\cG_{\!\!(t)}$ in \eqref{cG_t}, to eventually obtain a simple formula for $\delta \cR$. The latter should be a tensor, leading us to identify a set of terms as the standard Riemann tensor, given in \eqref{Riemann}. Using its properties, we finally get
\beq
\delta \cR  = 2\cN^n\na_p (\tg_{nr} \delta \b^{pr}) - \cG_{\!\!(t)}{}_r^{mn} \na_m (\tg_{pn} \delta \b^{pr}) + \delta \b^{pr} \Big( 2 \tg_{pn} \na_r \na_q \b^{nq} + \b^{ms} \tg_{np} \R^n{}_{rms} \Big) \ . \label{dcRfinal1}
\eeq
It may look surprising to get the Riemann tensor in the equation of motion of $\b$. We understand it because this tensor appears within the variation of $\cR$ which is a curvature scalar, whose connection $\cG$ contains the standard connection $\G$, as can be seen easily in \eqref{cG_t}. One may still think that the Riemann tensor could be traded for (the commutator of) covariant derivatives acting on $\b$. But it turns out not to be the case; we will still manage to rewrite it in terms of the Ricci tensor. From \eqref{dcRfinal1}, we perform ``integrations by parts'', and rewrite the result in such a way that it combines nicely with the variation of the dilaton term \eqref{dilvarb}. This allows in addition to make the Ricci tensor appear. We get
\bea
& \delta \left( \cR +  4 (\b^{mp}\del_p \tp - \T^m)^2 \right) + e^{2d} \del(\dots) \label{dcRfinal4} \\
&= \delta \b^{pr} \tg_{np} \Big(\frac{1}{2} \tg_{rq} \tg^{sm} e^{2\tp} \na_m (e^{-2\tp} \na_s \b^{nq} ) + 2 \R_{sr} \b^{ns} - e^{-2\tp} \na_q ( e^{2\tp} \na_r \b^{nq}) + 4 \na_r (\b^{nq} \del_q \tp ) \Big) \ .\nn
\eea

Finally, we consider the variation of the $R$-flux term. Starting with its first expression \eqref{fluxes} in terms of $\na$ is possible, but as detailed in appendix \ref{ap:eombeta}, we prefer to use its expression \eqref{RfluxcN} in terms of $\cN$. We obtain, analogously to the $b$-field equation of motion,
\beq
-\frac{1}{2} e^{-2d} \delta R^2  = -\frac{1}{2} e^{-2d} \delta \b^{pr} \tg_{ms} \tg_{ru} \tg_{pv} \left( e^{2\tp} \cN^m (e^{-2\tp}  R^{suv}) - 2 \T^m R^{suv}\right) + \del(\dots) \ .\label{dR22}\\
\eeq
From this result and \eqref{dcRfinal4}, we finally obtain for \eqref{dL2db}
\bea
\!\! \delta \tL_{\b} + \del(\dots) &= e^{-2d} \delta \b^{pr} \tg_{np} \Bigg(-\frac{1}{2} \tg_{ms} \tg_{ru}  \left( e^{2\tp} \cN^m (e^{-2\tp}  R^{sun}) - 2 \T^m R^{sun}\right) \label{dL2bfinal}\\
& \phantom{=  } + \frac{1}{2} \tg_{rq} \tg^{sm} e^{2\tp} \na_m (e^{-2\tp} \na_s \b^{nq} ) + 2 \R_{sr} \b^{ns} - e^{-2\tp} \na_q ( e^{2\tp} \na_r \b^{nq}) + 4 \na_r (\b^{nq} \del_q \tp )  \Bigg) \nn
\eea
from which one can read the equation of motion for $\b$, given in \eqref{beom}.

\section{Beyond $\tL_{\b}$}\label{sec:BeyondLbeta}

\subsection{Dimensional reduction, new solutions, and extension of the theory}\label{sec:4dnewsol}

In the following sections \ref{sec:4d} and \ref{sec:NSNSsol}, we relate $\tL_{\b}$ to four-dimensional theories, and discuss the possibility of finding new interesting ten-dimensional solutions. Both require an usual compactification ansatz for the space-time, in particular having an internal manifold. So we assume to work with a standard underlying differential geometry, governed by the metric $\tg_{mn}$. We come back to the question of non-geometry in section \ref{sec:nongeo}.

\subsubsection{Dimensional reduction to four dimensions}\label{sec:4d}

A dimensional reduction of the ten-dimensional Lagrangian $\tL_0$ was performed in \cite{Andriot:2011uh, Andriot:2012an} as discussed below \eqref{diagram}, and this lead to non-geometric terms in the four-dimensional scalar potential. However, this did not allow a precise identification of the ten-dimensional non-geometric fluxes (by matching their four-dimensional counterparts upon reduction), especially for the $Q$-flux. As explained in the Introduction, we now have candidate expressions in flat indices for ten-dimensional $Q$- and $R$-fluxes, \eqref{Q} and \eqref{RfQf}, and a Lagrangian $\tL_{\b}$ expressed in terms of them \eqref{Lflat}, so using here $\tL_{\b}$ should make this identification more straightforward. To achieve this, one should derive from $\tL_{\b}$ a four-dimensional potential in terms of these fluxes, and then compare it with that of gauged supergravity. We perform here the first step, and comment on the second.

To get a four-dimensional scalar potential, let us come back to the dimensional reduction performed in \cite{Andriot:2011uh, Andriot:2012an}, that follows the procedure of \cite{Andriot:2010ju}. We start with a ten-dimensional action, taken here to be $\frac{1}{2 \kappa^2} \int \d x^{10} \tL_{\b}$. Its dimensional reduction will only consider the dependence on two scalar fields, the volume $\rho$ and the dilaton $\sigma$. Interestingly enough, this dependence is model-independent, and allows in addition to distinguish geometric from non-geometric terms in the potential. To perform the reduction, we first need to consider the ten-dimensional space-time as split into a four-dimensional space-time and six-dimensional compact (internal) manifold. We then pick a compactification ansatz for our fields accordingly: the metric factorizes into two parts $\tg_{(4)}$ and $\tg_{(6)}$, and the four-dimensional metric $\tg_{(4)}$ depends only on the four-dimensional coordinates (there is in particular no warp factor). The
same product structure is assumed for the vielbeins. $\b$ is chosen to have only components on the internal space, and only dependence on the internal coordinates. This makes the non-geometric fluxes to have purely internal components as well. Given this ansatz, the two scalars are defined as fluctuations around background valued fields, denoted with an index ${}^{(0)}$
\beq
\tg_{(6)mn} = \rho\ \tg_{(6)mn}^{(0)} \ , \ e^{-\tp}= e^{-\varphi}\ e^{-\tp^{(0)}} \ , \ \sigma = \rho^{\frac{3}{2}} e^{-\varphi} \ , \ \b^{mn}= \b^{(0)mn} \ .\label{scalars}
\eeq
The vacuum value of the scalars is by definition $1$. We further consider that $\tg_{(6)mn}^{(0)}$ depends purely on internal coordinates, and that $e^{\tp^{(0)}}=g_s$ is constant, while the two scalars depend a priori on four-dimensional coordinates. Finally, we define the background internal volume as $v_0=\int \d x^6 \sqrt{|\tg_{(6)}^{(0)}|}$, and the four-dimensional Planck mass $M_4$ as $M_4^2= v_0/(2 \kappa^2 g_s^2)$.

We now consider this compactification ansatz within the ten-dimensional Lagrangian $\tL_{\b}$. To start with, all indices in $\cG^{mn}_p$, $\cR^{mn}$, $\T^m$ and $R^{mnp}$ are purely internal, and these quantities do not depend on $\rho$. The corresponding terms in the Lagrangian thus only get dependence on $\rho$ by scaling. The ten-dimensional $\R(\tg)$ gets split into three terms: a four-dimensional one $\R_{(4)}$, a scaled six-dimensional one $\rho^{-1} \R_{(6)}^{(0)}$, and one depending on derivatives of $\rho$; the latter enters the scalar kinetic terms that we will denote ``${\rm kin}$''. Finally, we get interestingly that $\b^{mn} \del_n \tp =0$, while $(\del \tp)^2$ only contributes to ${\rm kin}$. Having determined the dependence on the scalars, we can now reduce to four dimensions. There, we go to the Einstein frame by scaling the components of the four-dimensional metric as $\tg_{(4)} = \sigma^{-2} \tg_{(4)}^E$. We get generically the four-dimensional action
\beq
S_E= M^2_4 \int \d x^4 \sqrt{|g^E|} \left(\R^E_{(4)} + {\rm kin} - \frac{1}{M^2_4} V(\rho, \sigma) \right)\ .
\eeq
The scalar potential has been argued \cite{Hertzberg:2007wc} to be generically given, for the NSNS sector, by
\beq
V(\rho,\sigma)= \sigma^{-2} \left(\rho^{-3}\ V_H + \rho^{-1}\ V_f + \rho\ V_Q + \rho^{3}\ V_R \right) \ .\label{potgen}
\eeq
where the four different terms should get contributions from the flux in their index. From $\tL_{\b}$ \eqref{L2}, we obtain here
\bea
& V_f =-\frac{M^2_4}{v_0} \int \d x^6 \sqrt{|\tg_{(6)}^{(0)}|}\ \ \R_{(6)}^{(0)} \ , \ \ \qquad \qquad \qquad V_H=0 \ ,\label{potcurved}\\
& V_Q =-\frac{M^2_4}{v_0} \int \d x^6 \sqrt{|\tg_{(6)}^{(0)}|}\ \left( \cR^{(0)} + 4 (\T^{(0)})^2 \right) \ , \ V_R =\frac{M^2_4}{2 v_0} \int \d x^6 \sqrt{|\tg_{(6)}^{(0)}|}\ {R^{(0)}}^2 \ .\nn
\eea
As mentioned in \cite{Andriot:2011uh, Andriot:2012an}, getting non-geometric terms $V_Q$ and $V_R$ is already an interesting result: these cannot be obtained from the standard NSNS Lagrangian $\L_{{\rm NSNS}}$ \eqref{LNSNS}.

Let us now rewrite these potential terms using flat indices, so that we see directly the dependence on the fluxes. This is simply done thanks to the formulas \eqref{Ricflat}, \eqref{defRQ} and \eqref{cRRqRf}. It is however customary in four dimensions \cite{Shelton:2005cf} to require the tracelessness of $f$ and $Q$
\beq
\forall b\ ,\ f^a{}_{ab}=0 \ , \ Q_a{}^{ab}=0 \ , \label{traceless}
\eeq
where the former, also known as the unimodularity of the corresponding Lie algebra, can be justified from a higher dimensional perspective by requiring the compactness of the internal manifold. This condition \eqref{traceless} has two interesting effects: first, derivatives of fluxes disappear from the potential terms \eqref{potcurved}, without having assumed the former to be constant; in addition, the explicit dependence on $\b$ disappears from $\cR$, which is of interest if we want the potential to be given purely in terms of fluxes. There remains however one explicit dependence on $\b$ within $\T^a$, as can be seen from the formula \eqref{Ta}. We could therefore consider the further restriction
\beq
\forall b\ , \ \T^b=0 \ .\label{T=0}
\eeq
This is also justified by the role of $\T^a$ as a conformal weight, similarly to the dilaton, discussed below \eqref{xiT} and \eqref{tracecN}. As the dilaton terms do not contribute to the potential, we may require the same for $\T^a$. We come back to this condition in footnote \ref{foot:T=0}. The two simplifications \eqref{traceless} and \eqref{T=0} are equivalent to the following conditions on $\b$, as can be read from the various definitions
\beq
\forall b\ ,\ f^a{}_{ab}=0 \ , \ Q_a{}^{ab}=0 \ , \ \T^b=0 \Leftrightarrow \forall b\ ,\ f^a{}_{ab}=0 \ , \ \b^{cd} f^b{}_{cd} = 0 \ , \ \del_a \b^{ab} = 0 \ . \label{simplif}
\eeq
Thanks to those, all explicit dependence on $\b$ in the potential \eqref{potcurved} is removed, and we finally obtain
\bea
& V_f =\frac{M^2_4}{4 v_0} \int \d x^6 \sqrt{|\tg_{(6)}^{(0)}|}\ \left( \eta_{ad} \eta^{be} \eta^{cg} \ f^a{}_{bc} f^d{}_{eg} + 2 \eta^{cd}\ f^a{}_{bc} f^b{}_{ad} \right) \ , \label{potflat} \\
& V_Q =\frac{M^2_4}{4 v_0} \int \d x^6 \sqrt{|\tg_{(6)}^{(0)}|}\ \left( \eta^{ad} \eta_{be} \eta_{cg} \  Q_a{}^{bc} Q_d{}^{eg} + 2 \eta_{cd}\ Q_a{}^{bc} Q_b{}^{ad} +2 \eta_{ab}\ R^{acd} f^b{}_{cd} \right) \ ,\nn\\
& V_R =\frac{M^2_4}{4 v_0} \int \d x^6 \sqrt{|\tg_{(6)}^{(0)}|}\ \ \frac{1}{3} \eta_{ad}\eta_{be} \eta_{cg}\  R^{abc} R^{deg}\ ,\nn
\eea
where all fluxes are purely internal, and taken in their background value. Let us make a few comments on that result. To start with, these formulas are fairly simple, with respect to those of \cite{Andriot:2012an} for instance. They simplify even more if the fluxes are constants, since the integral becomes just a factor $v_0$. As this potential \eqref{potflat} is expressed purely in terms of fluxes, it should allow a simple comparison to four-dimensional gauged supergravities. It looks rather likely to match the scalar potential obtained from the ``STU-models'' considered for instance in \cite{Shelton:2005cf, Aldazabal:2006up, deCarlos:2009qm, Danielsson:2012by, Damian:2013dq} and references therein. This would provide a definite identification of the ten-dimensional non-geometric fluxes. Another interesting feature of this result is the somehow unexpected $R^{acd} f^b{}_{cd}$ term in $V_Q$: this contribution does not depend on the $Q$-flux. But viewing this scalar potential as generated from (squares
of) a superpotential in a standard manner, the presence of such a mixed term can be understood. Given the generic scaling with $\rho$ in \eqref{potgen}, it is actually here the only possible mixed term, and only place, where it could appear. This term will also play an important role when looking for pure NSNS solutions in section \ref{sec:NSNSsol}. We leave to future work a precise verification of our potential matching the one derived from a superpotential of gauged supergravity; let us though give a further argument in favor of our scalar potential being the correct one.

The work of \cite{Aldazabal:2011nj, Geissbuhler:2011mx, Grana:2012rr, Geissbuhler:2013uka} starts with a generic generalized vielbein in DFT. Upon few assumptions, among which the dependence in these vielbeins on the (doubled) coordinates, a four-dimensional action of gauged supergravity including non-geometric fluxes is recovered by a dimensional reduction from DFT. An essential object in this procedure is a generalized structure constant, related to the generalized vielbein in a similar fashion as the standard one \eqref{fabc}. According to its indices being up or down (in the same sense as \eqref{updownnot}), this object generates one of the four NSNS fluxes appearing in \eqref{potgen}. Interestingly, this reproduction from DFT of the correct four-dimensional action, and of the Bianchi identities for the fluxes \cite{Shelton:2005cf}, only relies on that generalized structure constant. It is somehow independent of the actual content of the generalized vielbein. On the contrary, specifying the latter
corresponds to distinguishing its various components, such as $e$, $b$, $\b$, and giving accordingly a relation between the (non)-geometric fluxes and these more fundamental fields. But doing so is only a ``ten-dimensional matter'', because these fields only play a role at that level. Going from DFT, expressed in terms of a generic generalized vielbein, to four dimensions, one does not need to look at the ``intermediate step'' that is the ten-dimensional theory. In \cite{Aldazabal:2011nj, Geissbuhler:2013uka}, a form of the generalized vielbeins is nevertheless specified. Choosing $b=0$ there leads to the same generalized vielbein as our $\teee$ \eqref{genvielb} (we come back in section \ref{sec:beyond} to the case with $b \neq 0$). Applying in their DFT context the strong constraint $\tilde{\del}=0$ should then reduce their formalism to our ten-dimensional theory. As mentioned in the Introduction, the expressions for their non-geometric fluxes $Q_a{}^{bc}$ and $R^{abc}$ then also reduce to the ones
considered here. Therefore, our ten-dimensional theory is in some sense contained as a subcase in their DFT approach, where one chooses a particular explicit form for the generalized vielbein. Since they claim to reproduce the correct four-dimensional gauged supergravity, we conclude that reducing our ten-dimensional theory should provide the same result. As a consequence, the above potential should be the correct one.\footnote{This discussion and the work done here seem to be in agreement with the results of \cite{Blumenhagen:2013hva}, that appeared during the completion of the present paper.}

\subsubsection{Pure NSNS solutions}\label{sec:NSNSsol}

We now turn to the question of finding new ten-dimensional solutions, satisfying the compactification ansatz described around \eqref{scalars}. Interestingly, having a theory in ten dimensions expressed in terms of $Q$-  and $R$-fluxes allows, for the first time, to look directly there for solutions with non-geometric fluxes, especially some that fit the standard compactification ansatz. The NSNS sector alone, that we consider in this paper, may though be too restricted to get such solutions. Indeed, considering the above compactification ansatz, together with the standard ten-dimensional NSNS Lagrangian $\L_{{\rm NSNS}}$ \eqref{LNSNS}, only leads to trivial solutions. In other words, taking the above ansatz, with in particular no Ramond-Ramond or gauge flux contribution, no brane or orientifold plane (no warp factor), and a constant dilaton, leads to a solution with a vanishing $H$-flux, a flat internal manifold and a flat four-dimensional space-time. A way to reach this conclusion is to follow the
analogous reasoning \cite{Andriot:2010ju} to the one made below, where we essentially combine conditions obtained from the Einstein and dilaton equations of motion. The more general framework of \cite{Gautason:2013zw} gives the same result.\footnote{In the absence of any brane or orientifold plane, and of any Ramond-Ramond field, the integrated Einstein equation gives the cosmological constant in terms of the on-shell bulk action $S_{{\rm bulk}}$ \cite{Gautason:2013zw}. Two other equations of that paper then play a role: a first one relates $S_{{\rm bulk}}$ to the square of the $H$-flux, while a second one makes $S_{{\rm bulk}}$ vanish for this set of fields and sources. One therefore gets a four-dimensional Minkowski space-time and no flux.}

Despite this analogy, let us indicate a possibility to get pure NSNS solutions of $\b$-supergravity. Such solutions should satisfy the equations of motion given in \eqref{dileom}, \eqref{Einstein} and \eqref{beom}. As in the above ansatz, we consider a constant dilaton $\tp$. Motivated by the discussion around \eqref{T=0}, we additionally look for solutions satisfying $\T^m=0$.\footnote{\label{foot:T=0}The condition $\T^m=\na_p \b^{mp} =0$ is not too constraining for the fluxes. Indeed, this trace of $\na \b$ does not appear directly in the fluxes; equivalently in flat indices, the combinations appearing in the right-hand side of \eqref{simplif} do not enter directly the definitions of the (flat) fluxes \eqref{Q} and \eqref{RfQf}. There is also no combination of (components of) the fluxes that gives $\T^a$. So it appears like an independent quantity, that we then fix to a desired value; this is consistent with its interpretation as a
conformal weight. Note also $\T^m=0$ is an interesting intermediate condition between no assumption and the simplifying assumption of \cite{Andriot:2011uh}: the latter implied not only $\T^m=0$ but also a vanishing $R$-flux, while we can still have here a non-zero $R$-flux.} These two conditions simplify the ten-dimensional equations of motion to
\bea
& \R + \cR -\frac{1}{2} R^2  =\ 0 \ ,\label{dileomsimple}\\
& \R_{pq} - \tg_{m(p} \tg_{q)n} \cR^{mn} + \frac{1}{4} \tg_{pm} \tg_{qn} \tg_{rs} \tg_{uv} R^{mru} R^{nsv} = 0 \ ,\label{Einsteinsimple}\\
& \!\!\!\!\! -\frac{1}{2} \tg_{ms} \tg_{ru} \tg_{np}  \cN^m  R^{sun}  + \frac{1}{2} \tg_{np} \tg_{rq} \tg^{sm} \na_m \na_s \b^{nq} + 2 \tg_{n[p} \R_{r]s} \b^{ns} - \na_q ( \tg_{n[p} \na_{r]} \b^{nq}) = 0\ . \label{beomsimple}
\eea
The compactification ansatz leads to having a vanishing four-dimensional Ricci tensor from the Einstein equation (there is no contribution to the four-dimensional energy momentum tensor), so in particular
\beq
\R_{(4)} = 0 \ .
\eeq
For a maximally symmetric four-dimensional space-time, this condition makes it Minkowski. Taking the ten-dimensional trace of the Einstein equation \eqref{Einsteinsimple}, one obtains
\beq
\R_{(4)} + \R_{(6)} - \cR + \frac{3}{2} R^2 = 0 \ .
\eeq
These two conditions, together with the dilaton equation of motion \eqref{dileomsimple}, are solved by the following constraints on the internal quantities
\beq
-2 \R_{(6)} = R^2 = \cR \ . \label{solution}
\eeq
On the contrary to the situation with standard NSNS fields as in \eqref{LNSNS}, the internal quantities here can a priori be found non-vanishing! This is essentially due to the presence of three types of fluxes, instead of two for the standard NSNS case (see for instance the potential \eqref{potgen}). This asymmetry may look surprising when simply counting the degrees of freedom from the fundamental fields: $b$ and $\b$ have the same number. An asymmetry nevertheless appears when looking at the placement of indices of these two fields, with respect to that of the derivative $\del_m$. This difference allows at the supergravity level to define two fluxes from $\b$ and only one from $b$, as clearly seen from their definitions. A related question is that of the independence of $Q$ and $R$. At least, we see from \eqref{RfQf} that an $R$-flux can be present without a $Q$-flux, the other way round being obvious. Thanks to the three fluxes and associated quantities in \eqref{solution}, the system is not over
constrained, on the contrary to the standard NSNS case, so we conclude that interesting pure NSNS solution could here in principle be found.

Solving the condition \eqref{solution} is nevertheless not simple. As $R^2 \geq 0$, a non-trivial solution would have a negatively curved internal manifold: $\R_{(6)} < 0$. All nilmanifolds (except the torus), as well as some solvmanifolds, verify this requirement; they additionally satisfy by definition the unimodularity condition \eqref{traceless} on $f$. So this is an interesting set to look for solutions. A review on solvmanifolds can be found in \cite{Andriot:2010ju}, and more examples are present in \cite{Fino:2010dx, Console:2012wp}. A larger set of interesting Lie group based manifolds is described in \cite{Danielsson:2011au}. The condition \eqref{solution} however implies as well that $\cR \geq 0$. Imposing in addition the tracelessness condition \eqref{traceless} gives, from \eqref{defRQ} and \eqref{cRRqRf}
\beq
\cR= -\frac{1}{4} \left( \eta^{ad} \eta_{be} \eta_{cg} Q_a{}^{bc} Q_d{}^{eg} + 2 \eta_{cd} Q_a{}^{bc} Q_b{}^{ad} +2 R^{acd} f^b{}_{cd} \eta_{ab}\right)\ . \label{cRsol}
\eeq
Getting the above positive is not easy, as the first term is negative. The last term in $R^{acd} f^b{}_{cd}$ could certainly help, so it should better be non-vanishing. This simple analysis already leads to non-trivial constraints on the field configuration. We tried to solve the $\cR > 0$ condition on a few manifolds of \cite{Andriot:2010ju}, namely one or two Heisenberg manifolds (the twisted torus of section \ref{sec:torex}) denoted by the algebra $\Gg_{3.1}$, and those associated to $\Gg_{3.4}^{-1}$, $\Gg_{3.5}^{0}$, as well as $\Gg_{5.17}^{0,0,r}$ also called $s\ 2.5$ (the last two manifolds can be negatively curved for appropriately chosen radii). With reasonable ans\"atze for the fluxes, the sum of the last two terms in \eqref{cRsol} was either zero or positive, or when it was not the case, $\cR$ still failed to be positive. Finding pure NSNS solutions with this compactification ansatz therefore looks difficult, even if a priori possible.

Let us finally say a word on the equation of motion of $\b$ \eqref{beomsimple}, which would bring additional constraints. Given the compactification ansatz, this equation is non-trivial only for $p,r$ being internal indices. An additional scalar condition can be obtained by contracting these indices with a $\b^{pr}$, but the result is not enlightening. Similarly, replacing the Ricci tensor by its expression in the Einstein equation does not seem to help for solving the equations. Turning this equation to flat indices may on the contrary bring interesting information, as the fluxes would appear explicitly. Doing so is however not as straightforward as for the two other equations; we thus leave this to future work.

\subsubsection{Beyond $\tg, \b, \tp$: extension to a complete $\b$-supergravity and more}\label{sec:beyond}

While an extension to a complete $\b$-supergravity requires future work, the NSNS sector alone considered here could already provide a lift to (the NSNS sector of) four-dimensional solutions with non-geometric fluxes. Of particular interest are the four-dimensional de Sitter solutions found in \cite{deCarlos:2009qm, Danielsson:2012by, Blaback:2013ht, Damian:2013dq, Damian:2013dwa}. Unfortunately, all those share the property of having, together with a non-geometric flux, a non-vanishing $H$-flux. We believe this to be due to the orientifold projection considered there, which is of the $O3$ type. An absence of $H$-flux (that would fit better with our work) typically happens with an $O5$ or $O6$ projection. This leads us to the question, already discussed in \cite{Andriot:2011uh}, of having an additional $H$-flux in our ten-dimensional theory. This would imply at first to have a $b$-field, together with our $\b$. This gives however too many degrees of freedom (d.o.f.) in the NSNS sector. Such a situation could
therefore only occur in a democratic-like formalism, where one first introduces twice the standard number of d.o.f., and then imposes a constraint to get back the right amount. A particular case of such a scenario is for instance to have $b$ along some directions and $\b$ along others, a ``constraint'' that could be written as
\beq
b_{mn} \b^{np} = 0 \quad {\rm without}\ {\rm sum} \ . \label{bbetaconstraint}
\eeq
With a few more assumptions, this leads to having the $H$-flux only along directions orthogonal to the non-geometric fluxes. The corresponding ten-dimensional theory would then simply be the sum of the standard $H$-flux term and of the new $\tL_{\b}$. The de Sitter solutions mentioned above nevertheless always have the $H$-flux and a non-geometric flux sharing a common direction. So such a simple extension of our present work is not enough to lift the NSNS sector of these solutions. More involved constraints on the fields or fluxes could certainly be considered.

Having $b$ and $\b$ together could also be rephrased \cite{Andriot:2011uh} as choosing another generalized vielbein, which would have a different form than the two $\eee$ and $\teee$ \eqref{genvielb} (probably avoiding the zero blocks present there). The simple constraint \eqref{bbetaconstraint} would then correspond to $b$ and $\b$ being non-zero in orthogonal blocks. Given another generalized vielbein and a constraint, one could go through the same procedure as described in section \ref{sec:GGderiv} with the GG formalism, and derive the corresponding ten-dimensional Lagrangian. From this point of view, it is clear that the standard supergravity and the $\b$-supergravity will always be two limits of any such ``NSNS democratic formalism''. This brings further motivation to the present work.

In \cite{Aldazabal:2011nj, Geissbuhler:2013uka}, a generalized vielbein parametrised by some $b$ and $\b$ is considered, and formulas are deduced for the four NSNS fluxes in terms of these fields, as discussed at the end of section \ref{sec:4d}. As we explained there, the precise form of the generalized vielbein has no impact on whether the four-dimensional theory is correctly reproduced, as the latter only depends on the fluxes, and not the fundamental ten-dimensional fields. The meaning of the precise parametrisation used for the generalized vielbein is unclear to us at the ten-dimensional level. In particular, no constraint on $b$ and $\b$ is considered. As a consequence, we are not sure how to interpret the corresponding formulas giving the (non)-geometric fluxes in terms of $b$ and $\b$. But they might of course be helpful to go beyond the two limit cases.\\

As mentioned in the Introduction, a complete $\b$-supergravity requires to have, on top of the fermions, a Ramond-Ramond (RR) sector (type II) or a gauge flux sector (heterotic). Non-geometric RR fluxes have been considered from a four-dimensional perspective \cite{Aldazabal:2006up, Aldazabal:2008zza, Dibitetto:2010rg}. It is likely that ten-dimensional counterparts of those would allow the completion towards a type II $\b$-supergravity. Similarly to the $Q$- and $R$-fluxes in terms $\b$, we would expect a ten-dimensional non-geometric RR flux to have an expression in terms of a new type of gauge potential. This new field would be related by some redefinition to the standard RR gauge potentials, analogously to $\b$ with respect to $b$.\footnote{A similar situation could hold for the heterotic case. The introduction there of a generalized vielbein and metric as in \cite{Andriot:2011iw} should help to read the field redefinition of the gauge potentials, as in \eqref{fieldredefH}.} The introduction of a
trivector in the context of M-theory \cite{Malek:2012pw, Malek:2013sp} goes in this direction. Another possibility would be that the $b$-field, usually entering the relation between RR fluxes and gauge potentials, should now be turned into a $\b$, and this would provide new types of fluxes. We hope to come back to these ideas in future work.

Finally, RR non-geometric fluxes should have sources: those would be the non-geometric counterparts of the $D$-branes at least, if not also the orientifold planes. Such objects should as well be explored. The recent work on exotic branes is probably related to that question (see \cite{deBoer:2010ud, deBoer:2012ma} and references therein). The NSNS sector should have analogous objects, and the counterparts to the $NS5$-brane, namely the $Q$- and $R$-branes, were recently proposed and studied in \cite{deBoer:2010ud, Bergshoeff:2011se, deBoer:2012ma, Hassler:2013wsa, Kimura:2013fda}. As for the non-geometric fluxes, all these new objects could be interesting ingredients for phenomenology.

Let us conclude this discussion with a comment on branes world-volume theory. The standard DBI action for a $D$-brane with world-volume $\Sigma$, in absence of gauge flux, is given by
\beq
- T \int_{\Sigma} e^{-\p} \sqrt{|P_{\Sigma}(g+b)|} \ ,
\eeq
where $T$ is the tension of the brane and $P_{\Sigma}$ is the pullback to the world-volume, applied on the ten-dimensional metric and $b$-field. Using the field redefinition \eqref{fieldredef1} and \eqref{fieldredef2}, we reformulate this action into
\beq
- T \int_{\Sigma} e^{-\tp} \sqrt{|\tg|} \sqrt{|\tg^{-1}+\b|} \sqrt{|P_{\Sigma}((\tg^{-1}+\b)^{-1})|} \ ,
\eeq
which would further simplify if the ten-dimensional space-time splits into $\Sigma$ and an orthogonal space. This rewritten DBI action may play a role for the world-volume theory of the non-geometric counterparts to $D$-branes. The new gauge potentials mentioned above should then enter the other parts of such a theory.

\subsection{From non-geometry to geometry}\label{sec:nongeo}

So far, we have either worked locally with our Lagrangian $\tL_{\b}$, or assumed an underlying differential geometry governed by the metric $\tg_{mn}$, together with possible non-geometric fluxes. This last situation is certainly the one of interest for many applications. Although we have not discussed this point so far, a ten-dimensional geometry with non-geometric fluxes could be related to a non-geometry in the sense of \cite{Hellerman:2002ax, Dabholkar:2002sy, Flournoy:2004vn}. This question was addressed in \cite{Andriot:2011uh, Andriot:2012an, Andriot:2013txa} and we do not repeat here the detailed arguments, but rather bring some new material. We first present a toroidal example of \cite{Kachru:2002sk, Lowe:2003qy}: it provides a good illustration of various aspects further discussed, in particular the supergravity limit. We then tackle the question of the (non)-geometry underlying our theory by studying the gauge symmetries of $\tL_{\b}$, and analysing whether they can realise the transition functions to be used (those preserving the form of the generalized vielbein $\teee$). For the standard NSNS description, using other symmetries for the transition functions is the typical signal of non-geometry. This discussion is related to the notion of the generalized cotangent bundle that we introduce.

\subsubsection{The toroidal example and supergravity limit}\label{sec:torex}

This example is made of three different NSNS field configurations, that are T-dual to each other thanks to Buscher T-dualities. These NSNS field configurations can either be completed with ingredients of other sectors to get supergravity solutions \cite{Marchesano:2007vw}, or be background themselves within a dilute flux approximation \cite{Andriot:2012vb}. The last situation was used to study classical and quantum properties of a closed string of these approximated backgrounds, and non-commutativity was shown to appear on the non-geometric one \cite{Andriot:2012vb}. We use here the conventions of that paper (in particular $\alpha'=\frac{1}{2}$ and $2\pi H$ is quantized), and detail these three field configurations in table \ref{tab:torex}. The first one is a three-torus along $x^1=x,\ x^2=y,\ x^3=z$ with radii $R_{m=1,2,3}$, together with a dilaton $\p_0$, and a $b$-field linear in the coordinate $z$, giving a constant $H$-flux. The second one is a twisted torus, where the circle along $x$ is fibered over the base torus along $y,\ z$. It is the Heisenberg manifold, the simplest example of nilmanifold; we come back to its fibration in section \ref{sec:ET*}. Additionally, there is no $b$-field and $H$-flux, but a non-zero structure constant $f$, that can be computed using \eqref{frametorex} and \eqref{Cartan}. This field configuration is T-dual to the first one along the $x$ direction, which is an isometry. Performing a T-duality along the other isometry, namely the $y$ direction, one gets the third field configuration, that is non-geometric. To see the latter, one should study how the fields patch when going around the base circle along $z$. For this, it is convenient to compute the generalized metric $\hhh$ given in \eqref{fieldredefH}, from which one obtains
\beq
T_{{\cal C}}^T\ \hhh \big|_{z=0}\ T_{{\cal C}} = \hhh \big|_{z=2\pi}\ , \quad {\rm where}\ \ T_{{\cal C}}=\begin{pmatrix} \id_3 & \varpi \\ 0 & \id_3 \end{pmatrix} \ , \ \varpi= \begin{pmatrix} 0 & 2\pi H & 0 \\ -2 \pi H & 0 & 0 \\ 0 & 0 & 0 \end{pmatrix} \ . \label{TCtorex}
\eeq
The transition matrix $T_{{\cal C}}$ is the one needed to patch the generalized metric around this circle (we come back in section \ref{sec:ET*} to such matrices). $T_{{\cal C}}$ can be viewed as a trivial embedding in $O(3,3)$ of an element of the T-duality group $O(2,2)$. It does not take the form of a diffeomorphism or a $b$-field gauge transformation (see section \ref{sec:ET*}). It is therefore a purely stringy symmetry, required for the global completion of the geometry and field configuration: this makes the latter an example of a non-geometry. Moving away from a standard differential geometry can be problematic when using this solution, especially for a compactification: the space is not a manifold anymore, preventing it from serving as the internal compact piece. Another indication of this point is given by the volume form which is not globally well-defined anymore: indeed the factor $\sqrt{|g|}$ depends on the non single-valued function $f_0$. This can also be seen through the dilaton being ill-defined.

\begin{table}
\begin{center}
\begin{tabular}{|c||c|c|}
\hline
Configuration &  Fields & Flux (flat indices) \\
\hline
\hline & &\\[-0.5ex]
Torus $\&$ $H$-flux & $g = \begin{pmatrix} R_1^2 & 0 & 0 \\ 0 & R_2^2 & 0 \\ 0 & 0 & R_3^2 \end{pmatrix}\ , \ b = \begin{pmatrix} 0 & H z & 0 \\ - H z & 0 & 0 \\ 0 & 0 & 0 \end{pmatrix}\ , \ e^{-2\p}= e^{-2\p_0}$ & $H_{123}=\frac{H}{R_1R_2R_3}$ \\[5ex]
\hline & &\\[-0.5ex]
Twisted torus &  $g = \begin{pmatrix} \frac{1}{R_1^2} & - \frac{H z}{R_1^2} & 0 \\ - \frac{H z}{R_1^2} & R_2^2 + \left(\frac{H z}{R_1} \right)^2 & 0 \\ 0 & 0 & R_3^2 \end{pmatrix} \ , \ b = 0 \ , \ e^{-2\p}= e^{-2\p_0} R_1^2$ & $f^1{}_{23}=- \frac{H}{R_1R_2R_3}$ \\[6.5ex]
\hline & & \\[-0.5ex]
Non-geometry & $g = f_0 \begin{pmatrix} \frac{1}{R_1^2} & 0 & 0 \\ 0 & \frac{1}{R_2^2} & 0 \\ 0 & 0 & \frac{R^2_3}{f_0} \end{pmatrix} \ , \ b =  f_0 \begin{pmatrix} 0 & -\frac{H z}{R_1^2 R_2^2} & 0 \\ \frac{H z}{R_1^2 R_2^2} & 0 & 0 \\ 0 & 0 & 0 \end{pmatrix}\ , $ & \\
 & $ e^{-2\p}= e^{-2\p_0} R_1^2 R_2^2\ f_0^{-1} \ , \ {\rm with}\ f_0=\left(1 + \left(\frac{H z}{R_1 R_2} \right)^2 \right)^{-1} $ & \\[3.5ex]
\hline
\end{tabular}
\caption{The toroidal example, with the standard NSNS fields and fluxes}\label{tab:torex}
\end{center}
\end{table}

These properties of the non-geometric field configuration are also known to be problematic for the supergravity limit. The function $f_0$ makes the radii go from below the string scale to above, spoiling a possible large volume limit. Indeed, $f_0(z=0)=1$, giving a large volume limit in the regime $R_1 \sim R_2 \ll 1$ (choosing for simplicity the two fiber radii to be of the same order) so that $g_{11} \sim g_{22} \gg 1$. But one gets $f_0(z=2\pi)=$ {\small $1 \Big/ \left(1 + \left(\frac{2\pi H}{R_1 R_2} \right)^2 \right)$}, where $2\pi H$ is quantized, when going around the base circle, leading to $g_{11} \sim g_{22} \sim \left(\frac{R_1}{2\pi H}\right)^2 \ll 1$: the large volume limit is then lost. This variation of $f_0$ within the dilaton prevents us furthermore from defining a (small) string coupling constant as done below \eqref{scalars}. Note that despite these two issues with the supergravity limit, this non-geometric configuration is thought to lead to an admissible string background, because it is T-dual
to standard geometric situations \cite{Hull:2004in}.

The field redefinition \eqref{fieldredef1} and \eqref{fieldredef2}, or extensions of it as described in section \ref{sec:beyond}, has been proposed \cite{Andriot:2011uh} to cure the problems of non-geometry, by restoring a standard geometry and introducing new fluxes. For the toroidal example, the new fields, computed from the standard NSNS ones of the non-geometric configuration, are given by
\beq
\! \tg = \begin{pmatrix} \frac{1}{R_1^2} & 0 & 0 \\ 0 & \frac{1}{R_2^2} & 0 \\ 0 & 0 & R_3^2 \end{pmatrix}\ , \ \b = \begin{pmatrix} 0 & H z & 0 \\ - H z & 0 & 0 \\ 0 & 0 & 0 \end{pmatrix}\ , \ e^{-2\tp}= e^{-2\p_0} R_1^2 R_2^2\ , \ Q_3{}^{12}= \frac{H}{R_1R_2R_3} \ . \label{tgbtorex}
\eeq
A standard geometry of a three-torus is restored, together with a well-defined dilaton. When going around the circle along $z$, the $\b$ patches with a constant shift, so the $Q$-flux is globally well-defined. There are further non-trivial checks indicating that this field configuration is a good one to consider. The dependence in radii of the metric and the dilaton are the expected (T-dual) ones. Also, the ten-dimensional $Q$-flux, computed from \eqref{Q}, has the same value as the fluxes of the other field configurations, as expected from the four-dimensional T-duality chain \cite{Shelton:2005cf}. Additionally, it is worth emphasising that we restore, together with a standard geometry and the well-definedness of fields, a supergravity limit. A large volume limit is certainly possible in the regime $R_1 \sim R_2 \ll 1$, which is the (expected) T-dual regime to the other two geometric field configurations. This regime is also compatible with a small string coupling constant given by the new dilaton. We believe
that these interesting properties should hold for more examples: for instance, the two inverse powers involved in the redefinition of the metric from $g$ to $\tg$ \eqref{fieldredef1} are generically responsible for the correct dependence in radii, and the large volume limit. This toroidal example therefore illustrates the role of $\tL_{\b}$, and $\b$-supergravity, in providing a ten-dimensional geometric description of some non-geometries (on the contrary to $\L_{{\rm NSNS}}$).\footnote{In \cite{Andriot:2011uh, Andriot:2012an, Andriot:2013txa} was underlined the role of the total derivative difference between $\L_{{\rm NSNS}}$ and the new Lagrangian. This $\del(\dots)$ being ill-defined would allow to have only one of the two Lagrangians well-defined, and thus a preferred description. Considering the condition $\T^m=0$ discussed in section \ref{sec:4dnewsol}, we see in \eqref{relLag} that the total derivative between $\L_{{\rm NSNS}}$ and $\tL_{\b}$ simplifies but does not vanish: it is
 given by $\del \ln \frac{|\tg|}{|g|}$, which as argued above, is typically ill-defined.} Our formalism may also describe more.

We have discussed how a large volume limit can be restored using $\tL_{\b}$ and its fields. For a complete supergravity limit, one should though consider all higher order corrections in $\alpha'$ to such an effective theory, and verify that they are subdominant. This could be worked-out from a world-sheet perspective. Performing the field redefinition \eqref{fieldredef1} on the standard bosonic string $\sigma$-model gives the following action
\beq
\frac{2}{\pi} \int \d^2 \sigma  \left( \left(\tg^{-1}+ \b \right)^{-1} ({\cal X})\right)_{mn} \ \del_{\sigma_-} {\cal X}^m\ \del_{\sigma_+} {\cal X}^n \ , \label{sigmamodel}
\eeq
where we use the conventions of \cite{Andriot:2012vb}. This action may play a role in such a verification. Also, the vanishing $\b$-functionals are usually given by the standard supergravity equations of motion. As $\L_{{\rm NSNS}}$ and $\tL_{\b}$ only differ by a total derivative, their equations of motion should be the same, up to the field redefinition. Therefore, the vanishing $\b$-functionals of \eqref{sigmamodel}, completed with the dilaton $\tp$, should be given by the equations of motion derived in this paper, namely \eqref{dileom} - \eqref{beom}.

\subsubsection{$\b$ gauge transformation and generalized cotangent bundle $E_{T^*}$}\label{sec:ET*}

Let us now discuss the symmetries of $\tL_{\b}$ and some related aspects. Diffeomorphisms are its first gauge symmetry: $\tL_{\b}$ is manifestly diffeomorphism covariant, as commented in section \ref{sec:calc}. The field redefinition \eqref{fieldredef1} and \eqref{fieldredef2} relates tensors, so the notion of a diffeomorphism remains the same through this procedure. The other gauge symmetry of the standard $\L_{{\rm NSNS}}$ is given by the $b$-field gauge transformation, written for convenience with a shift matrix $s$
\beq
\begin{cases} g \rightarrow g \\
b \rightarrow b + s
\end{cases} \ ,\ \ {\rm where}\ \ s_{mn}=\del_{[m} {\it \Lambda}_{n]} \ , \label{transfob}
\eeq
and ${\it \Lambda}_m$ is subject to further constraints (see for instance \cite{Coimbra:2011nw}). By performing the field redefinition \eqref{fieldredef1} on \eqref{transfob}, we can rewrite this transformation, using matrix notation, in terms of the new fields
\beq
\begin{cases} \tg \rightarrow \left(\id + (\tg^{-1} + \b) s \right)^T \ \tg\ \left(\id + (\tg^{-1} + \b) s \right) \\
\b \rightarrow \left(\id + (\tg^{-1} + \b) s \right)^{-1} \left( \b - (\tg^{-1} + \b) s (\tg^{-1} + \b)^T \right) \left(\id + (\tg^{-1} + \b) s \right)^{-T}
\end{cases} \ . \label{transfo1}
\eeq
As $\tL_{\b}$ differs only by a total derivative from $\L_{{\rm NSNS}}$, the transformation \eqref{transfo1} should be a gauge symmetry of our Lagrangian $\tL_{\b}$: we call it the $\b$ gauge transformation. The novelty, with respect to \eqref{transfob}, is that $\tg$ changes as well as $\b$ under this gauge transformation. This is somehow expected, as the redefinition \eqref{fieldredef1} mixes the two fields. Although not problematic for the theory, this behaviour of $\tg$ is troublesome for the {\it interpretation} of this field, that we called so far a metric. A metric of a standard manifold should only transform under diffeomorphisms; if it has an additional transformation, the underlying geometry may differ from a conventional one, or the interpretation of the field is not the correct one. We come back to this point further down.\\

Let us now turn to transition functions: comparing them to the symmetries of the theory is crucial to distinguish geometry from non-geometry, as discussed recently in \cite{Blumenhagen:2013aia}. We introduced with the Generalized Geometry formalism in sections \ref{sec:preldisc} and \ref{sec:GGOdd} the notion of generalized bundle $E$ defined over a set of patches. Its structure group $O(d,d)$ is by definition made of transition functions that relate the generalized frames when going from one patch to the other. An element of the structure group, viewed as a transition matrix, therefore acts on the generalized flat index $A$. For a generalized frame allowing for a splitting, one can consider locally a generalized vielbein, as given in \eqref{genframes}. A transition matrix $T$ then relates two generalized vielbeins on patches $\zeta$ and $\vartheta$ as
\beq
T^B{}_A\ \reee^A{}_M \big|_{\zeta} = \reee^B{}_M \big|_{\vartheta} \ .\label{transflat}
\eeq
As discussed in sections \ref{sec:preldisc} and \ref{sec:GGOdd}, preserving a specific form of the generalized frame (in particular with a splitting) reduces generically the structure group to $G_{{\rm split}}$. The bundle gets reduced accordingly.\footnote{More precisely, as discussed in section \ref{sec:GGOdd}, $G_{{\rm split}}$ is a subgroup of $O(d,d)\times \mathbb{R}^+$ and it is the conformal extension of $E$ that gets reduced.} Preserving the form of the frame with $b$-field \eqref{Bsplit} amounts to maintain the block structure of the generalized vielbein $\eee$ given in \eqref{genvielb}; one then has $G_{{\rm split}}= GL(d,\mathbb{R}) \ltimes \mathbb{R}^{d(d-1)/2}$, where schematically $GL(d,\mathbb{R})$ gives the diagonal transformation on the vielbein $e$, and $\mathbb{R}^{d(d-1)/2}$ is the antisymmetric lower off-diagonal block shifting the $b$-field in $\eee$ \cite{Coimbra:2011nw}. The bundle then gets reduced to the generalized tangent bundle $E_T$, that can be viewed as a fibration of the
cotangent bundle over the tangent bundle, denoted as
\beq
\begin{array}{ccc} T^*\mmm & \hookrightarrow & E_T \\
 & & \downarrow \\
 & & T \mmm
\end{array} \label{ET}
\eeq
This particular ordering in the fibration can be understood when comparing with a standard fibration, such as the one of the twisted torus considered in section \ref{sec:torex}. The Cartan one-forms of the latter can be read from the metric in table \ref{tab:torex}, and correspond to the co-frame. These and the frame are given for that example by
\beq
\!\!\!\!\!\!\! e^a= e^a{}_m \d x^m = \begin{cases} e^1=\frac{1}{R_1} (\d x - H z \d y)\\
e^2 = R_2\ \d y \\
e^3 = R_3\ \d z
\end{cases} \!\!\!\!\!\! , \quad
\del_a= (e^{-T})_a{}^m \del_m = \begin{cases} \del_1 = R_1\ \del_x \\
\del_2=\frac{1}{R_2} (\del_y + H z \del_x)\\
\del_3 = \frac{1}{R_3}\ \del_z
\end{cases} \label{frametorex}
\eeq
The non-trivial one-form $e^1$ is given by the sum of a (local) one-form $\d x$ along the fiber and the connection one-form living typically on the base. This changes for the frame, where the non-trivial one is now $\del_2$, given by the sum of a (local) base vector $\del_y$, and the fiber vector $\del_x$ multiplied by the connection one-form component. If we compare this frame structure to that of the generalized frame with $b$-field \eqref{Bsplit}, we deduce for the latter that the one-forms $e^a$ are along the fiber, and the base directions are given by $\del_a$. The connective structure is given by the $b$-field $b_{ab}$, as mentioned in section \ref{sec:preldisc}. So we recover the structure \eqref{ET} of the generalized tangent bundle.

If we make the same comparison to the generalized frame with $\b$ \eqref{betasplit}, we obtain the opposite situation: the base directions are given by the one-forms $\te^a$, the fiber by $\del_a$ and the connective structure by $\b^{ab}$. This formula \eqref{betasplit} is only a local expression, but if there is a global completion that preserves this local form of the frame, then the corresponding bundle should be ``a generalized cotangent bundle'' $E_{T^*}$, i.e.
\beq
\begin{array}{ccc} T \mmm & \hookrightarrow & E_{T^*} \\
 & & \downarrow \\
 & & T^* \mmm
\end{array} \label{ET*}
\eeq
The associated structure group $G_{{\rm split}}$ should by definition preserve the form of the frame \eqref{betasplit}. Therefore, it is again given by $GL(d,\mathbb{R}) \ltimes \mathbb{R}^{d(d-1)/2}$, with the difference that $\mathbb{R}^{d(d-1)/2}$ now matches the antisymmetric upper off-diagonal block shifting $\b$ in the generalized vielbein $\teee$ of \eqref{genvielb}. First hints on such a bundle and its structure group were given in \cite{Grana:2008yw}.

Having presented bundles and structure groups, we now study the corresponding transition matrices. To ease their comparison to the symmetries discussed above, it is useful, when possible, to go to generalized curved indices. Having at least locally generalized vielbeins, one can define from \eqref{transflat} a transition matrix $T_{{\cal C}}$ with curved indices
\beq
(T_{{\cal C}})^M{}_N \equiv (\reee^{-1})^M{}_B \big|_{\zeta} \ T^B{}_A\ \reee^A{}_N \big|_{\zeta} =  \reee^M{}_B \big|_{\zeta}\ \reee^B{}_N \big|_{\vartheta} \ .\label{transcurv}
\eeq
The generalized vielbein, and metric $\hhh= \reee^T\ \mathbb{I}\ \reee$ as in \eqref{fieldredefH}, then transform as
\beq
\reee^A{}_M \big|_{\zeta}\ (T_{{\cal C}})^M{}_N = \reee^A{}_N \big|_{\vartheta} \ , \quad T_{{\cal C}}^T\ \hhh\big|_{\zeta}\ T_{{\cal C}} = \hhh\big|_{\vartheta} \ , \label{glueH}
\eeq
an example of which was thus given in \eqref{TCtorex}. If the form of the vielbein $\teee$ in \eqref{genvielb} is preserved by $T$ as it should be for $E_{T^*}$, then we can write $T_{{\cal C}}$ as follows
\bea
& \teee= \begin{pmatrix} \te & \te \b \\ 0 & \te^{-T} \end{pmatrix}= \begin{pmatrix} \te & 0 \\ 0 & \te^{-T} \end{pmatrix} \begin{pmatrix} \id & \b \\ 0 & \id \end{pmatrix} \ , \ \teee^{-1}= \begin{pmatrix} \te^{-1} & - \b \te^T \\ 0 & \te^T \end{pmatrix} \ ,\label{TCteee} \\
& T_{{\cal C}} = \teee^{-1} \big|_{\zeta}\ \teee \big|_{\vartheta} = \begin{pmatrix} \Delta & 0 \\ 0 & \Delta^{-T} \end{pmatrix} \begin{pmatrix} \id & \varpi_{\b} \\ 0 & \id \end{pmatrix} \ {\rm with}\ \Delta=\te^{-1} \big|_{\zeta}\ \te \big|_{\vartheta} \ , \ \varpi_{\b}=\b \big|_{\vartheta} - \Delta^{-1}\ \b \big|_{\zeta}\ \Delta^{-T} \ .\nn
\eea
It is worth noting that $T_{{\cal C}}$ has the same form as $\teee$, in particular with $\varpi_{\b}$ being antisymmetric. We can now compare these transition functions with the symmetries of the theory.\\

We consider a field configuration given by $\tg$ and $\b$ on a set of patches, together with the transition functions relating them on the overlaps; for instance, going from $\zeta$ to $\vartheta$, one has some $\Delta$ and $\varpi_{\b}$ read from \eqref{TCteee}. For this configuration to be a geometric one (in the sense of $\tL_{\b}$), these transition functions have at least to be realised as symmetries of the theory. This would hold in either of the following two cases:
\begin{itemize}
\item The first possibility is that the transition functions are realised by the gauge symmetries discussed above. For example, if an $s$ exists such that
    \bea
& \Delta = \id + (\tg^{-1} + \b)\big|_{\zeta}\  s \ , \label{exTCs}\\
& \varpi_{\b} = - \Delta^{-1} (\tg^{-1} + \b)\big|_{\zeta}\  s\ (\tg^{-1} + \b)^T\big|_{\zeta}\  \Delta^{-T} \nn \\
& \phantom{\varpi_{\b}} = - \left((\tg^{-1} + \b)^{-1}\big|_{\zeta}\  + s \right)^{-1} s \left((\tg^{-1} + \b)^{-1}\big|_{\zeta}\  + s \right)^{-T} \nn \ ,
\eea
then the transition function is completely realised by the $\b$ gauge transformation \eqref{transfo1}. Diffeomorphisms could as well be considered to realise, at least part of, the transition functions. There is actually an interesting combination of $\b$ gauge transformation and diffeomorphism. Suppose that on the overlap of two patches, given a matrix $s$ and the local expressions of $\tg$ and $\b$, one finds a diffeomorphism such that
\beq
\frac{\del x'}{\del x} = \left( \id + (\tg^{-1} + \b)\big|_{\zeta}\ s \right)(x) \ . \label{conddiff1}
\eeq
Then, transforming the fields under the $\b$ gauge transformation \eqref{transfo1} and further under (the inverse of) the diffeomorphism \eqref{conddiff1} gives on that overlap of two patches the effective transformation
\beq
\begin{cases} \tg \rightarrow \tg \\
\b \rightarrow \b - (\tg^{-1} + \b) s (\tg^{-1} + \b)^T
\end{cases} \ . \label{transfo2}
\eeq
For the concrete field configuration, the transformation of $\tg$ under the $\b$ gauge transformation is compensated by a diffeomorphism, while $\b$ is only shifted.\footnote{This effective transformation could be related to the $\b$-diffeomorphism of \cite{Blumenhagen:2012nt}, that acts in a similar fashion; studying the analogue to the condition \eqref{conddiff1} would then be interesting.} This effective transformation has the advantage of avoiding an undesired transformation of the metric discussed below \eqref{transfo1}. If the transition functions are realised in that manner, not only the field configuration is geometric in the sense of $\tL_{\b}$, but we have a standard differential geometry described by the metric $\tg$. The differential conditions \eqref{conddiff1} could be an important constraint for any field configuration that should be used in a compactification for instance, and it would be interesting to study this requirement in more details.

The constraint \eqref{conddiff1} may also be a condition to construct a cotangent generalized cotangent bundle $E_{T^*}$ over {\it a manifold} with metric $\tg$. Restricting transition functions to be part of the set of gauge transformations already allows a priori to construct the bundles corresponding to the theory. For the generalized vielbein $\eee$ with $b$-field, the elements of the transition matrix can be restricted to give only diffeomorphisms and gauge transformations \eqref{transfob}, i.e. the symmetries of the theory, and the generalized tangent bundle $E_T$ then provides a geometric picture of it. Similarly here for $\teee$ and $\tL_{\b}$, we have just discussed how the transition functions (of $E_{T^*}$) could be restricted to the symmetries of the theory, for instance through \eqref{exTCs}. The difference however with the $b$-field case is the transformation of the metric. To define a generalized cotangent bundle $E_{T^*}$ over a manifold with metric $\tg$, the restriction discussed around \eqref{conddiff1} might be necessary. We leave this point to future investigations.

Even if a field configuration $\tg$ and $\b$ is patched as discussed above through the gauge symmetries, so well described by the Lagrangian $\tL_{\b}$, there is a drawback to such a situation, pointed out in \cite{Blumenhagen:2013aia}. It is then easy to translate this whole set-up back into the standard $g$ and $b$. The transition functions, initially realised by the symmetry \eqref{transfo1} and diffeomorphisms, then translate into the symmetry \eqref{transfob} and diffeomorphisms, i.e. the gauge symmetries of the standard $\L_{{\rm NSNS}}$. This implies that the field configuration is also geometric in standard NSNS terms. Such a situation is not what was aimed at while introducing the field redefinition. Rather, an interesting case would be a field configuration non-geometric in one set of fields becoming geometric in the other set, as in the toroidal example. So a situation where transition functions are realised by the two gauge symmetries of $\tL_{\b}$ could occur, but would not be of physical interest.

The toroidal example is not realised this way, as it is non-geometric for standard NSNS fields. More precisely, cutting the base circle into two patches, one can study the transition functions of $g$ and $b$ given in table \ref{tab:torex} on the overlaps. One sees that no diffeomorphism on the metric can reproduce the change in the function $f_0$, a reason being that $f_0$ is not periodic in $z$. It follows that even if the new fields $\tg$ and $\b$ of \eqref{tgbtorex} are simpler, their transition functions are not realised by diffeomorphisms and $\b$ gauge transformation (see also \cite{Blumenhagen:2013aia}). Interestingly, the generalized metric $\hhh$ is generically unchanged by our field redefinition (see \eqref{fieldredefH}), so the transition matrix $T_{{\cal C}}$ patching $\hhh$ as in \eqref{glueH} is the same for both choices of generalized vielbein and fields. It is given here by \eqref{TCtorex}. This $T_{{\cal C}}$ has the same form as the ones \eqref{TCteee} admissible with $\teee$; more precisely, it simply shifts $\b$ by a constant. As argued above, this constant shift cannot be realised by the gauge symmetries of the theory. This brings us to the second option.

\item A second possibility is the presence of an additional symmetry, through which the transition functions are realised. The symmetries of $\L_{{\rm NSNS}}$, so of $\tL_{\b}$, are known to be only the gauge symmetries. Another symmetry would then appear only if we specify to a subcase, as a symmetry enhancement. A good example is the case studied in \cite{Andriot:2011uh}, where an additional constraint on any field $\b^{mn} \del_n \cdot = 0$ was imposed. This restricts the set of field configurations that can  be described. The Lagrangian $\tL_{\b}$ got reduced to a simpler expression where $\b$ enters only through $\del \b$. A new symmetry of the theory then appeared: the constant shifts of $\b$. The toroidal example fits well with that subcase, as it satisfies automatically the constraint. This allows to use the restricted $\tL_{\b}$ to describe it. Moreover, we just explained that its transition functions are given by a constant shift of $\b$, now a symmetry of the theory. This example can then be considered as geometric, in the sense of $\tL_{\b}$. It would be interesting to generalize the situation of \cite{Andriot:2011uh} via a more general constraint on the fields. One could think of $\T^m=0$ discussed in section \ref{sec:4dnewsol}, or the Bianchi identities on the fluxes \cite{Shelton:2005cf, Blumenhagen:2012ma, Blumenhagen:2012pc}. The consequences on the construction of a generalized cotangent bundle are also interesting. We hope to come back to this discussion in a future publication.

\end{itemize}

\vspace{0.4in}

\subsection*{Acknowledgments}

We would like to thank R. Blumenhagen, A. Deser, D. L\"ust, T. G. Pugh, F. Rennecke and C. Strickland-Constable for helpful discussions, as well as P. Patalong who was involved at an early stage of this project.

\newpage

\begin{appendix}

\section{Conventions}\label{ap:conv}

The flat indices are $a \dots l$ and the curved ones are $m \dots z$. $|\tg|$ denotes the absolute value of the determinant of the metric $\tg$. The squares introduced are defined as
\bea
& (\del \p)^2 \equiv g^{mn} \del_m \p\ \del_n \p\ , \ H^2 \equiv \frac{1}{3!} H_{mnp} H_{qrs} g^{mq} g^{nr} g^{ps} \ , \ R^2 \equiv \frac{1}{3!} R^{mnp} R^{qrs} \tg_{mq} \tg_{nr} \tg_{ps} \ ,\label{squares} \\
& (\del \tp)^2 \equiv \tg^{mn} \del_m \tp\ \del_n \tp\ , \ (\b^{mp}\del_p \tp - \T^m)^2 \equiv \tg_{mn} (\b^{mp}\del_p \tp - \T^m) (\b^{nq}\del_q \tp - \T^n) \ .\nn
\eea
For a generic metric $\tg_{mn}$ with Levi-Civita connection, the connection coefficients, covariant derivative, and Ricci scalar, are given by
\bea
& \G^m_{np}=\frac{1}{2} \tg^{mq} \left(\del_n \tg_{qp}+\del_p \tg_{qn}-\del_q \tg_{np}\right) \ ,\label{def}\\
& \na_m V^n = \del_m V^n + \G^n_{mp} V^p \ , \ \na_m V_n = \del_m V_n - \G^p_{mn} V_p \ ,\nn\\
& \R(\tg)= \tg^{mn} \R_{mn}\ , \ \R_{mn}= \del_p \G^p_{mn} - \del_n \G^q_{qm} + \G^{p}_{mn} \G^q_{qp} - \G^{p}_{mq} \G^q_{np} \ .\nn
\eea
The Riemann tensor is generically given as follows; for the Levi-Civita connection, it satisfies the following properties
\bea
& \R^n{}_{rms} = \del_m \G^n_{sr} - \del_s \G^n_{mr} + \G^q_{sr} \G^n_{mq} - \G^q_{mr} \G^n_{sq} \ , \ \R_{rs}=\R^n{}_{rns} \label{Riemann} \ ,\\
& \tg_{np} \R^n{}_{rms} = \tg_{nm} \R^n{}_{spr} = - \tg_{ns} \R^n{}_{mpr} = \tg_{ns} \R^n{}_{mrp} \ , \ \R^n{}_{[rms]} = 0 \ .\nn
\eea
Going to flat indices, we use the vielbein $\te^a{}_m$ and its inverse $\te^n{}_b$, associated to the metric as $\tg_{mn}= \te^a{}_m \te^b{}_n \eta_{ab}$, with $\eta_{ab}$ the components of the flat metric $\eta_d$. Tensors with flat indices are simply defined with the multiplication by the appropriate (inverse) vielbein(s), for instance $\b^{ab}=\te^a{}_m \te^b{}_n \b^{mn}$, and we also denote $\del_a= \te^m{}_a \del_m$. Going to matrix notation, one should be careful that the matrix product reproduces the correct index contraction. With the line index always on the left and the column on the right, whatever up or down, one then has sometimes to take the transpose. For instance, $\tg=\te^T \eta_d \te$, $\b_{{\rm flat}}= \te \b \te^T$ and $b_{{\rm flat}}=e^{-T} b e^{-1}$, where $(\te^T)_m{}^a=(\te)^a{}_m =\te^a{}_m$ and $(\te^{-T})_b{}^n=(\te^{-1})^n{}_b=\te^n{}_b$. From the vielbeins, one defines the structure constant $f^{a}{}_{bc}$ (or so-called geometric flux) as
\bea
& f^{a}{}_{bc} = 2 \te^a{}_m \del_{[b} \te^m{}_{c]} = - 2 \te^m{}_{[c} \del_{b]} \te^a{}_{m} \label{fabc} \ ,\\
& 2 \del_{[a} \del_{b]}=f^c{}_{ab} \del_c \ . \label{Cartan}
\eea
The spin connection coefficient is defined as in \eqref{defo} by $\o^a_{bc}\equiv \te^n{}_b\ \o_n{}^a{}_c \linebreak  \equiv \te^n{}_b \te^a{}_m \left(\del_n \te^m{}_c + \te^p{}_c \G^m_{np} \right)$. For the Levi-Civita connection, one can show, as given in \eqref{defof}, that $\o^a_{bc} = \frac{1}{2} \left(f^{a}{}_{bc} + \eta^{ad} \eta_{ce} f^{e}{}_{db} + \eta^{ad} \eta_{be} f^{e}{}_{dc} \right)$. One then has the following properties
\beq
\eta^{dc} \o^a_{bc} = - \eta^{ac} \o^d_{bc} \ , \ f^{a}{}_{bc} = 2 \o^a_{[bc]} \ ,\ f^a{}_{ab}=\o^a_{ab} \ .\label{prop}\\
\eeq
The Ricci scalar in flat indices is then given by $\R(\tg)=2 \eta^{bc} \del_a \o^a_{bc} + \eta^{bc} \o^a_{ad} \o^d_{bc} - \eta^{bc} \o^a_{db} \o^d_{ac}$, as in \eqref{Ricflat}.

Finally, we consider (constant) matrices $\gamma^a$ with flat indices, satisfying the standard Clifford algebra and the following associated properties \cite{Candelas:1984yd}
\bea
& \{ \gamma^a , \gamma^b \}= 2 \eta^{ab} \ , \ [ \gamma^a , \gamma^b ]= 2 \gamma^{ab} \ {\rm with} \ \gamma^{a_1 a_2 \dots a_p} \equiv \gamma^{[a_1} \gamma^{a_2} \dots \gamma^{a_p]} \  , \  \gamma^a \gamma^b = \eta^{ab} + \gamma^{ab} \label{g1}\\
& \gamma^a \gamma^{bc} = \gamma^{abc} + 2 \eta^{a[b} \gamma^{c]}\ ,\ \{ \gamma^a, \gamma^{bcd} \} = 6 \eta^{a[b} \gamma^{cd]}\ , \ \gamma^a \gamma^{bcd}=\gamma^{abcd}+3\eta^{a[b} \gamma^{cd]} \label{g2} \\
&[\gamma^{ab},\gamma^{cd}]=-8\delta^{[c}_{[g}\g_{h]}{}^{d]}\eta^{ag}\eta^{bh} ,\ \{\gamma^{ab},\gamma^{cd}\}= 2\gamma^{abcd}-4 \eta^{c[a}\eta^{b]d} \label{acomga2}\\
&\{\gamma^{abc},\gamma^{def}\}= 18\delta^{[d}_{[g}\gamma_{hi]}{}^{ef]}\eta^{ag}\eta^{bh}\eta^{ci}-12\delta_{[g}^{d} \delta_{h}^{e}\delta_{i]}^{f}\eta^{ag}\eta^{bh}\eta^{ci}  \ .\label{acomga3}
\eea

\section{Writing $\tL_{\b}$ and $\tL_0$ with flat indices and relating them}\label{ap:tLflat}

In this appendix, we write $\tL_{\b}$ given \eqref{L2} and $\tL_0$ given in \eqref{L0} in terms of fields with flat indices. In particular, we introduce the $Q$-flux defined in \eqref{Q}. We also prove the equality of these two Lagrangians, up to a total derivative that we work-out. This realises explicitly the second line of the diagram \eqref{diagram}.\\

We start by rewriting $\tL_{\b}$. It essentially amounts to writing $\cR$ in terms of objects with flat indices. Indeed, the two dilaton terms and the $R$-flux term only involve tensors, for which the rewriting with flat indices is straightforward: one simply multiplies by vielbeins. In addition, the standard Ricci scalar $\R(\tg)$ can be expressed in terms of $f$ as in \eqref{Ricflat}. So let us focus on $\cR$: it is defined in terms of the Ricci tensor $\cR^{mn}$, itself expressed in terms of the connection $\cG$ in \eqref{cR}. One can first write this $\cG$ in terms of quantities with flat indices. Using its definition \eqref{cG}, and the definitions \eqref{Q} and \eqref{defoQQ} of $Q$ and $\o_Q$, one can show
\beq
\cG_p^{mn}= - \te^a{}_p \te^m{}_b \te^n{}_c {\o_Q}_a^{bc} - \te^a{}_p \b^{mr} \del_r \te^n{}_a \ ,\label{cGoQ}
\eeq
which is equivalent to \eqref{defoQcG}. Using this expression within $\cR^{mn}$, one gets
\bea
\cR^{mn}=& \left( 2 \b^{c[g}\partial_c \b^{a]d} -2{\o_Q}_{b}^{[ag]}\b^{bd} \right) \te^{m}{}_{g} \partial_d \te^{n}{}_{a} + 2 \b^{c[g} \b^{a]d} \te^{m}{}_{g} \partial_c\partial_d \te^{n}{}_{a}\\
+& \left( \b^{bc}\partial_c {\o_Q}_{b}^{ga} -\b^{gc}\partial_c {\o_Q}_{b}^{ba} + {\o_Q}_{b}^{ga}{\o_Q}_{h}^{hb} -{\o_Q}_{b}^{hg}{\o_Q}_{h}^{ba} \right) \te^{m}{}_{g} \te^{n}{}_{a}\ . \nn
\eea
Using then the identities (analogous to the following \eqref{trick1} and \eqref{trick2})
\bea
&\ \ 2 \b^{c[g} \b^{a]d} \te^{m}{}_{g} \partial_c\partial_d \te^{n}{}_{a} = \b^{ac} \b^{gb}  f^d{}_{cb}\te^{m}{}_{g} \partial_d \te^{n}{}_{a}\ , \\
& \left( 2 \b^{c[g}\partial_c \b^{a]d}  - Q_b{}^{ag} \b^{bd} \right) \te^{m}{}_{g} \partial_d \te^{n}{}_{a} + \b^{ac} \b^{gb}  f^d{}_{cb}\te^{m}{}_{g} \partial_d \te^{n}{}_{a} = -R^{agd} \te^{m}{}_{g} \partial_d \te^{n}{}_{a} \ ,
\eea
one gets
\beq
\cR^{ab}= \b^{cd}\partial_d {\o_Q}_{c}^{ab} -\b^{ac}\partial_c {\o_Q}_{d}^{db} + {\o_Q}_{d}^{ab}{\o_Q}_{c}^{cd} -{\o_Q}_{d}^{ca}{\o_Q}_{c}^{db}  - \frac{1}{2} R^{acd} f^b{}_{cd} \ .
\eeq
Making use of some of the following properties of $\o_Q$ (analogous to those of $\o$ and $f$ in \eqref{prop})
\beq
\eta_{dc} {\o_Q}_a^{bc} = - \eta_{ac} {\o_Q}_d^{bc} \ , \ Q_{a}{}^{bc} = 2 {\o_Q}_a^{[bc]} \ , \ {\o_Q}_a^{ad} = Q_a{}^{ad}  \ ,\ \eta_{bc} {\o_Q}_a^{bc} = \eta_{ad} Q_b{}^{db} \ , \ \label{troQ}
\eeq
one forms $\R_Q$ defined in \eqref{defRQ} and obtains
\beq
\cR= \R_Q - \frac{1}{2} R^{acd} f^b{}_{cd} \eta_{ab}\ ,
\eeq
given also in \eqref{cRRqRf}.\footnote{\label{foot:cRRQap}As pointed out in footnote \ref{foot:cRRQ}, the difference between $\cR$ and $\R_Q$ is somehow unexpected. It could come from the placement of $m,n$ indices in the terms $\b^{mq}\del_q \cG_p^{pn} - \cG_p^{qm} \cG_q^{pn}$ of $\cR^{mn}$: the analogous terms in the generic $\R_{mn}$ \eqref{def} have different placements. Another possible explanation can be found in the commutator of two $\cN$: it is not only given by a Riemann tensor, but also by an $R$-flux \cite{Andriot:2012wx, Andriot:2012an}, that may correspond to the one appearing here in $\cR - \R_Q$. A consequence of this difference is on the tensorial properties: the term in $R^{acd} f^b{}_{cd}$ is a scalar with respect to diffeomorphisms, but not Lorentz transformations (similarly to $f^a{}_{bc}$). The same therefore holds for $\R_Q$, while $\cR$ is always a scalar.} This ends the rewriting of $\tL_{\b}$ with flat indices, as given in \eqref{Lflat}.

Let us now turn to $\tL_0$. It is convenient to introduce the following quantity $E^a{}_{bc}$, and show these various properties
\bea
& E^a{}_{bc} \equiv \te^a{}_m \del_b \te^m{}_c \ , \ f^{a}{}_{bc}= 2 E^a{}_{[bc]} \label{tricks}\\
& \te^m{}_c \te^a{}_p \te^b{}_q \del_m \b^{pq} = Q_c{}^{ab} - 2 \b^{d[a} E^{b]}{}_{dc} \ , \ \te^m{}_c \te^a{}_p \te^b{}_q \del_m \tg^{pq} = 2 \eta^{d(a} E^{b)}{}_{cd} \ .\nn
\eea
These formulas make it simpler to rewrite $\tL_0$ into flat indices. In particular, to introduce $Q_c{}^{ab}$, one just replaces all $\del \b$ as in \eqref{tricks}. This way, one rewrites three lines of $\tL_0$ as
\bea
& -\frac{1}{4} \tg_{mp} \tg_{nq} \tg^{rs} \ \del_r \b^{pq} \ \del_s \b^{mn}+\frac{1}{2} \tg_{mn} \del_{p} \b^{qm}\ \del_q \b^{pn} \label{L1a}\\
& + \tg_{nq} \tg_{rs} \b^{nm} \big(\del_p \b^{qr}\ \del_m \tg^{ps} + \del_p \tg^{qr}\ \del_m \b^{ps} \big) \nn \\
& -\frac{1}{4} \tg_{mp} \tg_{nq} \tg_{rs} \big(\b^{ru}\beta^{sv} \del_u \tg^{pq}\ \del_v \tg^{mn} -2 \b^{mu}\b^{nv} \del_u \tg^{qr}\ \del_v \tg^{ps} \big) \nn\\
= & -\frac{1}{4} \eta_{ab} \eta_{cd} \eta^{ef} Q_e{}^{ac} Q_f{}^{bd} + \frac{1}{2} \eta_{ab} Q_c{}^{da} Q_d{}^{cb} \nn\\
& +\frac{1}{2} \eta_{ab} \left( \b^{ac} Q_c{}^{gd} f^b{}_{gd} + 2 \b^{gc} Q_g{}^{be} f^a{}_{ec} + 4 \b^{ae} Q_g{}^{bc} E^g{}_{ec} \right) \nn\\
& +\frac{1}{2} \eta_{ab} \left( 2 \b^{ag} \b^{bd} E^f{}_{gc} E^c{}_{df} + \b^{fg} \b^{dc}  f^a{}_{gc} f^b{}_{fd} + 4 \b^{ag} \b^{cd} f^b{}_{fd} E^f{}_{(cg)} \right) \ ,\nn
\eea
where antisymmetry of $\b$, $Q$, or $f$ has been used. We then turn to the remaining line of interest in $\tL_0$. It includes the dilaton through $d$ that was defined in \eqref{fieldredef2}. For this reason the following identities are useful
\beq
\del_m d = \del_m \tp + \frac{1}{4} \tg_{pq} \del_m \tg^{pq} \ , \ \del_m \ln \sqrt{|\tg|} = -\frac{1}{2} \tg_{pq} \del_m \tg^{pq} \ ,
\eeq
as well as the trace of the connection $\T^n$ defined in \eqref{tracecG}, recognised to be a tensor
\beq
\T^n \equiv \cG_p^{pn} = \del_p \b^{np} - \frac{1}{2} \b^{nm} \tg_{pq} \del_m \tg^{pq}  = \frac{1}{\sqrt{|\tg|}} \del_p \left(\b^{np} \sqrt{|\tg|} \right) = \na_p \b^{np} \ ,\label{Tapp}
\eeq
with the standard covariant derivative defined in \eqref{def}. This $\T^n$ appears in the second dilaton term in $\tL_{\b}$. Because the structure of $\tL_{\b}$ (in both curved and flat indices) is fairly simple compared to that of $\tL_0$, it seems interesting to reach such a structure here. To do so, let us make this second dilaton term appear. So we first rewrite the line of interest of $\tL_0$ as follows
\bea
& 4 \tg_{mn}\beta^{mp} \beta^{nq} \del_p d \ \del_q d -2 \del_p d \ \del_q \left( \tg_{mn} \b^{mp}\b^{nq} \right)\\
= &\ e^{2d}\ \del_m \left(\frac{e^{-2d}}{|\tg|} \del_n \big(\tg_{pq} \b^{pm} \b^{qn} |\tg| \big) - 4 e^{-2d} \b^{pm} \tg_{pq} \T^q \right) \nn\\
& + 4 (\b^{mp} \del_p \tp - \T^m)^2 + \tg_{mn} \b^{mp} \b^{nq} \left(\frac{2}{\sqrt{|\tg|}} \del_p \del_q \sqrt{|\tg|} - 3 \del_p \ln \sqrt{|\tg|}\ \del_q \ln \sqrt{|\tg|} \right) \nn\\
& + \b^{mr} \b^{ns} \left( \tg_{mp} \tg_{nq} \del_r \del_s \tg^{pq} - 2 \tg_{nq} \tg_{mu} \tg_{pv} \del_{(r} \tg^{pq} \del_{s)} \tg^{uv} \right) \nn\\
& - 2 \tg_{mp} \tg_{nq} \del_r \tg^{pq} \left(\b^{mr} \del_s \b^{ns} - \b^{ns} \del_s \b^{mr} + \b^{mr} \b^{ns} \del_s \ln \sqrt{|\tg|} \right)\nn\\
& - \tg_{mn} \left(\del_p \b^{mp} \del_q \b^{nq} + \del_p \b^{mq} \del_q \b^{np} - 2 \b^{mp} \del_p \del_q \b^{nq} \right) \nn\\
& + \tg_{mn} \left( 2 \b^{mp} \del_p \b^{nq} \del_q \ln \sqrt{|\tg|} - 2 \b^{mp} \del_q \b^{nq} \del_p \ln \sqrt{|\tg|} \right) \ , \nn
\eea
where we essentially introduced total derivatives to treat the terms linear in $\del \tp$. One then uses \eqref{tricks} to go to flat indices, being careful with second order derivatives. The combination $E^a{}_{ba}= - \del_b \ln \sqrt{|\tg|}$ is also useful. One finally obtains
\bea
& 4 \tg_{mn}\beta^{mp} \beta^{nq} \del_p d \ \del_q d -2 \del_p d \ \del_q \left( \tg_{mn} \b^{mp}\b^{nq} \right) \label{L1b}\\
= &\ e^{2d}\ \del_m \left(\frac{e^{-2d}}{|\tg|} \del_n \big(\tg_{pq} \b^{pm} \b^{qn} |\tg| \big) - 4 e^{-2d} \b^{pm} \tg_{pq} \T^q \right) \nn\\
& + 4 (\b^{ab} \del_b \tp - \T^a)^2 + 2 \eta_{ab} \b^{ac} \del_c Q_d{}^{bd} - \eta_{ab} \left(Q_c{}^{ac} Q_d{}^{bd} + Q_c{}^{bd} Q_d{}^{ac} \right)\nn\\
&+ 2 \eta_{ab} Q_c{}^{bd} \left(\b^{fa} E^c{}_{fd} + \b^{cf} f^a{}_{fd} \right) + \eta_{ab} \b^{cd} \b^{fe} f^a{}_{fd} f^b{}_{ec} - \eta_{ab} \b^{ac} \b^{bd} E^f{}_{cg} E^g{}_{df} \nn\\
&+\eta_{ab} \b^{ac} \left(\b^{ef} \del_f E^b{}_{ce} - \b^{ef} \del_c E^b{}_{fe} + \frac{1}{2} \b^{fe} E^b{}_{cg} f^g{}_{fe} + \b^{ge} E^b{}_{fe} E^f{}_{gc} + 3 \b^{gf} E^e{}_{cg} f^b{}_{fe} \right) \ .\nn
\eea
Using in addition the identity
\beq
\del_f E^b{}_{ce} - \del_c E^b{}_{fe} = E^b{}_{cd} E^d{}_{fe} - E^b{}_{fd} E^d{}_{ce} + E^b{}_{de}f^d{}_{fc} \ , \label{trick1}
\eeq
and summing the two non-trivial contributions \eqref{L1a} and \eqref{L1b} of $\tL_0$, one eventually gets
\bea
e^{2d}\tL_0 =&\ \R(\tg) +4(\del \tp)^2 -\frac{1}{2} R^2 + 4 (\b^{ab} \del_b \tp - \T^a)^2 \label{L1fa}\\
& + 2 \eta_{ab} \b^{ac} \del_c Q_d{}^{bd} -\frac{1}{4} \eta_{ab} \eta_{cd} \eta^{ef} Q_e{}^{ac} Q_f{}^{bd} - \eta_{ab} \left(Q_c{}^{ac} Q_d{}^{bd} +\frac{1}{2} Q_c{}^{ad} Q_d{}^{bc} \right)\nn\\
& + \eta_{ab} \left(\frac{1}{2} \b^{ac} Q_c{}^{fd} f^b{}_{fd} - \b^{fc} Q_c{}^{ad} f^b{}_{fd} \right) + \frac{1}{2} \eta_{ab} \b^{fc} \b^{de} f^a{}_{fd} f^b{}_{ec} + \eta_{ab} \b^{ac} \b^{ef} f^b{}_{fd} f^d{}_{ce} \nn\\
& + e^{2d}\ \del_m \left(\frac{e^{-2d}}{|\tg|} \del_n \big(\tg_{pq} \b^{pm} \b^{qn} |\tg| \big) - 4 e^{-2d} \b^{pm} \tg_{pq} \T^q \right) \ .\nn
\eea
As already mentioned, the first row is simple to turn into flat, as mostly tensors are involved. In particular, it is worth noting for the $R$-flux, using \eqref{prop}
\bea
R^{abc} & = \te^a{}_m \te^b{}_n \te^c{}_p 3 \b^{q[m} \na_q \b^{np]} = 3 \b^{d[a} \na_d \b^{bc]} \nn\\
& = 3 \b^{d[a} \del_d \b^{bc]} - 3 \b^{d[a} f^b{}_{de} \b^{c]e} = 3 \b^{d[a} Q_d{}^{bc]} + 3 \b^{d[a} f^b{}_{de} \b^{c]e} \ ,\label{RfQf}
\eea
from which it is easy to show that
\beq
\!\!\!\!\!\!\!\!\! R^{abc} f^e{}_{bc} \eta_{ae} = -\eta_{ab} \left(\b^{ac} Q_c{}^{fd} f^b{}_{fd} -2 \b^{fc} Q_c{}^{ad} f^b{}_{fd} + \b^{fc} \b^{de} f^a{}_{fd} f^b{}_{ec} +2 \b^{ac} \b^{ef} f^b{}_{fd} f^d{}_{ce} \right) \label{trick2}
\eeq
allowing to rewrite the last but one row of \eqref{L1fa}. The second line of \eqref{L1fa} can be rewritten using $\R_Q$ defined in \eqref{defRQ}, the properties of $\o_Q$ given in \eqref{troQ}, and the following
\beq
\eta_{bc} {\o_Q}_a^{db} {\o_Q}_d^{ac} =  \frac{1}{4} \eta_{ab} \eta_{cd} \eta^{ef} Q_e{}^{ac} Q_f{}^{bd} + \frac{1}{2} \eta_{ab} Q_c{}^{ad} Q_d{}^{bc}  \ .
\eeq
Using these results, one rewrites \eqref{L1fa} as
\bea
e^{2d}\tL_0 =&\ \R(\tg) +4(\del \tp)^2 -\frac{1}{2} R^2 + 4 (\b^{ab} \del_b \tp - \T^a)^2 + \R_Q - \frac{1}{2} R^{abc} f^e{}_{bc} \eta_{ae} \label{L1fb}\\
& + e^{2d}\ \del_m \left(\frac{e^{-2d}}{|\tg|} \del_n \big(\tg_{pq} \b^{pm} \b^{qn} |\tg| \big) - 4 e^{-2d} \b^{pm} \tg_{pq} \T^q \right) \ .\nn
\eea
We recognise $\tL_{\b}$ given in \eqref{Lflat}, up to a total derivative.

From this result and \eqref{L0}, we obtain the final relation between the various Lagrangians
\bea
& \L_{{\rm NSNS}} - \del_m\left(e^{-2d}\big(\tg^{mn}\tg^{pq}\del_n\tg_{pq} - g^{mn} g^{pq} \del_n g_{pq} + \del_n(\tg^{mn}-g^{mn})\big)\right) \label{relLag}\\
= & \ \tL_0 \nn\\
= & \ \tL_{\b} + \del_m \left(\frac{e^{-2d}}{|\tg|} \del_n \big(\tg_{pq} \b^{pm} \b^{qn} |\tg| \big) - 4 e^{-2d} \b^{pm} \tg_{pq} \T^q \right) \ ,\nn
\eea
that realises the second line of the diagram \eqref{diagram}. The total derivative between $\L_{{\rm NSNS}}$ and $\tL_{\b}$ can be simplified a little by noticing at first, as in \cite{Andriot:2011uh, Andriot:2012an}, that $\tg^{mn}-g^{mn} = -\tg_{pq} \b^{pm} \b^{qn}$.\\

As a final digression, let us give a first step in an attempt to relate directly $\L_{{\rm NSNS}}$ and $\tL_{\b}$ in flat indices. Doing so would amount to translate the various quantities with flat indices appearing in the former. To start with, one can show (see \eqref{defF} and \eqref{ete}) from the field redefinition of the metric \eqref{fieldredef1} that the vielbeins are related by $e^a{}_m=k^a{}_b (F^{-1})^b{}_c\ \te^c{}_m$ where $F^a{}_b= \delta^a_b + \b^{ac} \eta_{cb}$ and $k \in O(d-1,1)$. From this relation, one can compute the structure constant $\hat{f}^a{}_{bc}$ for the vielbein $e^a{}_m$ with \ref{fabc}, in terms of the new fields. Assuming $k$ constant, one gets
\bea
\hat{f}^a{}_{bc} = (k^{-1})^d{}_b (k^{-1})^f{}_c\ k^a{}_h (F^{-1})^h{}_i\ \big( & f^i{}_{df} + 2 \eta_{l[f} F^e{}_{d]} Q_e{}^{il} + 2 \b^{ij} \eta_{l[d} f^l{}_{f]j} \\
& + 2 \eta_{jd} \eta_{lf} \b^{gi} \b^{e[j} f^{l]}{}_{eg} + \eta_{jd} \eta_{lf} \b^{ej} \b^{gl} f^{i}{}_{ge} \big) \ .\nn
\eea
This formula could then be used to rewrite the Ricci scalar of $\L_{{\rm NSNS}}$. If we make the same simplifying assumptions as in \cite{Andriot:2011uh}, leading in particular to \eqref{Qabcsimplif}, and taking as in that paper $k^a{}_b= \delta^a_b$, the above expression reduces to
\bea
& (F^{-1})^a{}_i\ \big( f^i{}_{bc} + 2 \eta_{l[c} Q_{b]}{}^{il} + 2 \b^{ij} \eta_{l[b} f^l{}_{c]j} \big) \nn \\
= & f^a{}_{bc} + (F^{-1})^a{}_i\ \big(2 \eta_{l[c} Q_{b]}{}^{il} - \b^{ij}\eta_{jl} f^l{}_{bc} + 2 \b^{ij} \eta_{l[b} f^l{}_{c]j} \big) \ ,
\eea
that corrects the wrong equation (4.19) of \cite{Andriot:2011uh}. Analogously to \eqref{RfQf}, one also has $H_{abc}= 3 \del_{[a} b_{bc]} + 3 \hat{f}^d{}_{[ab} b_{c]d}$. Using this, one could also rewrite the corresponding term in $\L_{{\rm NSNS}}$.

\section{Derivation of $\tL_{\b}$ from an $O(d-1,1) \times O(1,d-1)$ structure}\label{ap:GG}

In this appendix, we provide details on computations mentioned in section \ref{sec:OOstruct}. These allow eventually to derive the Lagrangian $\tL_{\b}$ given in \eqref{Lflat} using the Generalized Geometry formalism. We also detail the claim that the field redefinition is an $O(d-1,1) \times O(1,d-1)$ transformation.

\subsection{Determination of the $O(d-1,1) \times O(1,d-1)$ derivative}\label{ap:OOder}

In section \ref{sec:OOstruct}, we explain how preserving an $O(d-1,1) \times O(1,d-1)$ structure leads generically to the derivative \eqref{DWbubgen}. We determine here the various pieces of this derivative for the frame \eqref{betasplit} and derivative \eqref{betacovder}, following the procedure described in that section. To start with, the derivatives $\del_A$ in the unbarred - barred notation can be read after a simple rotation \eqref{rot} from the up/down one
\beq
\del_A = \begin{cases} \del_a = \del_a + \eta_{ab} \b^{bc} \del_c \\ \del_{\ov{a}} = \del_{\ov{a}} - \ov{\eta_{ab}} \b^{\ov{bc}} \del_{\ov{c}} \end{cases} \ ,
\eeq
where in the right-hand sides we do not write the $\delta$'s and use the alignment of vielbeins. We now consider the connection. $\hO$ is made of the (former) $O(d,d)$ piece $\Omega$, and a piece due to the conformal weight; let us start with $\Omega$ alone. Its fully unbarred component is given by a rotation from up/down components, as follows
\bea
\Omega_a{}^b{}_c & = P_a{}^D P_c{}^F (P^{-T})^b{}_E\ \Omega_{(u/d)D}{}^E{}_F\\
& =\frac{1}{2} \left(\delta^d_a (\Omega_d{}^b{}_c + \eta^{be} \eta_{cf} \Omega_{de}{}^f) + \eta_{ad} ( \Omega^{db}{}_c + \eta^{be} \eta_{cf} \Omega^d{}_e{}^f)  + \eta_{ad} \eta_{cf} \delta^b_e \Omega^{def}  \right) \ ,
\eea
where we used the fixing discussed in section \ref{sec:GGOdd} $\Omega_d{}^{ef}= \Omega^d{}_{ef} = \Omega_{def} = 0$, that lead eventually to the derivative \eqref{betacovder}. We also identified there
\beq
\Omega_a{}^b{}_c= \omega^b_{ac} \ , \ \Omega^a{}_b{}^c  = {\omega_{Q}}_b^{ac} \ , \ \Omega^{abc}= \Omega^{[abc]}=\frac{1}{3} R^{abc} \ .
\eeq
We recall as well that $\Omega_{De}{}^f = - \Omega_D{}^f{}_e$. Using these results and the antisymmetry properties of $\o$ and $\o_Q$, we conclude
\beq
\Omega_a{}^b{}_c = \omega^b_{ac} - \eta_{ad} {\omega_{Q}}_c^{db} + \frac{1}{6} \eta_{ad} \eta_{cf} \delta^b_e R^{def} \ .
\eeq
One proceeds similarly for the other unbarred - barred components of $\O$. A subtlety occurs for the mixed components, because of the projection to the $O(d-1,1)\times O(1,d-1)$ structure. For instance, in $\Omega_{\ov{a}}{}^b{}_c$, one gets a piece given by $-\frac{1}{2} \ov{\eta_{ag}} \eta_{ch} \delta^{\ov{g}}_d \delta^b_e \delta^h_f  \Omega^{def}$ (we write all $\delta$'s to clarify the discussion). $\Omega^{def}$ has been identified in section \ref{sec:GGOdd} with its fully antisymmetrised part, itself related to the $R$-flux. However, because of the projection, one should be careful in the placement of unbarred and barred indices. As discussed above \eqref{DWbubgen}, the two indices on the right should be of the same type. Therefore, out of the decomposition $\O^{[def]}=3(\O^{d[ef]} + \O^{f[de]} + \O^{e[fd]})$, one should only keep the contribution of the first term. This leads to $-\frac{1}{2} \ov{\eta_{ag}} \eta_{ch} \delta^{\ov{g}}_d \delta^b_e \delta^h_f  R^{def}$.

Finally, let us consider the other piece of $\hO$, namely the contribution to be added due to the conformal weight. In \cite{Coimbra:2011nw}, it is changed from \eqref{tOOL} in the up/down notation to the following in the unbarred - barred\footnote{It is worth noting that this term $\hO_A{}^B{}_C - \O_A{}^B{}_C $ in \eqref{tOOLb} is automatically compatible with the metric $\eta_{AB}$, on the contrary to the one in \eqref{tOOL}. This is consistent with the fact that $\hO$ is the $O(d-1,1)\times O(1,d-1)$ connection, and that there is no conformal factor in the structure group anymore.}
\beq
\hO_A{}^B{}_C= \O_A{}^B{}_C - \frac{1}{9} (\delta^B_A \Lambda_C - \eta_{AC} \eta^{BE} \Lambda_E) \ , \label{tOOLb}
\eeq
where we believe that the normalisation factor $9$ can be understood as $\delta^a_a -1= \delta^{\ov{a}}_{\ov{a}} - 1$. The trace of the above remains the same as that of \eqref{tOOL}, i.e. given by
\beq
\hO_D{}^D{}_C= \O_D{}^D{}_C - \Lambda_C \ .
\eeq
This implies that the identification of $\Lambda$ \eqref{xi} made thanks to the torsion-free condition is in any case valid. So we follow here the same prescription \eqref{tOOLb}, and should only define the unbarred - barred components of $\Lambda$ from the up/down ones \eqref{Laud}. This is done again by a rotation
\beq
\Lambda_C = \begin{cases} \Lambda_c = \lambda_c + \eta_{cd} \xi^d \\
             \Lambda_{\ov{c}} = \lambda_{\ov{c}} - \ov{\eta_{cd}} \xi^{\ov{d}}
            \end{cases} \ .
\eeq
Combining all these contributions to \eqref{DWbubgen}, we obtain eventually the $O(d-1,1)\times O(1,d-1)$ derivative as given in \eqref{DWbubfinal}.

\subsection{The field redefinition is an $O(d-1,1) \times O(1,d-1)$ transformation}\label{ap:K}

We make here a short digression to comment on the transformation relating the generalized vielbeins $\eee$ and $\teee$ in \eqref{genvielb}. Let us first consider formally a $2d \times 2d$ matrix $K$ given in terms of generic $d \times d$ matrices $O_1$ and $O_2$ or the combinations $O_{\pm}$ as
\beq
K= \begin{pmatrix} O_1 & O_2\ \eta_d^{-1} \\ \eta_d\ O_2 & \eta_d\ O_1\ \eta_d^{-1} \end{pmatrix} \ ,\ O_{\pm} = O_1 \pm O_2 \ . \label{defKgen}
\eeq
Then, one can show the equivalence between the four following sets of conditions
\beq
\begin{array}{c|} K \in O(2d-2,2) \\ K^T\ \mathbb{I}\ K= \mathbb{I} \end{array}\ \Leftrightarrow\ \begin{array}{|c|} K \in O(d,d) \\ K^T\ \eta_{(u/d)}\ K= \eta_{(u/d)} \end{array}\ \Leftrightarrow\ \begin{array}{|c} O_+ \in O(d-1,1)\ , \ O_- \in O(1,d-1) \\ O_{\pm}^T\ (\pm \eta_{d})\ O_{\pm}= \pm \eta_{d} \end{array} \nn
\eeq
\beq
\Leftrightarrow\ \begin{array}{|l} O_1^T \ \eta_d\ O_1 + O_2^T \ \eta_d\ O_2 = \eta_d \\ O_1^T \ \eta_d\ O_2 + O_2^T \ \eta_d\ O_1 = 0 \end{array} \ , \label{propO1O2}
\eeq
with $\mathbb{I}$ defined in \eqref{genvielb} and $\eta_{(u/d)}$ in \eqref{etaud}.

Let us now show that such a matrix $K$ is the one allowing to transform one generalized vielbein into the other
\beq
\eee= K \teee \Leftrightarrow K=\eee \teee^{-1} = \begin{pmatrix} e\te^{-1} & -e\b \te^T \\ e^{-T} b \te^{-1} & e^{-T} \te^{T} - e^{-T} b \b \te^{T} \end{pmatrix}   \ . \label{defKgenviel}
\eeq
To do so, we need the information that the fields in \eqref{defKgenviel} are not independent but related by the field redefinition \eqref{fieldredef1}. We rewrite the latter in a more convenient way\footnote{The starting point to get \eqref{defF} is to rewrite \eqref{fieldredef1} as
\beq
g=(\tg^{-1}-\b)^{-1} \tg^{-1} \tg \tg^{-1} (\tg^{-1}+\b)^{-1} \ , \ b=-(\tg^{-1}-\b)^{-1} \b (\tg^{-1}+\b)^{-1}\ ,
\eeq
where the change of sign in front of $\b$ in the brackets with respect to \eqref{fieldredef1} is actually allowed: this sign can be chosen freely without affecting the field redefinition \cite{Andriot:2011uh}.}
\beq
\begin{array}{c|} e^T \eta_d e = \te^T F^{-T} \eta_d F^{-1} \te \\ b = - \te^T F^{-T} \eta_d \te \b \te^T \eta_d F^{-1} \te \end{array}\ \ ,\ \ {\rm with}\ F=\id + \te \b \te^T \eta_d \label{defF}
\eeq
\beq
\Leftrightarrow\ e = k F^{-1} \te \ , \ e^{-T} b e^{-1} = - k^{-T} \eta_d \te \b \te^T \eta_d k^{-1} \ ,\ \ {\rm with}\ k^T \eta_d k =\eta_d \ . \label{ete}
\eeq
A little algebra then allows to show that $K$ defined in \eqref{defKgenviel} can be written as in \eqref{defKgen}, with
\beq
O_1=k F^{-1} \ , \ O_2=k (F^{-1} - \id) \ . \label{O1O2}
\eeq
Interestingly, the field redefinition that we used to obtain this result is equivalent to having $K \in O(2d-2,2)$. Therefore, the properties \eqref{propO1O2} should be automatically satisfied with \eqref{O1O2}. It is indeed the case: when using \eqref{O1O2}, they boil down to the condition
\beq
2 F^{-T} \eta_d F^{-1} = \eta_d F^{-1} + F^{-T} \eta_d \ \Leftrightarrow\  2 \eta_d = F^{T} \eta_d + \eta_d F \ ,
\eeq
which is true given the definition of $F$. To conclude, we have shown that the transformation taking us from $\eee$ to $\teee$ and realising the field redefinition is given by the matrix $K$ in \eqref{defKgen} with the entries \eqref{O1O2}, and it satisfies the properties \eqref{propO1O2}.

The fact that $K \in O(d,d)$ is also important as it acts on the $O(d,d)$ index of the generalized vielbeins. As such, it can then be rotated as described in \eqref{rot}. One obtains the simple result
\beq
P^{-T} K P^{T} = \begin{pmatrix} O_+ & 0 \\ 0 & O_- \end{pmatrix} \ .
\eeq
This result makes it obvious that this transformation is an $O(d-1,1)\times O(1,d-1)$ \cite{Grana:2008yw}, thanks to the equivalence \eqref{propO1O2}. Additionally, it coincides with the $O(d-1,1)\times O(1,d-1)$ structure we want to preserve in section \ref{sec:OOstruct}. As a side remark, note though that it does not survive the alignment of vielbeins we impose there, as $O_+ \neq O_-$ a priori. This is expected because this transformation does not even preserve the form of the generalized vielbeins (by definition), i.e. one has $K \notin G_{{\rm split}}$ for either of the two frames \eqref{Bsplit} and \eqref{betasplit}. More precisely for this particular $K$,
\beq
O_+=O_- \ \Leftrightarrow\ F=\id \ \Leftrightarrow\ \b=0 \ \Leftrightarrow\ b=0 \ ,
\eeq
which is indeed the only case where the form of the generalized vielbeins is preserved (they are actually the same, up to $k$).

\subsection{Computation of $S$}\label{ap:S}

In this appendix, we compute explicitly the quantity $S$ as given in \eqref{S1}, using the definitions of section \ref{sec:OOstruct}, analogously to \cite{Hohm:2011nu}. As explained below \eqref{S1}, the first three lines of this expression should vanish: let us first detail the verification of this point. To start with, we compute, using \eqref{Cartan}, \eqref{g1}, and the alignment of vielbeins
\bea
& \gamma^a \gamma^b (\del_a + \eta_{ad} \b^{de} \del_e) (\del_b + \eta_{bc} \b^{cf} \del_f) - \ov{\eta^{ab}}(\del_{\ov{a}} - \ov{\eta_{ad}} \b^{\ov{de}} \del_{\ov{e}}) (\del_{\ov{b}} - \ov{\eta_{bc}} \b^{\ov{cf}} \del_{\ov{f}}) \\
=& \gamma^{ab} \Bigg( \frac{1}{2} f^f{}_{ab} + \eta_{ad} \b^{de} f^f{}_{eb} - \eta_{ad} \del_b (\b^{df}) + \eta_{ad} \eta_{bc} \b^{de} \left(\del_e(\b^{cf}) + \frac{1}{2} \b^{cg} f^f{}_{eg} \right) \Bigg) \del_f + 2 \del_c (\b^{cf}) \del_f\ . \nn
\eea
One should then verify that
\bea
\!\!\!\!\!\! 0&= 2 \del_c (\b^{cf}) \del_f + 2 \eta^{ac} X_c (\del_a + \eta_{ad} \b^{de} \del_e) - Z_{\ov{a}} \ov{\eta^{ab}} (\del_{\ov{b}} - \ov{\eta_{bc}} \b^{\ov{cf}} \del_{\ov{f}})\ , \label{0g}\\
\!\!\!\!\!\! 0&= \gamma^{ab} \Bigg( \frac{1}{2} f^f{}_{ab} \del_f + \eta_{ad} \b^{de} f^f{}_{eb} \del_f - \eta_{ad} \del_b (\b^{df}) \del_f + \eta_{ad} \eta_{bc} \b^{de} \left(\del_e(\b^{cf})  + \frac{1}{2} \b^{cg} f^f{}_{eg} \right) \del_f  \nn\\
& \phantom{= \gamma^{ab} \Bigg(}  + 6 \eta^{ce} X_{[eab]} (\del_c + \eta_{cd} \b^{df} \del_f) - 2 \ov{\eta^{de}} Y_{\ov{d}ab} (\del_{\ov{e}} - \ov{\eta_{ec}} \b^{\ov{cf}} \del_{\ov{f}}) \Bigg)\ . \label{0gg}
\eea
To prove \eqref{0g}, it is useful to recall that $\Lambda$ was given in \eqref{Lambda}, and that one can rewrite $\xi$ from \eqref{xiT} as
\beq
\xi^d = \b^{de}\lambda_e - 2 \T^d \ .
\eeq
Using \eqref{troQ}, \eqref{Ta}, and \eqref{prop}, one then verifies \eqref{0g}. To prove \eqref{0gg}, one can decompose it into the terms having no, one, two, or three $\beta$, and show that they vanish separately. The antisymmetry of the $a,b$ indices and the properties of $\o$ and $\o_Q$ are useful, together with the alignment of vielbeins and \eqref{RfQf}, to prove the cancellation in \eqref{0gg}.

We are then left with the last three lines of \eqref{S1}. Using the identities \eqref{g2}, \eqref{acomga3} and \eqref{acomga2}, one can rewrite these lines, and therefore $S$, as
\bea
- \frac{1}{4} & S \epsilon^+ \label{S2}\\
= & \Bigg[ (\gamma^{abcf}+3\eta^{a[b} \gamma^{cf]}) (\del_a + \eta_{ad} \b^{de} \del_e) (X_{bcf}) + \gamma^a \gamma^c (\del_a + \eta_{ad} \b^{de} \del_e) (X_c) \nn \\
& \!\!\! +\frac{1}{2}X_{ade}X_{bcf}(18\delta^{[b}_{[g}\gamma_{hi]}{}^{cf]}\eta^{ag}\eta^{dh}\eta^{ei}-12\delta_{[g}^{b} \delta_{h}^{c}\delta_{i]}^{f}\eta^{ag}\eta^{dh}\eta^{ei}) + X_{ade}X_c (6 \eta^{c[a} \gamma^{de]}) + X_aX_c \gamma^a\gamma^c\nn \\
& \!\!\! - \ov{\eta^{ab}}(\del_{\ov{a}} - \ov{\eta_{ad}} \b^{\ov{de}} \del_{\ov{e}}) (Y_{\ov{b}cf}) \gamma^{cf}  - \frac{1}{2}\ov{\eta^{ab}} Y_{\ov{a}de} Y_{\ov{b}cf} (2\gamma^{decf}-4 \eta^{c[d}\eta^{e]f}) - Z_{\ov{a}} \ov{\eta^{ab}} Y_{\ov{b}cf} \gamma^{cf} \Bigg] \epsilon^+ \nn \ .
\eea
To compute this expression, we decompose it into the various orders of antisymmetric products of $\gamma$ matrices. The zeroth order will give the scalar of interest, while the higher orders (two and four $\gamma$'s) will vanish. This is consistent with the idea of $S$ being a scalar. The following identities will be helpful to show the vanishing of the terms at order $\gamma^{ab}$, and $\gamma^{abcd}$
\bea
\del_{[a}f^{e}{}_{bf]}=&f^{e}{}_{d[a}f^{d}{}_{bf]}\ ,\label{delf}\\
\del_{[a}Q_{f]}{}^{de}-\b^{g[d}\del_g f^{e]}{}_{af}=&\frac{1}{2}Q_{g}{}^{de}f^{g}{}_{af}-2Q_{[a}{}^{g[d}f^{e]}{}_{f]g}\ ,\label{delQ}\\
\del_a R^{ghi} -3\b^{d[g}\del_d Q_{a}{}^{hi]}=&- 3 R^{d[gh}f^{i]}{}_{ad}+3Q_{a}{}^{d[g}Q_{d}{}^{hi]}\ ,\label{delR}\\
\b^{g[d}\del_g R^{abc]}=&-\frac{3}{2}R^{g[da}Q_{g}{}^{bc]}\ .\label{bdelR}
\eea

Let us start with the terms at order $\gamma^{ab}$. We proceed by using the explicit expressions for $X_{ade},\ X_{a},\ Y_{\ov{b}cf}\ \text{and}\ Z_{\ov{a}}$, the alignment of vielbeins, and computing separately the terms at each order in $\b$. At zeroth order in $\b$ the following equation holds thanks to \eqref{delf} with two indices contracted
\bea
0=&\frac{3}{4}\eta^{a[b} \gamma^{cf]} \eta_{ce}\del_a  \o^{e}_{bf}+\frac{1}{2} \gamma^{ac}\del_a  (\o^{g}_{gc}-\la_c)+ \frac{3}{4} \eta_{dh}\o^{h}_{ae}(\o^{g}_{gc}-\la_c)  \eta^{c[a} \gamma^{de]}\\
&- \frac{1}{4}\ov{\eta^{ab}}\eta_{ce}\del_{\ov{a}}  \o^{e}_{\ov{b}f}\gamma^{cf}- \frac{1}{4}(\o^{\ov{g}}_{\ov{ga}}-\la_{\ov{a}})\ov{\eta^{ab}} \eta_{ce} \o^{e}_{\ov{b}f} \gamma^{cf}\ .\nn
\eea
At first order in $\b$, we make use of \eqref{delf} and \eqref{delQ} with two indices contracted to show
\bea
0=&\frac{3}{4}\eta^{a[b} \gamma^{cf]} (-\eta_{ce}\eta_{bh}\del_a {\o_Q}^{he}_{f} + \eta_{ad}\eta_{cg} \b^{de} \del_e\o^{g}_{bf}) \\
&+ \frac{1}{2}\gamma^{ac}(\eta_{ce}\del_a ({\o_Q}^{de}_{d}-\xi^{e})+ \eta_{ad} \b^{de} \del_e (\o^{g}_{gc}-\la_c)) \nn\\
&+\frac{3}{4} (\eta_{dh}\o^{h}_{ae}\eta_{cg}({\o_Q}^{dg}_{d}-\xi^{g}) -\eta_{dg}\eta_{ah}{\o_Q}^{hg}_{e}(\o^{g}_{gc}-\la_c)) \eta^{c[a} \gamma^{de]}\nn\\
&- \frac{1}{4}(\eta_{ce}\del_{\ov{a}}{\o_Q}^{\ov{a}e}_{f} - \eta_{ce} \b^{\ov{be}} \del_{\ov{e}}\o^{e}_{\ov{b}f})  \gamma^{cf} \nn\\
&- \frac{1}{4}(\o^{\ov{g}}_{\ov{ga}}-\la_{\ov{a}}) \eta_{ce}{\o_Q}^{\ov{a}e}_{f} \gamma^{cf}+\frac{1}{4}({\o_Q}^{\ov{db}}_{\ov{d}}-\xi^{\ov{b}}) \eta_{ce}\o^{e}_{\ov{b}f} \gamma^{cf}\ .\nn
\eea
At second order in $\b$, we verify using \eqref{delQ} and \eqref{delR} with two indices contracted
\bea
0=&\frac{3}{24}\eta^{a[b} \gamma^{cf]} (\eta_{be}\eta_{cg}\eta_{fh}\del_a R^{egh}-6\eta_{cg}\eta_{bh} \eta_{ad} \b^{de} \del_e {\o_Q}^{hg}_{f}) \\
&+\frac{1}{2} \gamma^{ac} \eta_{ad} \b^{de} \del_e \eta_{cg}({\o_Q}^{dg}_{d}-\xi^{g}) \nn\\
&- \frac{1}{8}(6\eta_{dg} \eta_{ah} {\o_Q}^{hg}_{e}\eta_{cf}({\o_Q}^{bf}_{b}-\xi^{f})- \eta_{af}\eta_{dg}\eta_{eh}R^{fgh}(\o^{b}_{bc}-\la_c))  \eta^{c[a} \gamma^{de]} \nn\\
&+ \frac{1}{8}(\eta_{cg}\eta_{fh}\del_{\ov{a}}R^{\ov{a}gh} + 2\eta_{cg}\ov{\eta_{ad}} \b^{\ov{de}} \del_{\ov{e}} {\o_Q}^{\ov{a}g}_{f}) \gamma^{cf} \nn\\
&+ \frac{1}{8}(\o^{\ov{d}}_{\ov{da}}-\la_{\ov{a}}) \eta_{cg}\eta_{fh} R^{\ov{a}gh} \gamma^{cf}+ \frac{1}{4}\ov{\eta_{ag}}({\o_Q}^{\ov{dg}}_{\ov{d}}-\xi^{\ov{g}}) \eta_{ce} {\o_Q}^{\ov{a}e}_{f}\gamma^{cf}\ .\nn
\eea
The terms at third order in $\b$ vanish without using any of the above identities
\bea
0=&\frac{3}{24}\eta^{a[b} \gamma^{cf]}  \eta_{bg}\eta_{ch}\eta_{fi}\eta_{ad} \b^{de} \del_e R^{ghi}+\frac{1}{8} \eta_{af}\eta_{dg}\eta_{eh}R^{fgh}\eta_{ci}({\o_Q}^{bi}_{b}-\xi^{i}) \eta^{c[a} \gamma^{de]}\\
&-\frac{1}{8} \eta_{ch}\eta_{fi}\ov{\eta_{ad}} \b^{\ov{de}} \del_{\ov{e}} R^{\ov{a}hi} \gamma^{cf}- \frac{1}{8}\ov{\eta_{ag}}({\o_Q}^{\ov{dg}}_{\ov{d}}-\xi^{\ov{g}}) \eta_{cg}\eta_{fh} R^{\ov{a}gh} \gamma^{cf}\ , \nn
\eea
which concludes our verification that all terms in $\gamma^{ab}$ vanish.

We now turn to the terms coming with an antisymmetric product of four $\gamma$ matrices. For these, we first use
\beq
X_{abc}X_{def}9\delta^{[d}_{[g}\gamma_{hi]}{}^{ef]}\eta^{ag}\eta^{bh}\eta^{ci}=X_{abc}X_{def}(\eta^{ad}\g^{bcef}+4\eta^{ae}\g^{bcfd}+4\eta^{be}\g^{cafd}) \ , \label{gamma4}
\eeq
and the explicit expressions of $X_{bcf}\ \text{and}\ Y_{\ov{b}cf}$. We then show that the resulting expression vanishes order by order in $\b$, using the alignment of vielbeins. Starting at zeroth order in $\b$, we have to prove
\bea
0 & =\frac{1}{4}\eta_{ce}\del_a \o^{e}_{bf}\gamma^{abcf} \\
& \ + \frac{1}{16}\eta_{bg}\o^{g}_{ac}\eta_{eh}\o^{h}_{df}(\eta^{ad}\g^{bcef}+4\eta^{ae}\g^{bcfd}+4\eta^{be}\g^{cafd}) -\frac{1}{16}\ov{\eta^{ad}}\eta_{bg}\o^{g}_{\ov{a}c}\eta_{eh}\o^{h}_{\ov{d}f}\g^{bcef} \ .\nn
\eea
This can be verified, thanks to \eqref{delf}. At first order in $\b$, we use \eqref{delQ} to show that
\bea
0& = -\frac{1}{4}\eta_{ce}(\eta_{bg}\del_a {\o_Q}^{ge}_{f}-\eta_{ad} \b^{dg} \del_g \o^{e}_{bf})\gamma^{abcf} -\frac{1}{16}(\eta_{bg}\o^{g}_{\ov{a}c}\eta_{eh}{\o_Q}^{\ov{d}h}_{f}+\eta_{bh}{\o_Q}^{\ov{a}h}_{c}\eta_{eg}\o^{g}_{\ov{a}f})\g^{bcef} \nn\\
&\ -\frac{1}{16}(\eta_{bg}\o^{g}_{ac}\eta_{eh}\eta_{di}{\o_Q}^{ih}_{f}+\eta_{bh}\eta_{ai}{\o_Q}^{ih}_{c}\eta_{eg}\o^{g}_{df})(\eta^{ad}\g^{bcef}+4\eta^{ae}\g^{bcfd}+4\eta^{be}\g^{cafd}) \ .
\eea
At second order in $\b$, we verify using \eqref{delR}
\bea
0& = \frac{1}{24}(\eta_{bg}\eta_{ch}\eta_{fi}\del_a
R^{ghi}-6\eta_{ch}\eta_{bg} \eta_{ad} \b^{de}
\del_e{\o_Q}^{gh}_{f})\gamma^{abcf}\\
&\ +\frac{1}{96} \Big(6\eta_{bh}\eta_{ag}{\o_Q}^{gh}_{c}\eta_{ej}\eta_{di}{\o_Q}^{ij}_{f}\nn
\\
&\phantom{\ +\frac{1}{96} (}\  +\eta_{bg}\o^{g}_{ac}\eta_{dh}\eta_{ei}\eta_{fj}R^{hij} +\eta_{ag}\eta_{bh}\eta_{ci}R^{ghi}\eta_{ej}\o^{j}_{df}\Big) (\eta^{ad}\g^{bcef}+4\eta^{ae}\g^{bcfd}+4\eta^{be}\g^{cafd})\nn
\\
&\ -\frac{1}{32}(2\ov{\eta_{ad}}\eta_{bg}{\o_Q}^{\ov{a}g}_{c}\eta_{eh}{\o_Q}^{\ov{d}h}_{f}-\eta_{bg}\o^{g}_{\ov{a}c}\eta_{eh}\eta_{fi}R^{\ov{a}hi}-\eta_{bg}\eta_{ch}R^{\ov{a}gh}\eta_{ei}\o^{i}_{\ov{a}f})\g^{bcef}\nn
\ .
\eea
At third order in $\b$, we show using \eqref{bdelR} that
\bea
0& =\frac{1}{24} \eta_{be}\eta_{ci}\eta_{fh}\eta_{ad} \b^{dg} \del_g R^{eih}\gamma^{abcf}\\
&\ -\frac{1}{96}(\eta_{bh}\eta_{ag}{\o_Q}^{gh}_{c}\eta_{di}\eta_{ej}\eta_{fk}R^{ijk}+\eta_{ag}\eta_{bh}\eta_{ck}R^{ghk}\eta_{ej}\eta_{di}{\o_Q}^{ij}_{f})(\eta^{ad}\g^{bcef}+4\eta^{ae}\g^{bcfd}+4\eta^{be}\g^{cafd})\nn \\
&\ +\frac{1}{32}\ov{\eta_{ad}}(\eta_{bh}{\o_Q}^{\ov{a}h}_{c}\eta_{ej}\eta_{fg}R^{\ov{d}jg}+\eta_{bh}\eta_{cg}R^{\ov{a}hg}\eta_{ej}{\o_Q}^{\ov{d}j}_{f})\g^{bcef}\nn \ .
\eea
Finally, the forth order in $\b$ vanishes as follows
\bea
0& =\frac{1}{576}\eta_{ag}\eta_{bh}\eta_{ci}R^{ghi}\eta_{dj}\eta_{ek}\eta_{fl}R^{jkl}(\eta^{ad}\g^{bcef}+4\eta^{ae}\g^{bcfd}+4\eta^{be}\g^{cafd})\\
&\ -\frac{1}{64}\ov{\eta_{ad}}\eta_{bh}\eta_{ci}R^{\ov{a}hi} \eta_{eg}\eta_{fj}R^{\ov{d}gj}\g^{bcef}\nn \ .
\eea
We thus have shown that all terms in $\g^{abcd}$ vanish. From $\eqref{S2}$, we are then left only with terms without any $\gamma$. We compute them and finally get
\bea
-\frac{1}{4} S \epsilon^+ =& \Bigg[\frac{1}{2}\eta^{ac}\del_a \o^g_{gc}+\frac{1}{4}\eta^{ac}\o^d_{da}\o^g_{gc}-\frac{1}{4}\eta^{eb}\omega^h_{ae}\omega^a_{bh}-\frac{1}{2}\eta^{ac}\na_a \lambda_c+\frac{1}{4}\eta^{ac}\lambda_a\lambda_c \label{Sfinalap} \\
&+\frac{1}{2}\eta_{ac}\b^{ae} \del_e {\o_Q}_g^{gc} +\frac{1}{4}\eta_{hg}{\omega_{Q}}_e^{fh}  {\omega_{Q}}_f^{eg} +\frac{1}{4} \eta_{cg}{\o_Q}_d^{dc}  {\o_Q}_f^{fg} +\frac{1}{4}\eta_{fg}\omega^f_{ae}  R^{aeg}\nn \\
&+\frac{1}{2}\eta_{ab}\cN^a \xi^b+\frac{1}{4}\eta_{ab}\xi^{a}\xi^{b} -\frac{1}{48} \eta_{ec}\eta_{bh} \eta_{fg} R^{bfc} R^{heg} \Bigg] \epsilon^+ \nn \\
=&-\frac{1}{4} \Bigg(\R(\tg) + \R_Q - \frac{1}{2} R^{acd} f^b{}_{cd} \eta_{ab} -\frac{1}{2} R^2 \nn\\
& \phantom{-\frac{1}{4} \ \ } -4(\del \tp)^2 + 4\na^2 \tp - 4 (\b^{ab}\del_b \tp - \T^a)^2 -4\eta_{ab}\cN^a (\b^{bc}\del_c \tp - \T^b) \Bigg) \epsilon^+ \ , \nn
\eea
where the last line, given also in \eqref{Sfinalsec}, is obtained using \eqref{Ricflat}, \eqref{defRQ} and \eqref{xi}.

\section{Derivation of the equation of motion for $\b$}\label{ap:eombeta}

In this appendix, we detail the variation of $\tL_{\b}$ \eqref{L2} with respect to $\b$, leading to the equation of motion for this field, as discussed in section \ref{sec:beom}. This variation has three contributions, given in \eqref{dL2db}; while the dilaton term is studied in section \ref{sec:beom}, we consider here the other two, and start with $\delta \cR(\tg)$.

To work-out this first variation, the rewriting of $\cG$ as in \eqref{cG_t} is of great help. As a start, one gets from it
\beq
\delta \cG^{mn}_p= \delta \cG_{\!\!(t)}{}_p^{mn} + \G^n_{ps} \delta \b^{ms} \ , \ \delta \cG_{\!\!(t)}{}_p^{mn}= \frac{1}{2} \tg_{pq} \left( \tg^{rm} \na_r \delta \b^{nq} + \tg^{rn} \na_r \delta \b^{mq} - \tg^{qr} \na_r \delta \b^{mn} \right) \ , \label{dcG_t}
\eeq
from which one can obtain, as discussed in section \ref{sec:beom}, a first expression for $\delta \cR^{mn}$. It is given by \eqref{dcRmndb} that we repeat here for convenience
\bea
\delta \cR^{mn} & = \cN^p \left(\delta \cG_{\!\!(t)}{}_p^{mn}\right) - \cN^m \left(\delta \cG_{\!\!(t)}{}_p^{pn} \right) \label{dcRmndbap} \\
& -(\delta \b^{pr})\del_r \cG_p^{mn}+(\delta \b^{mr})\del_r \cG_p^{pn} + 2\cG_r^{pn} \delta \cG_p^{[mr]} \nn \\
& - \cG^{pm}_r \G^n_{ps} \delta \b^{rs} - \left(\cG^{pn}_r \G^r_{ps} - \cG^{rp}_r \G^n_{ps} \right) \delta \b^{ms} - \b^{pr} \del_r \left(\G^n_{ps} \delta \b^{ms} \right) \ . \nn
\eea
The terms in $\G$ are obtained in the variation of $\cR^{mn}$ by decomposing the variation of $\cG$ into $\delta \cG_{\!\!(t)}$ and the second term involving $\G$, as in \eqref{dcG_t}. Note also the presence of the antisymmetric part $\delta \cG_p^{[mr]}$, that we mentioned to be vanishing for the variation with respect to $\tg$ (see below \eqref{NderEinsteincR}), but not for $\b$. Here we obtain from \eqref{cG} that $\delta \cG_p^{[mr]}= -\frac{1}{2} \del_p \delta \b^{mr}$.

Using this, and decomposing $\cG$ as in \eqref{cG_t}, one can rewrite the second line of \eqref{dcRmndbap} as
\beq
-(\delta \b^{pr})\del_r \cG_{\!\!(t)}{}_p^{mn}+(\delta \b^{mr})\del_r \cG_{\!\!(t)}{}_p^{pn} -(\delta \b^{pr})\del_r (\b^{ms} \G^n_{ps}) - \cG_{\!\!(t)}{}_r^{pn} \del_p \delta \b^{mr} - \b^{ps} \G^n_{sr} \del_p \delta \b^{mr} \ .\label{2ndline}
\eeq
The last term of this expression cancels the $\del \delta \b$ coming from the very last term of \eqref{dcRmndbap}. The other terms in \eqref{2ndline} give one term in $\del \G$, and four others with derivatives of tensors. One should then complete these four derivatives into covariant derivatives. In particular, $\na_r \cG_{\!\!(t)}{}_p^{mn}$ contains a term $\G^n_{rs} \cG_{\!\!(t)}{}_p^{ms}$ that is worth rewriting as $\G^n_{rs} (\cG_{\!\!(t)}{}_p^{sm} + 2 \cG_{\!\!(t)}{}_p^{[ms]} )$. Indeed, one then has from \eqref{cG_t} that $2 \cG_{\!\!(t)}{}_p^{[ms]}=-\na_p \b^{ms}$. This last contribution then cancels the $\na \b$ obtained by completing the $\del \b$ in \eqref{2ndline}.

By decomposing the $\cG$ of the third line of \eqref{dcRmndbap} as in \eqref{cG_t}, then multiplying $\delta \cR^{mn}$ by $\tg_{mn}$ and using its symmetry, as well as the antisymmetry of some $\delta \b^{pr}$, simplifications occur. In addition, the terms in $\cN \delta \cG_{\!\!(t)}{}$ or $\na \cG_{\!\!(t)}{}$ become simple when multiplied by $\tg_{mn}$, using the compatibility of the metric with both covariant derivatives and the expressions \eqref{cG_t} and \eqref{dcG_t}. One is finally left with
\bea
\delta \cR & = 2\cN^p\na_n (\tg_{pq} \delta \b^{nq}) - \cG_{\!\!(t)}{}_r^{pn} \na_p (\tg_{mn} \delta \b^{mr}) \\
& + \delta \b^{pr} \Big( 2 \tg_{pn} \na_r \na_q \b^{nq} \nn\\
& \phantom{+ \delta \b^{pr} }\ - \b^{ms} (\tg_{mn} \del_r \G^n_{ps} + \tg_{pn} \del_s \G^n_{mr} ) + \tg_{mn} \G^n_{ps} \G^s_{rq} \b^{mq} - \tg_{pn} \G^n_{qs} \G^s_{mr} \b^{mq} \Big) \ .\nn
\eea
This expression should be a tensor; while it is clearly the case of the first two lines, the last line is not obviously tensorial. This leads us to look for the possible tensors which could fit there, and it turns out to involve the standard Riemann tensor, given in \eqref{Riemann}. Indeed, one gets using the antisymmetry of the two $\b$
\bea
& \delta \b^{pr} \left( - \b^{ms} (\tg_{mn} \del_r \G^n_{ps} + \tg_{pn} \del_s \G^n_{mr} ) + \tg_{mn} \G^n_{ps} \G^s_{rq} \b^{mq} - \tg_{pn} \G^n_{qs} \G^s_{mr} \b^{mq} \right) \\
& = \delta \b^{pr} \b^{ms} \frac{1}{2} \left(\tg_{np} \R^n{}_{rms} + \tg_{ns} \R^n{}_{mrp} \right) \ .\nn
\eea
Thanks to the properties of the Riemann tensor \eqref{Riemann}, the two terms turn out to be equal, giving finally \eqref{dcRfinal1} that we repeat here
\beq
\delta \cR  = 2\cN^n\na_p (\tg_{nr} \delta \b^{pr}) - \cG_{\!\!(t)}{}_r^{mn} \na_m (\tg_{pn} \delta \b^{pr}) + \delta \b^{pr} \Big( 2 \tg_{pn} \na_r \na_q \b^{nq} + \b^{ms} \tg_{np} \R^n{}_{rms} \Big) \ . \label{dcRfinal1ap}
\eeq
We comment on this Riemann tensor below \eqref{dcRfinal1}. The identity
\beq
(\na_m \na_n - \na_n \na_m) \b^{pq} = \R^p{}_{smn} \b^{sq} + \R^q{}_{smn} \b^{ps} \label{comnab}
\eeq
indicates that we cannot get completely rid of the Riemann tensor in $\delta \cR$. Using its properties \eqref{Riemann}, one can still show that
\beq
\R^n{}_{rms} = 2 \R^n{}_{[sm]r} \ , \ \b^{ms} \R^n{}_{rms} = - 2 \b^{sq} \R^n{}_{sqr} \ ,
\eeq
so using \eqref{comnab}, we get\footnote{One also has $(\na_r \na_q - \na_q \na_r) \b^{nq} = e^{2\tp}(\na_r \na_q - \na_q \na_r) (e^{-2\tp}\b^{nq})$. Two $\na$ on a scalar commute.}
\beq
\b^{ms} \R^n{}_{rms} = 2 (\na_r \na_q - \na_q \na_r) \b^{nq} + 2 \R_{sr} \b^{ns}  \ .\label{RieRic}
\eeq
Although \eqref{dcRfinal1ap} is not simplified by this identity, its combination with the dilaton term \eqref{dilvarb} will be. This will allow to trade the Riemann tensor for the Ricci one.

We now do an ``integration by parts'' on the first two terms of \eqref{dcRfinal1ap} multiplied by $e^{-2d}$. To do so, we use the Leibniz rule for both covariant derivatives, and \eqref{tracecN} to get total derivatives. We also use \eqref{identkindilT}, the definition \eqref{cG_t} of $\cG_{\!\!(t)}{}$, and symmetry properties. We get
\bea
e^{-2d} \delta \cR  &= \del(\dots) + e^{-2d} \delta \b^{pr} \tg_{np} \Big( \b^{ms} \R^n{}_{rms} + \frac{1}{2} \tg_{rq} \tg^{sm} e^{2\tp} \na_m (e^{-2\tp} \na_s \b^{nq} ) \label{dcRfinal2}\\
& \phantom{= \del(\dots) + e^{-2d} \delta \b^{pr} \tg_{np}} + e^{2\tp} \na_q (e^{-2\tp} \na_r \b^{nq} ) + 2 \na_r \na_q \b^{nq} + 4 e^{2\tp} \na_r (e^{-\tp} \na_q (e^{-\tp} \b^{nq} ) ) \Big) \ .\nn
\eea
It is worth noting that we can rewrite
\beq
4 e^{2\tp} \na_r (e^{-\tp} \na_q (e^{-\tp} \b^{nq} ) ) = 8 e^{\tp} \na_r \na_q (e^{-\tp} \b^{nq} ) - 4 \na_r \na_q \b^{nq} + 4 \na_r (\b^{nq} \del_q \tp ) \ .
\eeq
Indeed, the first term in the right-hand side above is the (opposite of) the one involved in \eqref{dilvarb}, giving
\bea
 \delta \left( \cR +  4 (\b^{mp}\del_p \tp - \T^m)^2 \right)  + & e^{2d} \del(\dots) = \delta \b^{pr} \tg_{np} \Big(\ \frac{1}{2} \tg_{rq} \tg^{sm} e^{2\tp} \na_m (e^{-2\tp} \na_s \b^{nq} ) \label{dcRfinal3} \\
&  + \b^{ms} \R^n{}_{rms} + e^{2\tp} \na_q (e^{-2\tp} \na_r \b^{nq} )  - 2 \na_r \na_q \b^{nq} + 4 \na_r (\b^{nq} \del_q \tp ) \Big) \ .\nn
\eea
This allows to use \eqref{RieRic} and trade the Riemann for the Ricci tensor, to finally get the expression \eqref{dcRfinal4} of this combination of two variations.

Eventually, we consider the variation of the $R$-flux term. Using the antisymmetry properties of the $R$-flux, and the fact we can use a $\na$ in its definition \eqref{fluxes}, one gets
\beq
\delta R^2 = R^{suv} \tg_{sp} \tg_{um} \left( (\delta \b^{rp}) \tg_{vq} \na_r \b^{mq} + \b^{qm} \tg_{vr} \na_q \delta \b^{rp} \right) \ .
\eeq
Multiplying this by $e^{-2d}$, one then performs the/an ``integration by parts'' on the second term. One gets after some rearranging
\bea
-\frac{1}{2} e^{-2d} \delta R^2 & = \frac{1}{2} e^{-2d} \delta \b^{pr} \Big(\b^{mq} e^{2\tp} \na_q (e^{-2\tp} \tg_{ms} \tg_{ru} \tg_{pv} R^{suv})\label{dR2}\\
& \phantom{= \frac{1}{2} e^{-2d} \delta \b^{pr}}\ + \tg_{pv} \tg_{ms} R^{suv} (\tg_{ur} \na_q \b^{mq} + \tg_{uq} \na_r \b^{mq} ) \Big) + \del(\dots) \ .\nn
\eea
An alternative derivation and formulation of this variation is given by the rewriting of the $R$-flux as in \eqref{RfluxcN}. Using it and \eqref{cG_t}, \eqref{relnacN}, one gets
\beq
\delta R^{mnp} = \frac{3}{2} \ \left( \cN^{[m} \delta \b^{np]} + (\delta \b^{q[m})\na_q \b^{np]} + \b^{q[m}\na_q \delta \b^{np]} \right) \ .
\eeq
From this, using the Leibniz rule and \eqref{tracecN}, one ends up with \eqref{dR22} that we repeat here
\beq
-\frac{1}{2} e^{-2d} \delta R^2  = -\frac{1}{2} e^{-2d} \delta \b^{pr} \tg_{ms} \tg_{ru} \tg_{pv} \left( e^{2\tp} \cN^m (e^{-2\tp}  R^{suv}) - 2 \T^m R^{suv}\right) + \del(\dots) \ .\label{dR22ap}\\
\eeq
One can verify that \eqref{dR22ap} matches with \eqref{dR2}, using \eqref{relnacN} to write
\beq
\tg_{ms} \cN^m R^{suv} = - \tg_{ms} \b^{mq} \na_q R^{suv} - \tg_{ms} \cG_{\!\!(t)}{}_n^{ms} R^{nuv} - 2 \tg_{ms} \cG_{\!\!(t)}{}_n^{m[v} R^{u]ns} \ ,
\eeq
and then from \eqref{cG_t} to get the symmetry properties of $\tg_{ms} \cG_{\!\!(t)}{}_n^{mv}$. We will use this formulation \eqref{dR22ap} of the variation. It may be simpler than \eqref{dR2} if the resulting equation is turned to flat indices, and the fluxes are identified. Note also that such a reformulation with $\cN$ cannot occur for the other variations, as they are linear in $\b$.

Compiling all results, namely \eqref{dcRfinal4} and \eqref{dR22}, one obtains for $\delta \tL_{\b}$ the expression \eqref{dL2bfinal}, and the equation of motion for $\b$ given in \eqref{beom}.\\

Using the simplifying assumption $\b^{pq} \del_q =0\ , \ \del_q \b^{pq}=0$, the equation \eqref{beom} should reduce to the one given in \cite{Andriot:2011uh}, namely
\beq
\del_m \left( \sqrt{|\tg|} e^{-2\tp} \tg^{mq} \tg_{nr} \tg_{ps} \del_q \b^{rs} \right) =0 \ . \label{beomassumption}
\eeq
Let us discuss this point for a constant dilaton. From the definition \eqref{fluxes}, one gets $R^{mnp}=0$ under the simplifying assumption. Therefore, the first line of \eqref{beom} would vanish. The equation \eqref{beomassumption} can clearly be recovered from the first term in the last line of \eqref{beom}. However this term generates other contributions. It is rather non-trivial that the latter cancel with the rest of this last line (which does not vanish by itself). We do not fully check this cancellation here. Nevertheless, a simpler manner to verify this is probably to use the alternative formulation of the last line of \eqref{beom} given by \eqref{dcRfinal3}. Under the simplifying assumption, one verifies from \eqref{tracecG} that $\T^n=\na_q \b^{nq}=0$, leaving essentially in \eqref{dcRfinal3} the study of the Riemann tensor term. Given its definition in \eqref{Riemann}, the contraction of this tensor with $\b$, and the assumption, cancel quite some terms; what is left to check looks then simpler.
Note that the two other equations of motion matching those of \cite{Andriot:2011uh} is much easier to verify.

The simplifying assumption of \cite{Andriot:2011uh} also reduces the expression \eqref{Q} of $Q_c{}^{ab}$ to
\beq
\te^p{}_c \te^a{}_m \te^b{}_n \del_p \b^{mn} \ . \label{Qabcsimplif}
\eeq
This observation helps to match the present results with the subcase treated in that paper.

\end{appendix}


\newpage

\providecommand{\href}[2]{#2}\begingroup\raggedright

\endgroup

\end{document}